\documentclass[pdftex,epjc3]{svjour3}

\usepackage{t1enc}
\usepackage[latin2,utf8]{inputenc}

\usepackage{graphicx}
\graphicspath{ {./figs/} }

\usepackage{soul}
\RequirePackage{graphicx}
\RequirePackage{mathptmx}      
\RequirePackage{flushend}
\RequirePackage[numbers,sort&compress]{natbib}
\RequirePackage[colorlinks,citecolor=blue,urlcolor=blue,linkcolor=blue]{hyperref}

\RequirePackage{amsfonts}
\RequirePackage{amsmath,amssymb}
\RequirePackage{bm}

\usepackage[utf8]{inputenc}

\usepackage[normalem]{ulem}

\usepackage[caption=false]{subfig}

\usepackage{amsmath}
\usepackage{commath}

\usepackage{graphicx,bm}
\usepackage{url}
\usepackage{float}
\usepackage{textcomp}
\usepackage{verbatim}
\usepackage{color}
\usepackage{wrapfig}
\usepackage{wasysym}

\usepackage{hyperref}
\hypersetup{colorlinks,urlcolor=black,citecolor=black,linkcolor=black,filecolor=black}
\usepackage{url}

\begin{document}
\markboth{}{}

%
%

\title{Observation of Odderon Effects at LHC energies -- \\
	A Real Extended Bialas-Bzdak Model Study}

\author{
        T. Cs\"{o}rg\H{o}\, \thanksref{e1,addr1,addr2}
        I. Szanyi\, \thanksref{e5,addr1,addr6}
}

\thankstext{e1}{e-mail: tcsorgo@cern.ch}
\thankstext{e5}{e-mail: istvan.szanyi@cern.ch}

\institute{
          Wigner FK, H-1525 Budapest 114, POB 49, Hungary\label{addr1}
          \and
           MATE Institute of Technology, K\'aroly R\'obert Campus, H-3200 Gy\"ongy\"os, M\'atrai \'ut 36, Hungary\label{addr2}
          \and
          E\"otv\"os University, H - 1117 Budapest, P\'azm\'any P. s. 1/A, Hungary\label{addr6}
}


\maketitle

\begin{abstract}
The unitarily extended Bialas-Bzdak model of elastic proton-proton scattering is applied, without modifications, to describe the differential cross-section of elastic proton-antiproton collisions in the TeV energy range, and to extrapolate these differential cross-sections to LHC energies. In this model-dependent study we find that the differential cross-sections of elastic proton-proton collision data at 2.76 and 7 TeV energies differ significantly from the differential cross-section of elastic proton-antiproton collisions extrapolated to these energies. The elastic proton-proton differential cross-sections, extrapolated to 1.96 TeV energy with the help of this extended Bialas-Bzdak model do not differ significantly from that of elastic proton-antiproton collisions, within the theoretical errors of the extrapolation. Taken together these results provide a model-dependent, but statistically significant evidence for a crossing-odd component of the elastic scattering amplitude at the at least 7.08 sigma level.  From the reconstructed Odderon and Pomeron amplitudes, we determine the $\sqrt{s}$ dependence of the corresponding total and differential cross-sections.
\end{abstract}



\section{Introduction\label{sec:introduction}}

Recently the TOTEM experiment measured  differential cross-sections of elastic proton-proton collisions in the TeV energy range, from $\sqrt{s} = 2.76$ through 7 and 8 to 13 TeV, together with the total, elastic and inelastic cross-sections and the real to imaginary ratio of the scattering amplitude at vanishing four-momentum transfer. These measurements provided surprizes and unexpected results. First of all, the shape of the differential cross-section of elastic scattering at $\sqrt{s} = 7$ TeV was different from all the predictions.

The total cross-section increases with increasing $\sqrt{s}$ according to theoretical expectations based on Pomeron-exchange, corresponding experimentally to the production of large rapidity gaps in high energy proton-proton and proton-antiproton collisions. These events correspond to large angular regions where no particle is produced. Their fraction, in particular the ratio of the elastic to the total proton-proton cross-section is increased above 25 \% at LHC energies.

In the language of quantum chromodynamics (QCD), the field theory of strong interactions, Pomeron-exchange
corresponds to the exchange of even number of gluons with vacuum quantum numbers.  In 1973, a crossing-odd counterpart to the Pomeron was
proposed by L. Lukaszuk and B. Nicolescu, the so-called Odderon~\cite{Lukaszuk:1973nt}. 
In QCD, Odderon exchange corresponds to the $t$-channel exchange of a color-neutral gluonic compound state 
consisting of an odd number of gluons, as noted by Bartels, Vacca and Lipatov in 1999~\cite{Bartels:1999yt}.

The Odderon effects remained elusive for a long time,  due to lack of a definitive and statistically significant
experimental evidence.

A direct way to probe the Odderon in elastic scattering is by comparing the differential 
cross-section of particle-particle and particle-antiparticle scattering at sufficiently high energies~\cite{Jenkovszky:2011hu,Ster:2015esa}.
Such a search was published at the ISR energy of $\sqrt{s}=53$ GeV in 1985~\cite{Breakstone:1985pe}, that
resulted in an indication of the Odderon, corresponding to a 3.35$\sigma$ significance level obtained from a simple
$\chi^2$ calculation, based on 
5 $pp$ and 5 $p\bar p$ data points in the 1.1 $\le $$ |t| $$ \le 1.5$ GeV$^2$ range (around the diffractive minimum).
This significance is smaller than the $5 \sigma$ threshold, traditionally accepted as a threshold for a discovery level
observation in high energy phyics.
Furthermore, the  colliding energy of $\sqrt{s} = 53$ GeV
was not sufficiently large so the  possible  Reggeon exchange effects were difficult to evaluate and control.
These difficulties rendered the Odderon search at the ISR energies rather inconclusive, but nevertheless inspiring and indicative,
motivating further studies.

In a series of recent papers, the TOTEM Collaboration published results with important implications for the Odderon search. These papers studied elastic proton-proton collisions in the LHC energy range between $\sqrt{s} = 2.76$ 
and $13$ TeV~\cite{Antchev:2017dia,Antchev:2017yns,Antchev:2018edk,Antchev:2018rec}. 
The total cross section, $\sigma_{\rm tot}(s)$ was found to increase, while the  real-to-imaginary ratio, $\rho_0(s)$, is found to decrease  with increasing energy, first identified 
at $\sqrt{s} = 13$ TeV \cite{Antchev:2017dia,Antchev:2017yns}.
These experimental results at vanishing four-momentum transfer were consistent with a possible 
Odderon effect and triggered an intense theoretical debate (see e.g.~Refs.~\cite{Khoze:2017swe,Samokhin:2017kde,Csorgo:2018uyp,Broilo:2018qqs,Pancheri:2018yhd,Goncalves:2018nsp,Selyugin:2018uob,Khoze:2018bus,Broilo:2018els,Troshin:2018ihb,Dremin:2018uwt,Martynov:2018nyb,Martynov:2018sga,Shabelski:2018jfq,Khoze:2018kna,Hagiwara:2020mqb,Contreras:2020lrh,Gotsman:2020mkd}). 
For example, Ref.~\cite{Gotsman:2018buo} demonstrated that such an indication at $t= 0$ is not a 
unique Odderon signal, as such a behaviour can be attributed to secondary Reggeon effects.
In spite of the rich experimental  results and the hot theoretical debate,
the Odderon remained rather elusive  at vanishing four-momentum transfer even in the TeV energy range ~\cite{Petrov:2018wlv} .

However, at larger four-momentum transfers, in the interference (diffractive dip and bump or minimum-maximum) region, the Odderon signals 
are significant at LHC energies. Let us mention here only two of them: the four-momentum transfer dependent nuclear slope parameter $B(t)$
and the scaling properties of elastic scattering at the TeV energy region. 

Two independent, but nearly simultaneous phenomenological papers suggested that
the four-momentum transfer dependence of the nuclear slope parameter, $B(t)$ is qualitatively different
in elastic proton-proton and proton-antiproton collisions~\cite{Csorgo:2018uyp,Martynov:2018sga}.
The TOTEM experiment has demonstrated in ref.~\cite{Antchev:2018rec} that indeed in elastic $pp$ collisions
at $\sqrt{s} = 2.76$ TeV, the nuclear slope $B(t)$ is increasing (swings) before it decreases and changes sign in the interference (diffractive dip and bump or minimum-maximum) region. After the diffractive maximum, the nuclear slope becomes positive again.
In contrast, elastic $p\bar p$ collisions measured by the D0 collaboration at the Tevatron energy of $\sqrt{s} = 1.96$ TeV
did not show such a pronounced diffractive minimum-maximum structure, instead an exponentially decreasing cone region at low $-t$
with a constant $B(t)$ is followed by a shoulder structure, without a pronunced diffractive minimum and maximum structure.
The TOTEM collaboration presented its results on  the elastic $pp$ differential cross-section at $\sqrt{s} = 2.76$ TeV and 
 concluded in ref.~\cite{Antchev:2018rec} that {\it ``under the condition that the effects due to the energy difference between TOTEM and D0 can be neglected,} these results
{\it provide evidence for a colourless 3-gluon bound state exchange in the t-channel of the proton-proton elastic scattering".}

This energy gap has been closed recently, in a model-independent way, based on a re-analysis of already published data using the scaling properties of elastic scattering in both $pp$ and $p\bar p$ collisions at TeV energies:
Refs.~\cite{Csorgo:2019ewn,Csorgo:2020msw,Csorgo:2020rlb} reported about a statistically significant Odderon signal in the comparison 
of the $H(x,s)$ scaling functions of elastic $pp$ collisions at $\sqrt{s} = 7.0 $ TeV to that 
of $p\bar p$ collisions at $\sqrt{s} = 1.96$ TeV.
 The difference between these scaling functions 
carries  an at least $6.26$ $\sigma$  Odderon signal, if all the vertical and horizontal, point-to-point fluctuating and point-to-point correlated
errors are taken into account. If the interpolation between the datapoints at $7$ TeV is considered as a theoretical curve, the significance of the Odderon
signal goes up to $6.55$ $\sigma$. Instead of  comparing the cross sections directly, this method removes the dominant $s$ dependent
quantities, by scaling out the $s$-dependencies of $\sigma_{\rm tot}(s)$, $\sigma_{\rm el}(s)$, $B(s)$ and $\rho_0(s)$, as well as
the normalization of the $H(x,s)$ scaling function, that also cancels the point-to-point correlated and $t$-independent normalization errors.

The model-independence of the results of refs.~\cite{Csorgo:2018uyp,Csorgo:2019ewn,Csorgo:2020msw,Csorgo:2020rlb} is an advantage when 
 a significant and model-independent Odderon signal is searched for. 
The domain of the signal region can also be determined with model-independent methods. 
Both the signal and its domain can be directly determined from the comparison of D0 and TOTEM data. However, a physical interpretation or a theoretical context is also desired, not only to gain a better understanding of the results, in order to have a more physical picture, but also to gain a predictive power and to be able to extrapolate the results to domains where experimental data are lacking, or, to regions where the scaling relations are violated.
To provide such a picture is one of the goals of our present manuscript. In this work, we continue a recent series of theoretical papers~\cite{Nemes:2012cp,CsorgO:2013kua,Csorgo:2013bwa,Nemes:2015iia}. These studies investigated the differential cross-section of elastic 
$pp$ collisions, but did not study the same effects in elastic $p\bar p$ collisions. The framework of these studies is the real extended and unitarized
Bialas-Bzdak model, based on refs.~\cite{Bialas:2006kw,Bzdak:2007qq,Bialas:2006qf,Bialas:2007eg}. This model considers protons as weakly bound states of constituent quarks and diquarks, or $p=(q,d)$ for short (for a more detailed summary of the model see \ref{sec:appendix-a}). In a variation on this theme, the diquark in the proton may also be considered
to be a weakly bound state of two constituent quarks, leading to the $p=(q,(q,q))$ variant of the Bialas-Bzdak model~\cite{Bialas:2006kw,Bzdak:2007qq}.
The model is based on Glauber's multiple scattering theory of elastic collisions~\cite{Glauber:1955qq,Glauber:1970jm,Glauber:2006gd}, assuming additionally, that all elementary distributions follow  a random Gaussian elementary process, and can be characterized by the corresponding $s$-dependent Gaussian radii.
These distributions include the parton distribution inside the quark, characterized by a Gaussian radius $R_q(s)$, the distributions of the partons inside the diquarks, characterized by the Gaussian radius $R_d(s)$ and the typical separation between the quarks and the diquarks 
characterized by the Gaussian radius $R_{qd}(s)$. In refs.~\cite{Nemes:2012cp,CsorgO:2013kua,Nemes:2015iia} it was shown that the $p=(q,(q,q))$ variant of the Bialas-Bzdak model gives too many diffractive minima, while experimentally only a single diffractive minimum is observed in $pp$ collisions.
This is a result that is consistent with the earlier detailed studies of elastic nucleus-nucleus collisions in ref.~\cite{Czyz:1969jg},
that observed that a single diffractive minimum occures only in elastic deuteron-deuteron or $(p,n)+(p,n)$ collisions, so the number of diffractive minima increases as either of the elastically colliding composite objects develops a more complex internal structure.

In the original version of the Bialas-Bzdak model, the scattering amplitude was assumed to be completely imaginary~\cite{Bialas:2006kw}.
This stucture resulted in a completely vanishing differential cross-section at the diffractive minima. This model was supplemented by
a real part, first perturbatively~\cite{Nemes:2012cp,CsorgO:2013kua,Csorgo:2013bwa}, subsequently in a non-perturbative and unitary manner~\cite{Nemes:2015iia}.
This way a new parameter called $\alpha(s)$ was introduced, that controls the value of the differential cross-section at the diffractive minimum (it is not to be confused with the strong coupling constant of QCD, that we denote in this work as $\alpha_s^{\rm QCD}$).  Our $\alpha(s)$ is a kind of opacity parameter, that measures the strength of the real part of the scattering amplitude, so it is responsible for both for filling up the dip region of the differential cross-sections and for the description of the real to imaginary ratio $\rho$ at vanishing four-momentum transfer.

The structure of this  unitary, Real Extended Bialas-Bzdak model (abbreviated as ReBB model) is thus very interesting as there are only four $s$-dependent  physical parameters:
$R_q$, $R_d$, $R_{qd}$ and $\alpha$.
However three out of these four parameters is a geometrical parameter, characterizing the $s$ dependence of parton distributions inside
the protons. Hence, it is natural to assume, that these distributions are the same inside protons and anti-protons, while the opacity parameter
$\alpha$ may be different in elastic  $pp$ and $p\bar p$ collisions.

So it is natural to expect, that this $\alpha(s)$ parameter may carry an Odderon signal
as its excitation function might be very different in elastic $pp$ collisions, that feature a pronounced dip at every measured
energy even in the TeV energy range~\cite{Antchev:2018rec}, while in elastic $p\bar p$ collisions, a significant dip is lacking even in measurements in the TeV energy range~\cite{Abazov:2012qb}.

In this manuscript, we thus extend the applications of the ReBB model from elastic $pp$ to elastic
$p\bar p$ collisions using the model exactly in the same form, as it was described in Ref.~\cite{Nemes:2015iia}. We fit exactly the same four physical parameters
to describe the differential cross-section of elastic proton-antiproton ($p\bar p$) scattering. Later we shall see that at the same energy, the geometrical
parameters in $pp$ and $p\bar p$ collisions are apparently consistent with one another, within the systematic errors of the analysis we obtain the
same $R_q(s)$, $R_d(s)$ and $R_{qd}(s)$ functions for $pp$ and $p\bar p$ reactions.

In this manuscript, we thus can investigate also the following independent questions: 
\begin{itemize}
    \item Is the Real Extended Bialas-Bzdak model of ref.~\cite{Nemes:2015iia} able to describe not only elastic $pp$ but also $p\bar p$ collisions?
    \item Is it possible to characterize the Odderon with only one physical parameter: the difference of the 
    opacity parameter $\alpha(s)$ in $pp$ and in $p\bar p$ collisions: $\alpha^{pp}(s) \ne \alpha^{p\bar p}(s)$?
\end{itemize}

We shall see that the answer to both of these questions is a definitive yes.

The structure of the manuscript is as follows.
In Section~\ref{s:formalism} we recapitulate the definition of the key physical quantities in elastic scattering and mention their main relations.
In Section~\ref{sec:fitdescription} we present the various error definitions and the evaluated $\chi^2$ formulae of both $pp$ and $p\bar p$ datasets. 
Subsequently,  in Section~\ref{sec:fit_results} we detail the optimization method and  summarize the fit results  in terms of four physical parameters determined at four different energies as listed in Table~\ref{tab:fit_parameters}, that form the basis of the determination of the energy dependencies of the model parameters  in  Section~\ref{sec:excitation_functions}.  The energy dependencies of both  proton-proton and proton-antiproton
elastic scattering in the TeV energy range are determined by a set of 10 physical parameters only, as listed in Table~\ref{tab:excitation_pars}.
As a next step for establishing the reliability of this  
$s$-dependence of the model parameters, we have performed also the so called validation or sanity tests in Section
~\ref{sec:sanity_tests}:
we have cross-checked that the obtained trends reproduce in a statistically acceptable manner each of the measured data also those, that were not utilized so far 
to establish the $s$-dependencies of the ReBB model parameters. 
After establishing that the excitation function of the ReBB model reproduces the measured data, we predict 
the experimentally not yet available large-$t$ differential cross-section of $pp$ collisions at $\sqrt{s} = 0.9$, $4$, $5$ and $ 8$ TeV
and we present the extrapolations
of the $pp$ differential cross-sections measured at the LHC energies of 2.76 and 7.0 TeV to the Tevatron energy of 1.96 TeV.
Vice versa,  we  also extrapolate the $p\bar p$ differential cross-sections from the SPS and Tevatron energies of $0.546$ and $1.96$ TeV
to the LHC energies of 2.76 and 7.0 TeV in Section~\ref{sec:extrapolations}. 
These results are discussed  in detail and put into context in Section~\ref{sec:discussion}.
 We summarize the results and conclude in Section~\ref{sec:summary}.

This work is closed with four Appendices.
For the sake of completeness, the unitary, real part extended Bialas-Bzdak model of ref.~\cite{Nemes:2015iia} is summarized in 
~\ref{sec:appendix-a}.
In \ref{sec:appendix-b} we derive and detail the relations between the opacity parameter $\alpha$ of the ReBB model and the real-to-imaginary ratio $\rho_0$. The main properties of Odderon and Pomeron exchange including the corresponding differential and total cross-sections in the TeV energy range are summarized in \ref{sec:appendix-c}.  Two small theorems are also given here: Theorem I indicates that if the differential cross-sections of elastic $pp$ and $p\bar p$ collisions are not the same in the TeV energy range, then the crossing-odd component of the elastic amplitude (Odderon) cannot vanish,
while Theorem II proves that in the framework of the ReBB model, this
is indeed due to the difference between the opacity parameters $\alpha(s)$ for $pp$ and $p\bar p$ collisions, linking  also mathematically the difference in the dip-filling property of the differential cross-sections of elastic scattering to the measurement of $\rho$  at the $t=0$ within the ReBB model. The non-linear corrections to the  linear in $\ln(s)$ excitation functions are also determined with the help of ISR $pp$ data at $\sqrt{s} = 23.5 $ GeV energy. These results are discussed in \ref{sec:appendix-d}, and found to have negligible effects on our results presented in the main body of the manuscript, corresponding to the TeV energy range.


\section{Formalism}
\label{s:formalism}

The elastic amplitude $T(s,t)$ (where $s$ is the squared central mass energy, and $t$ is the squared four-momentum transfer) is defined in Ref.~\cite{Nemes:2015iia} by Eq.~(6), Eq.~(9) and Eq. (29), furthermore summarized also in \ref{sec:appendix-a}. The experimentally measurable physical quantities, \textit{i.e.} the elastic differential cross section, the total, elastic and inelastic cross sections and the ratio $\rho_0$ are defined, correspondingly, as: 
\begin{equation}
\frac{{\rm d}\sigma}{{\rm d} t}(s,t)=\frac{1}{4\pi}\left|T\left(s,t\right)\right|^2\, ,
\label{eq:differential_cross_section}
\end{equation}
\begin{equation}
\sigma_{tot}(s)=2{\rm Im}T(s,t=0)\, ,
\label{eq:total_cross_section}
\end{equation}
\begin{equation}
\sigma_{el}(s)=\int \, dt \frac{d\sigma}{dt}(s,t),
\label{eq:elastic_cross_section}
\end{equation}
\begin{equation}
\sigma_{in}(s)=\sigma_{tot}(s)-\sigma_{el}(s)
\label{eq:inelastic_cross_section}
\end{equation}
and
\begin{align}
\rho_0(s)=\frac{\text{Re}\,T(s,t=0)}{\text{Im}\,T(s,t=0)}\,.
\label{eq:rho_parameter}
\end{align}

The earlier results show that the ReBB model gives statistically acceptable, good quality fits with CL $\ge 0.1 $ \% to the $pp$ differential cross section data at the ISR energies of 23.5 and 62.5 GeV as well as at the LHC energy of 7 TeV, in the $-t \ge 0.377$ GeV$^2$ kinematic region \cite{Nemes:2015iia}. Continuing that study, in this work we apply exactly the same formalism, without any change,  to the description of the differential cross-sections of proton-antiproton ($p\bar p$) scattering. 

This allows us to search for Odderon effects by comparing the $pp$ and $p\bar p$ differential cross sections at same energies and squared momentum transfer. Any significant difference between the $pp$ and $p\bar p$ processes at the same energy at the TeV scale provides an evidence for the Odderon exchange. In order to make this manuscript as self-contained and complete as reasonably possible, we have provided a derivation of this well-known property, in the form  of Theorem I of \ref{sec:appendix-c}.

\section{Fitting method \label{sec:fitdescription}}

Compered to the earlier ReBB study \cite{Nemes:2015iia}, in order to more precisely estimate the significance of a possible Odderon effect, here we use a more advanced form of $\chi^{2}$ definition which relies on a method developed by the PHENIX Collaboration and described in detail in Appendix A of Ref.~\cite{Adare:2008cg}. This method is based on the diagonalization of the covariance matrix, if the experimental errors can be separated
to the following  types of uncertainties:
\begin{itemize}
    \item Type A errors which are point-to-point fluctuating (uncorrelated) systematic and statistical errors;
    \item Type B errors which are point-to-point varying but 
correlated systematic uncertainties, for which the point-to-point correlation is 100 \%;
     \item Type C systematic errors which are point-independent, overall systematic uncertainties, that scale all the data points up and down by exactly the same, point-to-point independent factor.
\end{itemize}
In what follows we index these errors with the index of the data point as well as with subscripts $a$, $b$ and $c$, respectively.

In the course of the minimization of the ReBB model we use the following $\chi^{2}$ function:
\begin{eqnarray}
 \chi ^{2}&=&\left(\sum _{j=1}^{M} \left(\sum _{i=1}^{n_{j}}\frac{ \left( d_{ij}+ \epsilon _{bj} \widetilde\sigma _{bij}+ \epsilon _{cj}d_{ij} \sigma _{cj}-th_{ij} \right) ^{2}}{\widetilde{ \sigma }_{ij}^{2}}\right)+ \epsilon _{bj}^{2}+ \epsilon _{cj}^{2}\right) + \nonumber \\ 
 & &  \qquad \qquad 
 + \left( \frac{d_{ \sigma _{tot}}-th_{ \sigma _{tot}}}{ \delta  \sigma _{tot}} \right) ^{2}+ \left( \frac{d_{ \rho_{0}}-th_{ \rho _{0}}}{ \delta  \rho _{0}} \right) ^{2}.
 \label{eq:chi2-final}
\end{eqnarray}
This definition includes type A, point-to-point uncorrelated errors, type B point-to-point dependent but correlated errors and type C, point independent correlated errors. Furthermore, not only vertical, but the frequently neglected horizontal errors are included too.
Let us detail below the notation of this $\chi^2$ definition, step by step:
\begin{itemize}
    \item $M$ is the number of sub-datasets, corresponding to several, separately measured ranges of $t$, indexed with subscript $j$, at  a given energy $\sqrt{s}$. Thus $\sum_{j=1}^M n_j$ gives the number of fitted data points at a given center of mass energy $\sqrt{s}$;
    \item $d_{ij}$ is the $i$th measured differential cross section data point in sub-dataset $j$ and $th_{ij}$ is the corresponding theoretical value calculated from the ReBB model;
    \item$\widetilde{ \sigma }_{ij}$ is the type A, point-to-point fluctuating uncertainty of the data point $i$ in sub-dataset $j$,
    scaled by a multiplicative factor such that the fractional uncertainty is unchanged under multiplication by a point-to-point varying factor:
    \begin{equation}
    \widetilde{ \sigma }_{ij}^{2}= \widetilde\sigma _{aij} \left( \frac{d_{ij}+ \epsilon _{bj} \widetilde\sigma _{bij}+ \epsilon _{cj}d_{ij} \sigma _{cj}}{d_{ij}} \right)
     \end{equation}
     where the terms
     \begin{equation}
     \widetilde\sigma_{kij} =\sqrt{\sigma _{kij}^2+ (d^{\prime}_{ij} \delta_{k}t_{ij})^2}, \ \ k\in\{a,b\},
     \end{equation}
     include also the A and B type horizontal errors on $t$ following the propagation of the horizontal error to the $\chi^2$ as utilized by the so-called effective variance method of the CERN data analysis programme ROOT; $d^{\prime}_{ij}$ denotes the numerical derivative in point $t_{ij}$ with errors of type $k\in\{a,b\}$, denoted as $\delta_{k}t_{ij}$.  The numerical derivative is calculated as
     \begin{equation}
     d^{\prime}(t_{ij})=\frac{d_{(i+1)j}-d_{ij}}{t_{(i+1)j}-t_{ij}};
     \end{equation}
     \item The correlation coefficients for type B and C errors are denoted by $\epsilon_b$ and $\epsilon_c$, respectively. These numbers are free parameters to be fitted to the data, their best values are typically in the interval $(-1,1)$;
     \item The last two terms in Eq.~(\ref{eq:chi2-final}) are to fit also the measured total cross-section and ratio $\rho_{\rm 0}$ values along the differential cross section data points; $d_{\sigma_{\rm tot}}$ and $d_{\rho_{\rm 0}}$ denote the measured total cross section and ratio $\rho_{\rm 0}$ values, $\delta\sigma_{\rm tot}$ and $\delta\rho_{\rm 0}$ are their full errors, $\sigma_{\rm tot,th}$ and $\rho_{\rm 0,th}$ are their theoretical value calculated from the ReBB model;
\end{itemize}

This scheme has been validated by evaluating the $\chi^2$ from a full covariance matrix fit and from the PHENIX method of diagonalizing the covariance matrix of the differential cross-section of elastic $pp$ scattering measured by TOTEM at $\sqrt{s} = 13$ TeV~\cite{Antchev:2017dia}, using the L\'evy expansion method of Ref.~\cite{Csorgo:2018uyp}. The fit with the full covariance matrix results in the same minimum within one standard deviation of the fit parameters \cite{Csorgo:2020rlb}, hence in the same significance, as the fit with the PHENIX method. Based on this validation, we apply the PHENIX method in the data analysis
described in this manuscript.

Let us note also that in case of the $\sqrt{s} = 7 $ TeV TOTEM  data set, analysed below, the B type systematic errors, that shift all the data points together up or down with a $t$-dependent value are measured to be asymmetric~\cite{Antchev:2013gaa}. This effect is handled by using the up or down type B errors depending on the sign of the correlation coefficient $\epsilon_b$: for positive or negative sign  of  $\epsilon_b$, we utilized the type B errors upwards, or downwards, respectively. Note that the type A errors, that enter the denominator of the $\chi^2$ definition of eq.~(\ref{eq:chi2-final}), are symmetric even in the
case of this $\sqrt{s} = 7$ TeV $pp$ dataset.  The $\chi^2$ distribution assumes symmetric type A errors that enter the denominators of the $\chi^2$ definition. Thus, even in this case of asymmetric type B errors, that enter the numerators of eq.~(\ref{eq:chi2-final}) at $\sqrt{s} = 7$ TeV, the $\chi^2$ distribution can be utilized to estimate the significances and confidence levels of the best fits.


\section{Fit results\label{sec:fit_results}}

The ReBB model was fitted to the proton-proton differential cross section data measured by the TOTEM Collaboration
at $\sqrt{s}= 2.76$, $7.0$ and $13$ TeV, based on refs.~\cite{Antchev:2018rec,Antchev:2013gaa,Antchev:2017dia} as well as 
to   differential cross section data of elastic proton-antiproton scattering measured at $\sqrt{s} = 0.546$ and $1.96$ TeV in refs.~\cite{Battiston:1983gp,Bozzo:1985th,Abazov:2012qb}, respectively. 

Similarly to earlier studies of refs.~\cite{Bialas:2006kw,Bialas:2007eg,Csorgo:2013bwa,CsorgO:2013kua,Nemes:2015iia},
the model parameters $A_{qq}=1$  and $\lambda=\frac{1}{2}$
were kept at constant values throughout the fitting procedure.
Here $A_{qq}$ corresponds to  a normalization constant and $\lambda$ 
describes the mass ratio of constituent quarks to diquarks in the $p=(q,d)$ version of the Real Extended Bialas-Bzak model
of ref.~\cite{Nemes:2015iia}.
Thus the  number of free  parameters of this model, for a fixed $s$ and specific collision type is reduced to four: $R_{qd},\,R_{q},\,R_{d}$ and $\alpha$. 
It is natural to expect that 
$R_q(s)$, $R_d(s)$ and $R_{qd}(s)$ are the same functions of $s$, both for $pp$ and $p\bar p$ collisions,
as the distribution of partons inside protons at a given energy is expected to be the same as that of anti-partons inside anti-protons. In this section, this is however not assumed but tested and the
parameters of the ReBB model are determined at four different colliding energies in the TeV region, using    $pp$ data sets at $\sqrt{s} = 2.76$ and $7 $ TeV, and $p\bar p$ datasets at $\sqrt{s} = 0.546$ and $1.96$ TeV. These fits were performed in the diffractive interference or dip and bump region, with datapoints before  the diffractive minimum and after the maximum as well, in each case the limited range is not greater than $0.372 \le -t \le 1.2 $ GeV$^2$. In this kinematic range, the ReBB model provided a data description with a statistically acceptable fit quality, with  confidence levels CL $\ge 0.1$ \% in each case.

In this manuscript, our aim is to extrapolate the differential cross-section of elastic $pp$ and $p\bar p$ collisions 
to exactly the same energies,  in order to conclude in a model dependent way about the significance of a crossing-odd or Odderon effect in these data. For this purpose, a model that can be used to study the excitation function of the $pp$ and $p\bar p$ differential cross-sections in the $0.5 \le \sqrt{s} \le 7$ TeV domain is sufficient. The results of such kind of statistically acceptabe quality fits are summarized in Table~\ref{tab:fit_parameters} and detailed below.  Other data sets, that do not have sufficient amount of data in this interference region were utilized for cross-checks only, to test the extracted energy dependencies of the model parameters as detailed in Sec.~\ref{sec:sanity_tests}.
Additionally, we also describe the current  status of our fits to describe the differential cross-section at $\sqrt{s} = 13$ TeV at the end of this section.

\begin{figure}[H]
	\centering
\includegraphics[width=0.8\linewidth]{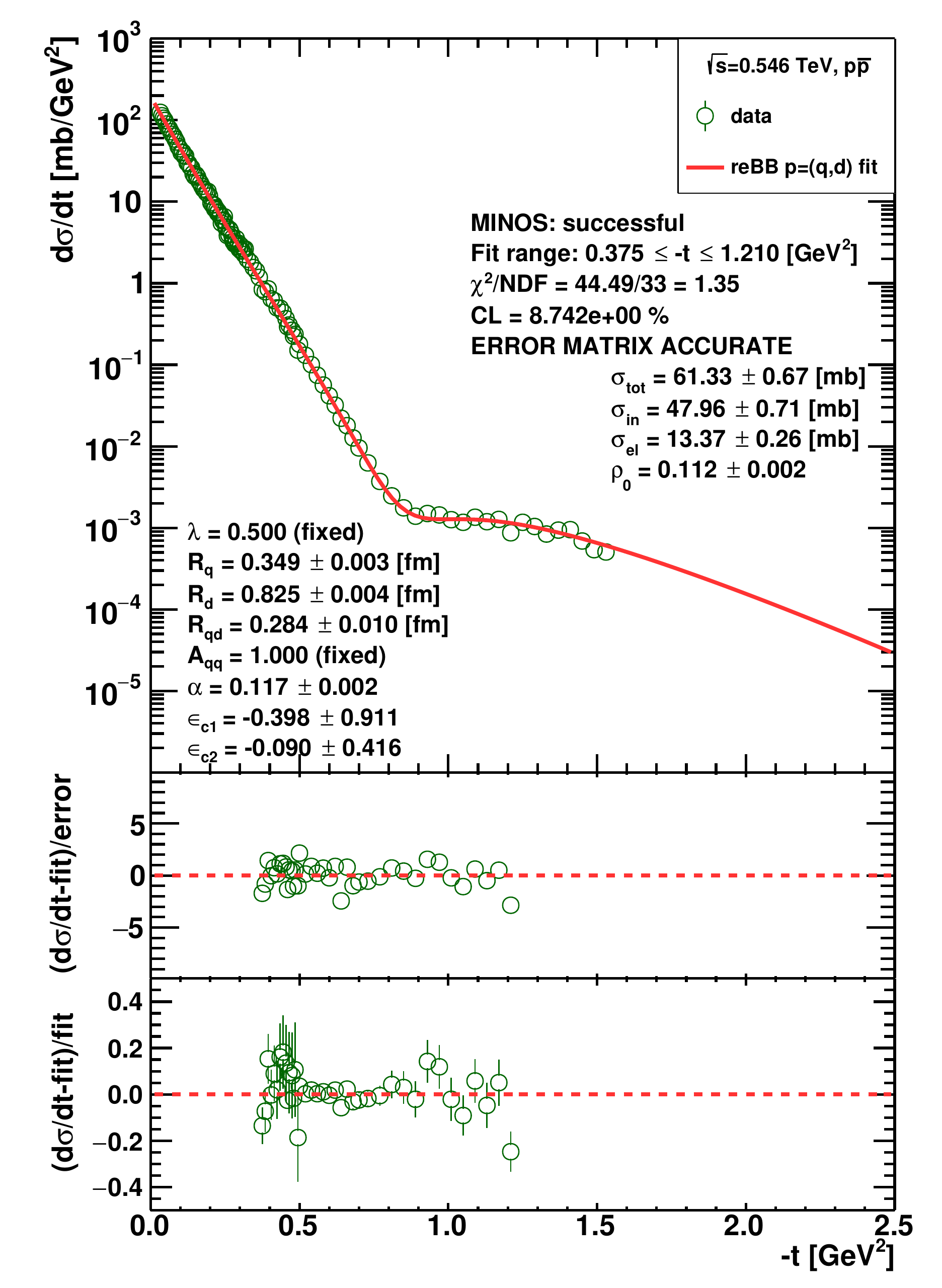}
	\caption{
	The fit of the ReBB model to the $p\bar p$ SPS $\sqrt{s}=0.546$~TeV data~\cite{Battiston:1983gp,Bozzo:1985th} in the range of $0.37 \leq -t \leq 1.2$ GeV$^2$. The fit includes the published errors, that are statistical (type A) and the normalization (type C) uncertainties, as well as the experimental value of the total cross section with its full error according to Eq.~(\ref{eq:chi2-final}).  The fitted parameters are shown in the left bottom corner and their values are rounded up to three decimal digits. The fit quality parameters and the values of the total, inelastic and elastic cross-sections as well as the value of the $\rho_0$ parameter are summarized in the top right part of the plot.
	}
	\label{fig:reBB_model_fit_0_546_TeV}
\end{figure}

\begin{figure}[H]
	\centering
\includegraphics[width=0.8\linewidth]{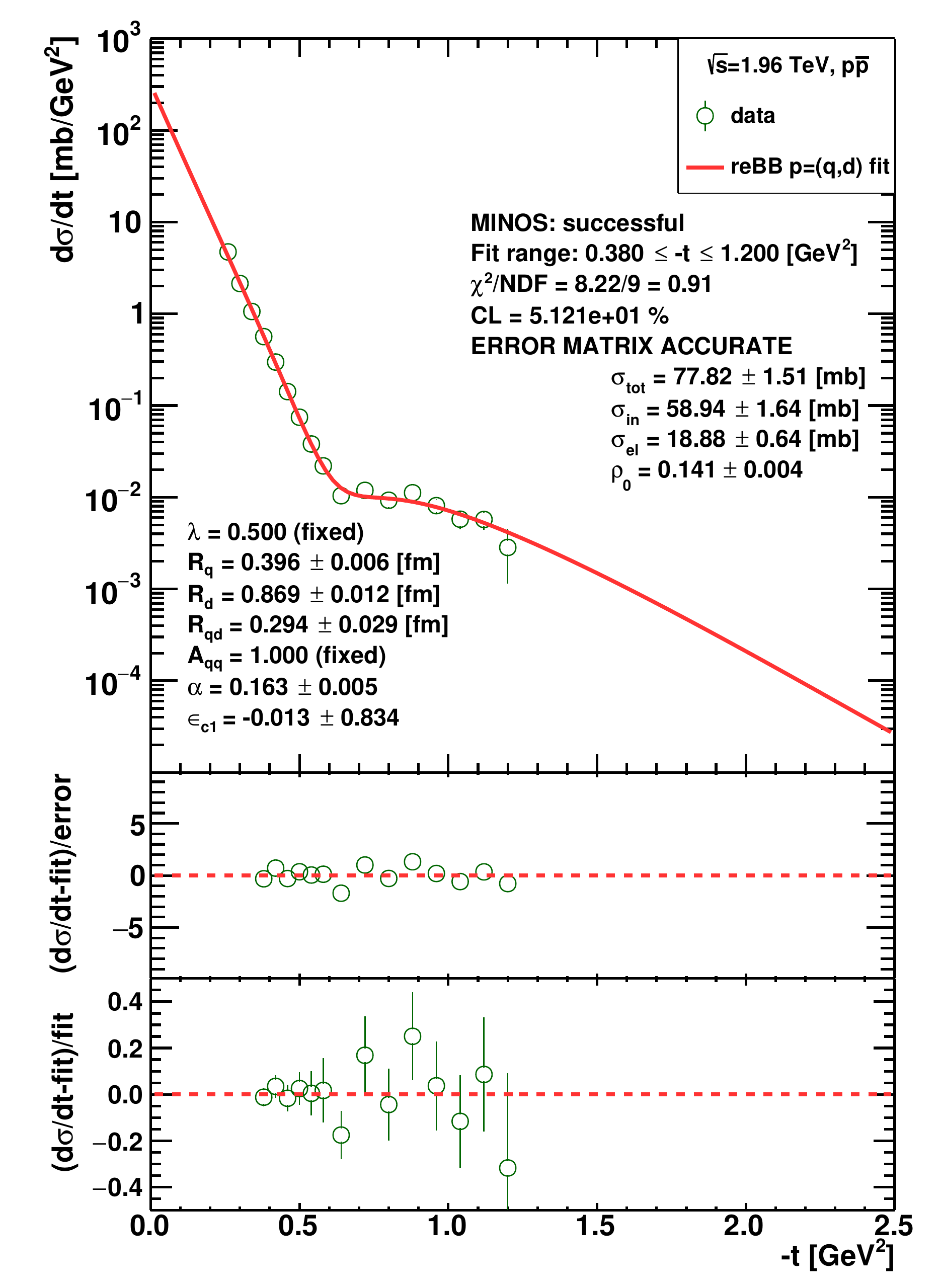}
	\caption{The fit of the ReBB model to the $p\bar p$ D0 $\sqrt{s}=1.96$~TeV data \cite{Abazov:2012qb} in the range of $0.37\leq-t\leq1.2 $ GeV$^2$. The fit includes the $t$-dependent statistical and systematic uncertainties added together quadratically and treated as type A errors as well as the normalization (type C) uncertainty according to Eq.~(\ref{eq:chi2-final}). The values of the total cross section and parameter
	$\rho_0$ used in the fit are the predicted values from the COMPETE Collaboration \cite{Cudell:2002xe}. 
	Otherwise, same as Fig.~\ref{fig:reBB_model_fit_0_546_TeV}.
	}
	\label{fig:reBB_model_fit_1_96_TeV}
\end{figure}

\begin{figure}[H]
	\centering
\includegraphics[width=0.8\linewidth]{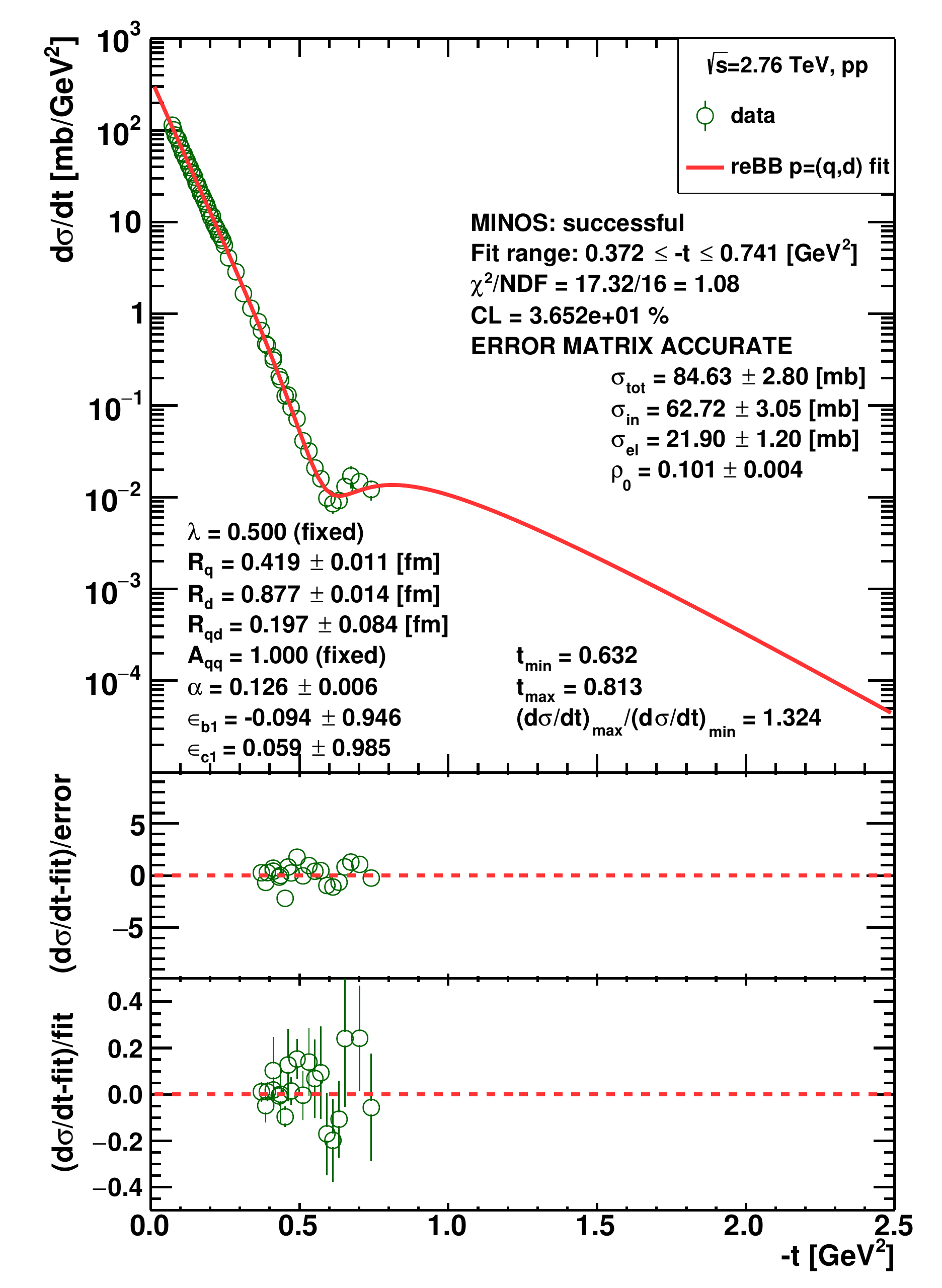}
	\caption{The fit of the ReBB model to the $pp$ TOTEM $\sqrt{s}=2.76$~TeV data in the range of $0.37\leq-t\leq0.74 $ GeV$^2$ \cite{Antchev:2018rec}. The fit includes the $t$-dependent statistical (type A) and systematic (type B) uncertainties, the normalization (type C) uncertainty and the experimental value of the total cross section with its full error according to Eq.~(\ref{eq:chi2-final}). 
	Otherwise, same as Fig.~\ref{fig:reBB_model_fit_0_546_TeV}.
%
}
	\label{fig:reBB_model_fit_2_76_TeV}
\end{figure}

\begin{figure}[H]
	\centering
\includegraphics[width=0.8\linewidth]{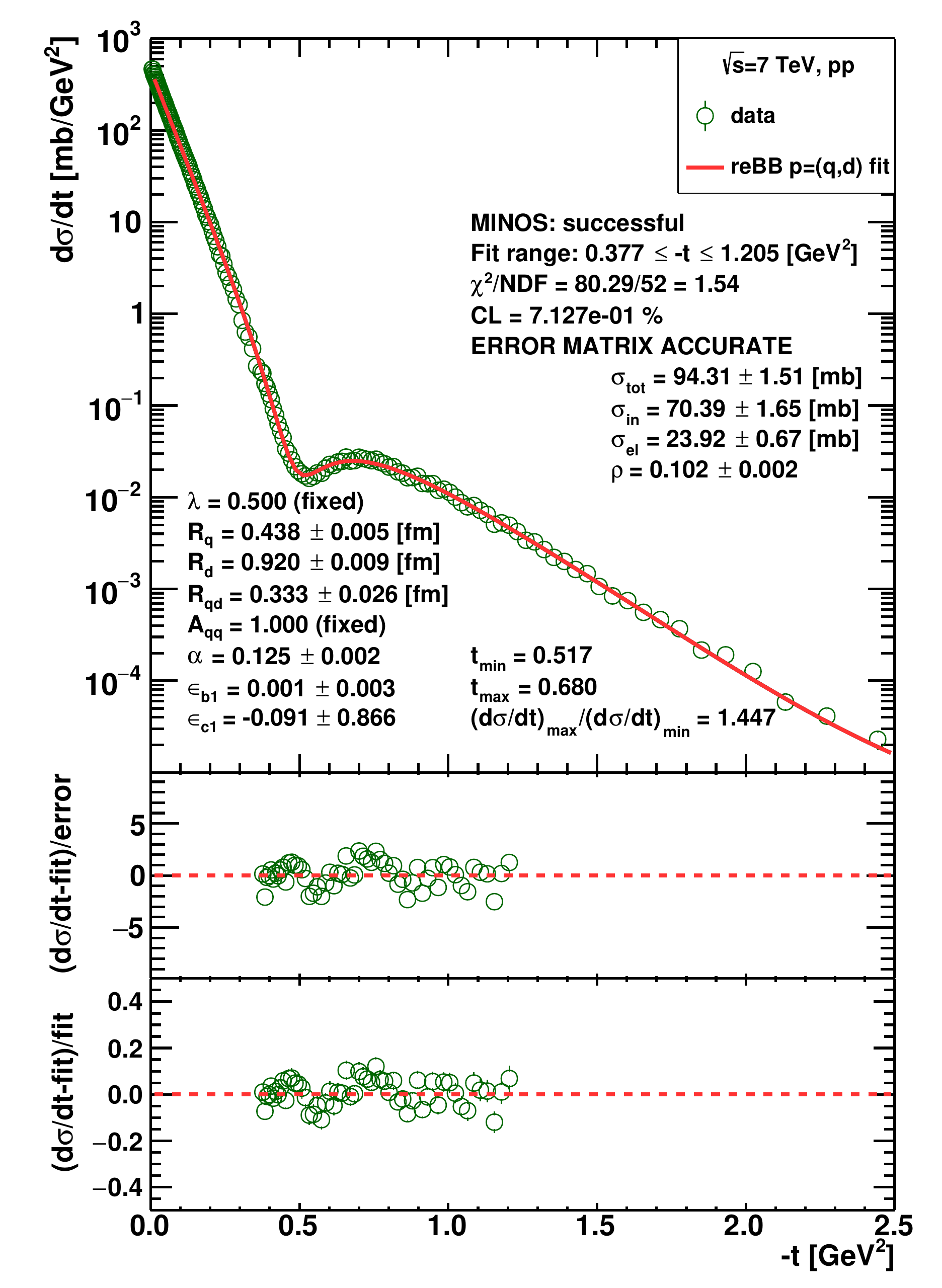}
	\caption{The fit of the ReBB model to the $pp$ TOTEM $\sqrt{s}=7$~TeV data in the range of $0.37\leq-t\leq1.2 $ GeV$^2$ \cite{Antchev:2013gaa}. The fit includes the $t$-dependent statistical (type A) and systematic (type B) uncertainties, the normalization (type C) uncertainty and the experimental values of the total cross section and parameter $\rho_0$ with their full error according to Eq.~(\ref{eq:chi2-final}).
	Otherwise, same as Fig.~\ref{fig:reBB_model_fit_0_546_TeV}.
}
	\label{fig:reBB_model_fit_7_TeV}
\end{figure}

\begin{figure}[H]
	\centering
	\includegraphics[width=0.8\linewidth]{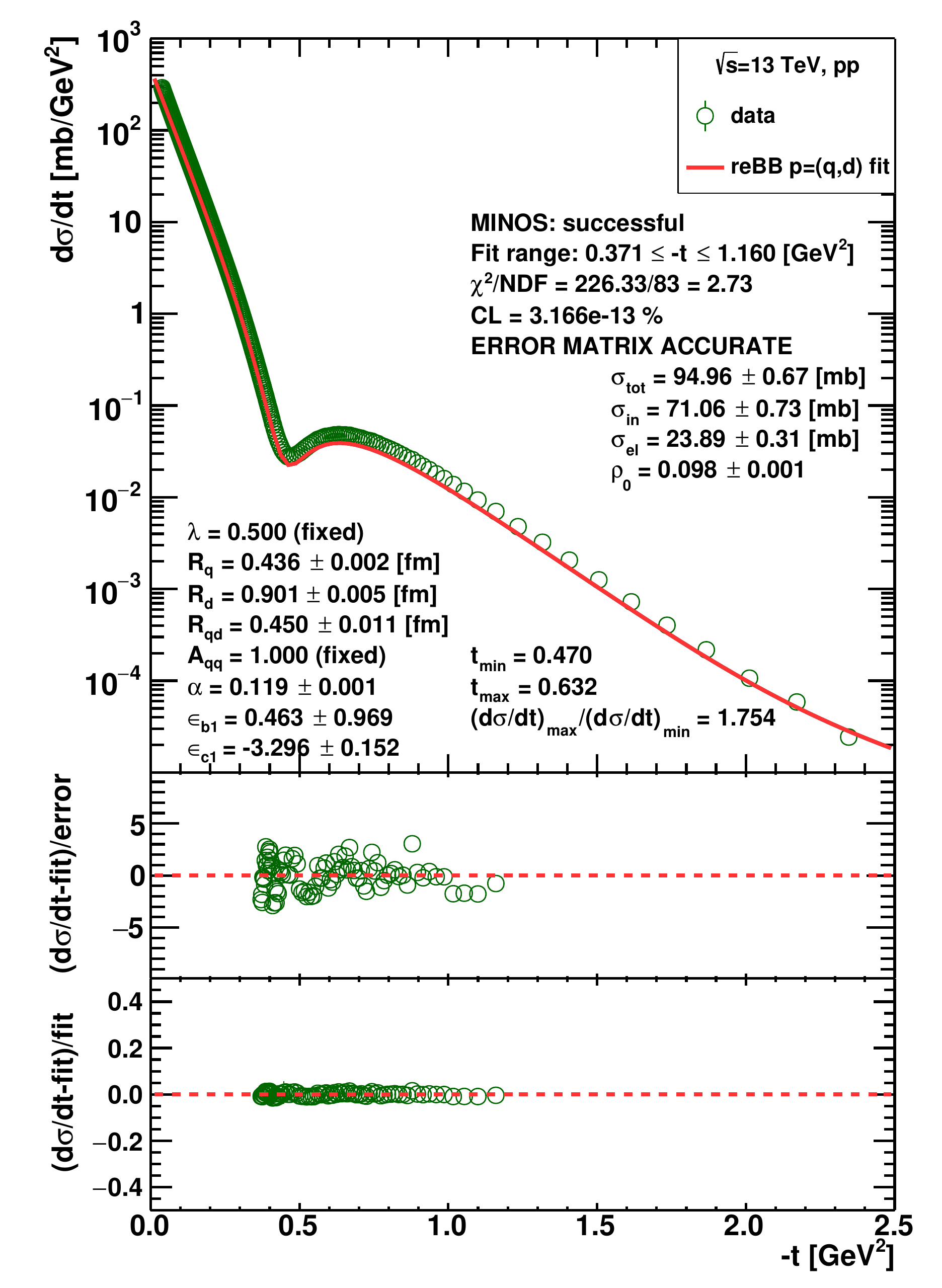}
	\caption{The fit of the ReBB model to the $pp$ TOTEM $\sqrt{s}=13$~TeV data in the range of $0.37\leq-t\leq1.2$ GeV$^2$ \cite{Antchev:2018edk}. The fit includes the $t$-dependent statistical (type A) and systematic (type B) uncertainties, the normalization (type C) uncertainty and the experimental values of the total cross section and parameter $\rho_0$ with their full error according to Eq.~(\ref{eq:chi2-final}). The fit parameters do not represent the data in a statistically acceptable manner,
	given that CL $\ll$ 0.1 \% .
		Otherwise, same as Fig.~\ref{fig:reBB_model_fit_0_546_TeV}.
	}
	\label{fig:reBB_model_fit_13_TeV}
\end{figure}

We thus describe three fits to  $pp$ differential cross section data sets at $\sqrt{s} = 2.76$,  $7$ and $13$ TeV as well as two fits to $p\bar p$ differential cross section datasets at $\sqrt{s} = 0.546$ and $1.96$ TeV, respectively.
Our fit results are graphically shown in Figs.~\ref{fig:reBB_model_fit_0_546_TeV}-\ref{fig:reBB_model_fit_13_TeV}.

The minimization of the $\chi^2$ defined by Eq.~(\ref{eq:chi2-final}) was done with Minuit and the parameter errors were estimated by using the MINOS algorithm which takes into account both parameter correlations and non-linearities.
We accept the fit as a successful representation of the fitted data under the condition that the fit status is converged, the error matrix is accurate and the confidence level of the fit, CL is $\ge 0.1$ \%, as indicated on Figs.~\ref{fig:reBB_model_fit_0_546_TeV}-\ref{fig:reBB_model_fit_7_TeV}. As these criteria are not satisfied on Fig.~\ref{fig:reBB_model_fit_13_TeV}, the parameters of this fit were not taken into account when determining the
excitation functions or the energy dependence of the physical fit parameters in the few TeV energy range.

Let us now discuss each fit  in a bit more detail. 

The S$p\bar p$S differential cross section data on elastic $p\bar p$ collisions~\cite{Battiston:1983gp,Bozzo:1985th} were measured in the squared momentum transfer range of $0.03\leq|t|\leq1.53 $ GeV$^2$ which in the fitted range has been subdivided into two sub-ranges with different normalization uncertainties (type C errors): for $0.37\leq|t|\leq0.495 $ GeV$^2$ $\sigma_{c}$ = 0.03 and for $0.46\leq|t|\leq1.2$ GeV$^2$ $\sigma_{c}$ = 0.1. In case of this data set, the vertical type A errors $\sigma_{ai}$ are available but the horizontal type A errors ($\delta_{a}t_i$) and the type B errors either vertical ($\sigma_{bi}$) or horizontal ($\delta_{b}t_i$) were not published. The measured total cross section with its total uncertainty is $\sigma_{\rm tot}=61.26\pm0.93$ mb \cite{Tanabashi:2018oca} while the $\rho_0=0.135\pm0.015$ value was measured at the slightly different energy of $\sqrt{s} = 0.541$ GeV. 
The total, elastic and inelastic cross sections and the parameter $\rho_0$ are calculated according to Eqs.~(\ref{eq:total_cross_section})-(\ref{eq:rho_parameter}).
The fit is summarized in Fig.~\ref{fig:reBB_model_fit_0_546_TeV}. The fit quality is satisfactory, CL = 8.74 \%. 
Compared to the available data in the literature \cite{Tanabashi:2018oca} ($\sigma_{in}=48.39\pm1.01$ mb and $\sigma_{el}=12.87\pm0.3$ mb) the model reproduces the experimental values of the forward measurables within one $\sigma$, thus these fit parameters represent the data in a statistically acceptable manner. 

The elastic $p\bar p$ differential cross section data is available at $\sqrt{s} = 1.96$ TeV in the range of $0.26\leq|t|\leq1.20 $ GeV$^2$, as published by the D0 Collaboration in  ref.~\cite{Abazov:2012qb}, with a type C normalization uncertainty of $\sigma_{c}$ = 0.144. For this data set, the vertical type A and type B errors were not published separately. Actually, the quadratically added statistical and systematic uncertainties were published, and as the statistical errors are point to-point fluctuating, type A errors, in our analysis the combined $t$ dependent D0 errors were handled as type A, combined statistical and systematic errors. Horizontal type A and type B errors were not published in ref.~\cite{Abazov:2012qb}. 
 At this energy, we do not find  published experimental  $\sigma_{\rm tot} $ and $\rho_0$ values. The values of the total cross section and parameter $\rho_0$ at this energy, that we utilized in the fitting procedure, are the predicted values from the COMPETE Collaboration \cite{Cudell:2002xe}: $\sigma_{\rm tot}=78.27\pm1.93$ mb and $\rho_0=0.145\pm0.006$. The quality of the corresponding fit, shown in  Fig.~\ref{fig:reBB_model_fit_1_96_TeV}, is satisfactory, CL = 51.12 \%, and the COMPETE values of forward measurables are reproduced within one standard deviation. We conclude that the corresponding ReBB model parameters represent the data in a statistically acceptable manner. 

Based on the successful description of these two $p\bar p$ datasets at $\sqrt{s} = 0.546$ and $1.96$ TeV,
we find that the form of the ReBB model as specified for $pp$ collisions in ref.~\cite{Nemes:2015iia} is able, without any modifications, to describe the differential cross-section of elastic $p\bar p$ collisions in the TeV energy range. Let us now discuss the new fits of the same model to elastic $pp$ collisions in the TeV energy range. 

At $\sqrt{s} = 2.76 $ TeV,  the differential cross section data of elastic $pp$ collisions was measured in the $t$ range of $0.072\leq -t\leq 0.74 $ GeV$^2$ by the TOTEM Collaboration~\cite{Antchev:2018rec}. Actually, this measurement was performed in two subranges: $0.072\leq|t|\leq0.462 $ GeV$^2$ and $0.372\leq|t|\leq0.74 $ GeV$^2$.  Both ranges had the same normalization uncertainty of $\sigma_c$ $=$ $0.06$. During the fit the $t$-dependent vertical statistical (type A) and vertical systematic (type B) errors (both horizontal and vertical ones), the normalization (type C) errors and the experimental value of the total cross section with its total uncertainty ($\sigma_{\rm tot}=84.7\pm3.3$ mb \cite{Antchev:2017dia}) were taken into account. Horizontal type A and type B errors are not published at this energy. The fit quality of the ReBB model is demonstrated on  Fig.~\ref{fig:reBB_model_fit_2_76_TeV}: the fit is satisfactory, with CL = 36.52 \%. The experimental values of the forward measurables ($\sigma_{in}=62.8\pm2.9$ mb, $\sigma_{el}=21.8\pm1.4$ mb \cite{Nemes:2017gut,Antchev:2017dia}) are reproduced within one standard deviations. Experimental data is not yet available for parameter $\rho_0$, however the value for $\rho_0$, calculated from the fitted ReBB model, is within the total error band of the COMPETE prediction \cite{Cudell:2002xe}. We thus conclude that the corresponding ReBB model parameters represent the $pp$ data at $\sqrt{s} = 2.76 $ TeV in a statistically acceptable manner. 

At $\sqrt{s} = 7$ TeV, the $pp$  differential cross section data was published by the TOTEM Collaboration~\cite{Antchev:2013gaa}, measured in the range of $0.005\leq|t|\leq2.443 $ GeV$^2$ . The measurement was performed in two subranges: $0.005\leq|t|\leq0.371$ GeV$^2$ and $0.377\leq|t|\leq2.443 $ GeV$^2$. Both ranges had the same normalization uncertainty of $\sigma_c$ = 0.042. The fit includes only the second subrange with the $t$-dependent (both vertical and horizontal) statistical (type A) and systematic (type B) errors, the normalization (type C) error and the experimental values of the total cross section and the parameter $\rho_0$ with their total uncertainties ($\sigma_{\rm tot}=98.0\pm2.5$ mb and $\rho_0=0.145\pm0.091$ \cite{Antchev:2013haa}). The quality of the corresponding fit, shown in Fig.~\ref{fig:reBB_model_fit_7_TeV}, is statistically acceptable with a CL = 0.71 \%. The experimental values of the forward measurables ($\sigma_{in}=72.9\pm1.5$ mb, $\sigma_{el}=25.1\pm1.1$ mb \cite{Antchev:2013haa}) are reproduced by the fitted ReBB model within one sigma (the experimental and
calculated values overlap within their errors). We thus conclude that the corresponding ReBB model parameters represent these $pp$ data at $\sqrt{s} = 7.0 $ TeV in a statistically acceptable manner, in the fitted range of $0.377\leq|t|\leq1.205 $ GeV$^2$,  
before and after the diffractive minimum.

At $\sqrt{s} = 8$ TeV, the TOTEM collaboration did not yet publish the final differential cross-section results in the range of the diffractive minumum and maximum. However, preliminary results were  presented at conferences~\cite{Kaspar:2018ISMD}, and the differential cross-section in the 
low $-t$ region was published in ref.~\cite{Antchev:2015zza}. We thus use this dataset for a cross-check only, but the lack of the data in the diffractive minimum prevents us to do a full ReBB model fit. Additional data at very low $-t$, in the Coulomb-Nuclear Interference region is also available from TOTEM at this particular energy~\cite{Antchev:2016vpy}, however, in the present study we do not discuss the kinematic range, where Coulomb effects may play any role.

At $\sqrt{s} = 13$ TeV, the  differential cross section data was measured by the TOTEM collaboration in the range of $0.03\leq|t|\leq3.8$ GeV$^2$ \cite{Antchev:2018edk} with a normalization (type C) uncertainty of $\sigma_c$ = 0.055. 
As far as we know, the only statistically acceptable quality fit with CL $\geq$ 0.1 \% to this dataset so far 
was obtained by some of us with the help of the model-independent L\'evy series in ref.~\cite{Csorgo:2018uyp}.
We also note that several new features show up in the soft observables of elastic scattering, with a threshold behaviour around
$\sqrt{s} = 5-7$ TeV, certainly below 13 TeV~\cite{Csorgo:2019fbf}.

We have cross-checked, if the ReBB model, that works reasonably well from $\sqrt{s} = 23.5$ GeV to $7$ TeV, is capable to describe
this data set at $\sqrt{s} = 13$ TeV in statistically acceptable manner, or not? The result was negative, as indicated in Fig.~\ref{fig:reBB_model_fit_13_TeV}. This fit includes the $t$-dependent statistical (type A) and systematic (type B) errors, the normalization (type C) error and the experimental values of the total cross section and the parameter $\rho_0$ with their total uncertainties ($\sigma_{\rm tot}=110.5\pm2.4$ mb and $\rho_0=0.09\pm0.01$ \cite{Antchev:2017yns}). The quality of the obtained fit (Fig.~\ref{fig:reBB_model_fit_13_TeV}) is not satisfactory, CL = 3.17$\times10^{-11}$ \% and neither the experimental values of the cross sections ($\sigma_{in}=79.5\pm1.8$ mb, $\sigma_{el}=31.0\pm1.7$ mb \cite{Antchev:2017dia} ) are reproduced by the fitted ReBB model within one sigma at 13 TeV. However, the value of $\rho_0$ was described surprisingly well.
This TOTEM dataset is very detailed and precise and changes of certain trends in $B(s)$
and the ratio $\sigma_{\rm el}(s)/\sigma_{\rm tot}(s)$ are seen experimentally~\cite{Csorgo:2019fbf}. Theoretically, a new domain of QCD may emerge at high energies, possibly characterised by hollowness or toroidal structure, corresponding to a black ring-like distribution of inelastic scatterings \cite{Dremin:2014dea,Dremin:2014spa,Albacete:2016pmp,Troshin:2017ucy}. A statistically significant, more than 5 $\sigma$ hollowness effect was found at $\sqrt{s} = 13$ TeV within a model-independent analysis of the shadow profile at these energies, using the technique of L\'evy series~\cite{Csorgo:2018uyp}. We conclude that the ReBB model needs to be generalized to have a stronger non-exponential feature at low $-t$ to accommodate the new features of the differential cross-section data at $\sqrt{s} = 13$ TeV or larger energies. This work is currently in progress, but goes well beyond the scope of the current manuscript. Most importantly, such a generalization is not necessary for a comparision of the differential cross-sections of elastic $pp$ and $p\bar p$ collisions in the few TeV range, as we have to bridge only a logarithmically small energy difference between the top D0 energy
of $\sqrt{s} = 1.96$ TeV and the lowest TOTEM energy of $\sqrt{s} = 2.76$ TeV.

We thus find, that the Real Extended Bialas - Bzdak model describes effectively and in a statistically acceptable manner
the differential cross-sections of elastic $pp$ and $p\bar p$ collisions in the few TeV range of $0.546 \le \sqrt{s} \le 7 $ TeV and in the squared four-momentum transfer range of $0.37\leq-t\leq1.2 $ GeV$^2$.
Its physical fit parameters represent the data and their energy dependence thus can be utilized to determine the excitation function of these model parameters, as detailed in Section~\ref{sec:excitation_functions}.

The values of the physical fit parameters and their errors obtained from the above discussed physically and statistically acceptable fits are summarized in Table~\ref{tab:fit_parameters}, 
where four datasets are analyzed and four different physical parameters are extracted at four different energies. 
These sixteen physical parameters form the basis of the determination of the energy dependencies, 
that are determined to be consistent with affine linear functions of $\ln(s)$. 
Three scale parameters are within errors the same in elastic $pp$ and $p\bar p$ collisions,
while the opacity parameters are different for $pp$ and $p\bar p$ collisions. Thus the excitation functions, the energy dependence of the differential cross-sections both for $pp$ and $p\bar p$ elastic scattering is determined by 5x2 = 10 physical parameters in this framework of calculations. These 10 parameters are summarized in Table ~\ref{tab:excitation_pars}. 

We thus conclude, that this Real Extended Bialas-Bzdak model is good enough to extrapolate the differential cross-section of elastic $pp$ collisions down to $\sqrt{s} = 0.546$ and $1.96$ TeV, and to extrapolate the same of elastic $p\bar p$ collisions up to $\sqrt{s} = 2.76$ and $7$ TeV. We duly note that, in order to evaluate similar observables at $\sqrt{s} = 13$ TeV or at even higher energies in a realistic manner, this model needs to be generalized and further developed.

\begin{table*}[htb]
		\caption{The values of the fitted ReBB model parameters to $pp$ and $p\bar p$ data from SPS to LHC energies. The errors and the values are rounded up to three valuable decimal digits. For 7 TeV, the parameter error values shown in parenthesis do not include the contribution from the parameter correlations, i.e., are less than the MINOS errors. }
			\label{tab:fit_parameters}
\begin{center}
	{\begin{tabular}{|c|c|c|c|c|} \hline\noalign{\smallskip}
	$\sqrt{s}$ [TeV]  &0.546 ($p\bar p$)  &1.960 ($p\bar p$)   &2.760 ($pp$)	      &7.000 ($pp$)          \\ 
	\noalign{\smallskip}\hline\noalign{\smallskip}
	$|t|$ [GeV$^{2}$] &(0.375, 1.210)     &(0.380, 1.200)     &(0.372, 0.741)     &(0.377, 1.205)    \\ 
	$\chi^{2}/NDF$	  &44.49/33           &8.22/9             &17.32/16           & 80.29/52         \\ 
	CL [\%] 	      &8.74               &51.12	          &36.52              & 0.713            \\ 
	$R_{q}$ [fm] 	  &0.349 $\pm$ 0.003  &0.396 $\pm$ 0.006  &0.419 $\pm$ 0.011  & 0.438 $\pm$ 0.005 ($\pm$ 0.001)\\ 
	$R_{d}$ [fm] 	  &0.825 $\pm$ 0.004  &0.869 $\pm$ 0.012  &0.877 $\pm$ 0.014  & 0.920 $\pm$ 0.009 ($\pm$ 0.002)\\ 
	$R_{qd}$ [fm]   &0.284 $\pm$ 0.010  &0.294 $\pm$ 0.029  &0.197 $\pm$ 0.084    & 0.333 $\pm$ 0.026 ($\pm$ 0.002)\\ 
	$\alpha$ 	      &0.117 $\pm$ 0.002  &0.163 $\pm$ 0.005  &0.126 $\pm$ 0.006  & 0.125 $\pm$ 0.002 ($\pm$ 0.001)\\ 
	$\epsilon_{b1}$   &--                 &--                 &-0.094 $\pm$ 0.946 &0.001 $\pm$ 0.003\\ 
	$\epsilon_{c1}$   &-0.398 $\pm$ 0.911 &-0.013 $\pm$ 0.834 & 0.059 $\pm$ 0.985 &-0.091 $\pm$ 0.866\\ 
	$\epsilon_{c2}$   &-0.090 $\pm$ 0.416 &--                 &--                 &--                \\ \hline 
	\end{tabular}}
	\end{center}
\end{table*}

\section{Excitation functions of the fit parameters \label{sec:excitation_functions}}

The values of the physical fit parameters and their errors obtained from the above discussed physically and statistically acceptable fits are summarized in Table~\ref{tab:fit_parameters}.
This table contains a list of five different physical parameters. Out of them the three scale parameters called
$R_q$, $R_d$ and $R_{qd}$ can be determined at four different energies, providing 12 numbers, while the opacity parameters
$\alpha^{pp}$ and $\alpha^{p\bar p}$ describing  
$pp$ and $p\bar p$ collisions can both be determined at two different energies only, providing additional 4 numbers, all-together 16 physical input parameters. These 16 physical parameters form the basis of the determination of the energy dependencies, that are determined to be consistent with affine linear functions of $\ln(s)$.

Namely, we fitted the $s$-dependence of the  model parameters one by one, using the
affine linear logarithmic function,
\begin{equation} 
P(s)=p_{0} + p_{1}\cdot\ln{(s/s_{0})}, \ \ P\in{\{R_{q},R_{d},R_{qd},\alpha\}},
\label{eq:parametrization_of_extrapolation_lin} 
\end{equation} 
where $p_0$ and $p_1$ are free parameters, $s_{0}$ is fixed at 1 GeV$^{2}$. 
We obtain good quality fits, with methods and results similar to that of ref.~\cite{Nemes:2015iia}, with confidence levels $CL \gg 0.1$ \%, as detailed in Table~\ref{tab:excitation_pars}. Three scale parameters are within errors the same in elastic $pp$ and $p\bar p$ collisions,
while the opacity parameters are different for $pp$ and $p\bar p$ collisions. Thus the excitation functions, the energy dependene of the differential cross-sections both for $pp$ and $p\bar p$ elastic scattering is determined by 5x2 = 10 phyiscal parameters in the framework of the ReBB model.

\begin{figure}[H]
	\centering
	\subfloat[Parameter $R_q$
	\label{fig:par_Rq_lin}]{%
		\includegraphics[width=0.5\linewidth]{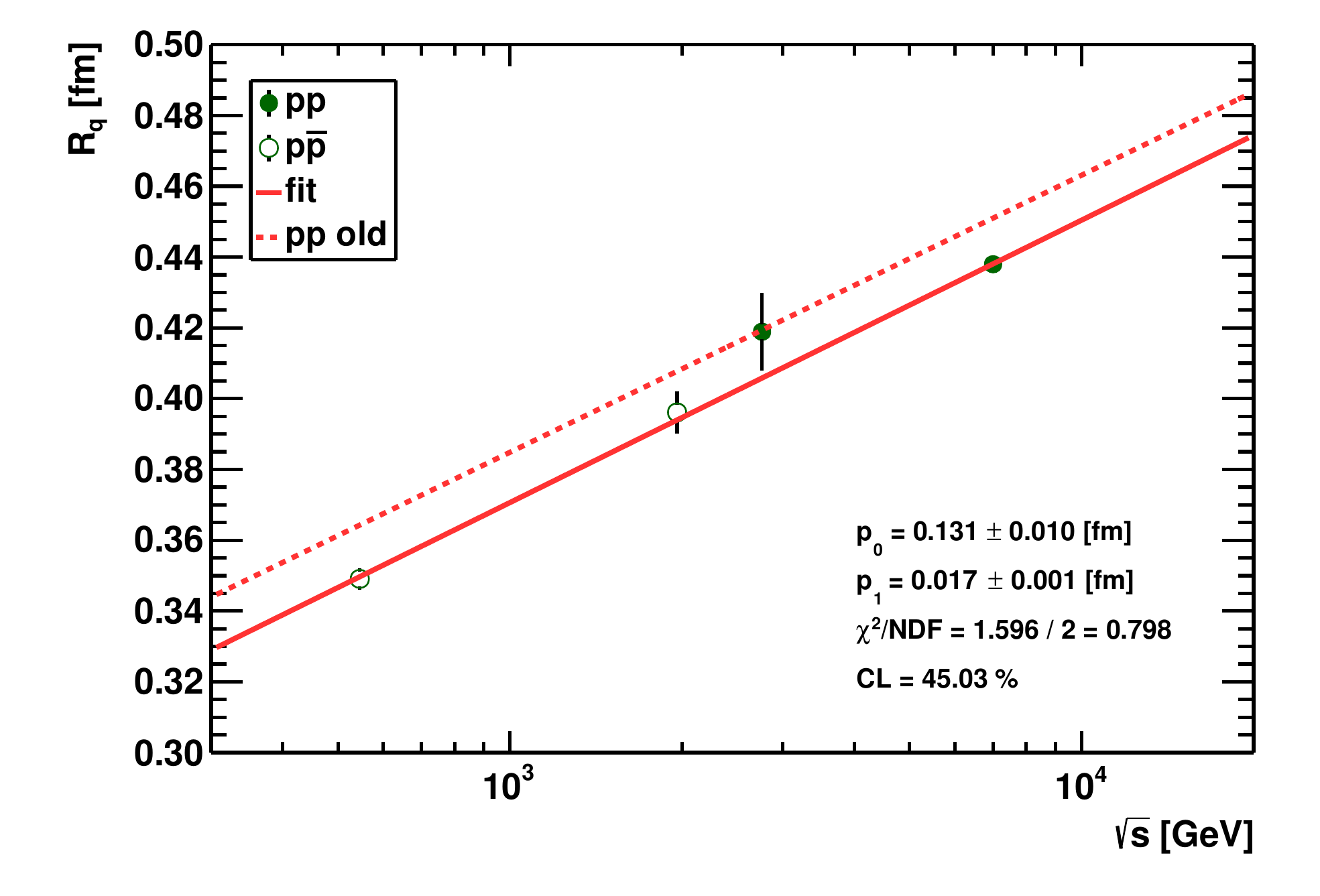}%
	}\hfill
	\subfloat[Parameter $R_d$\label{fig:par_Rd_lin}]{%
		\includegraphics[width=0.5\linewidth]{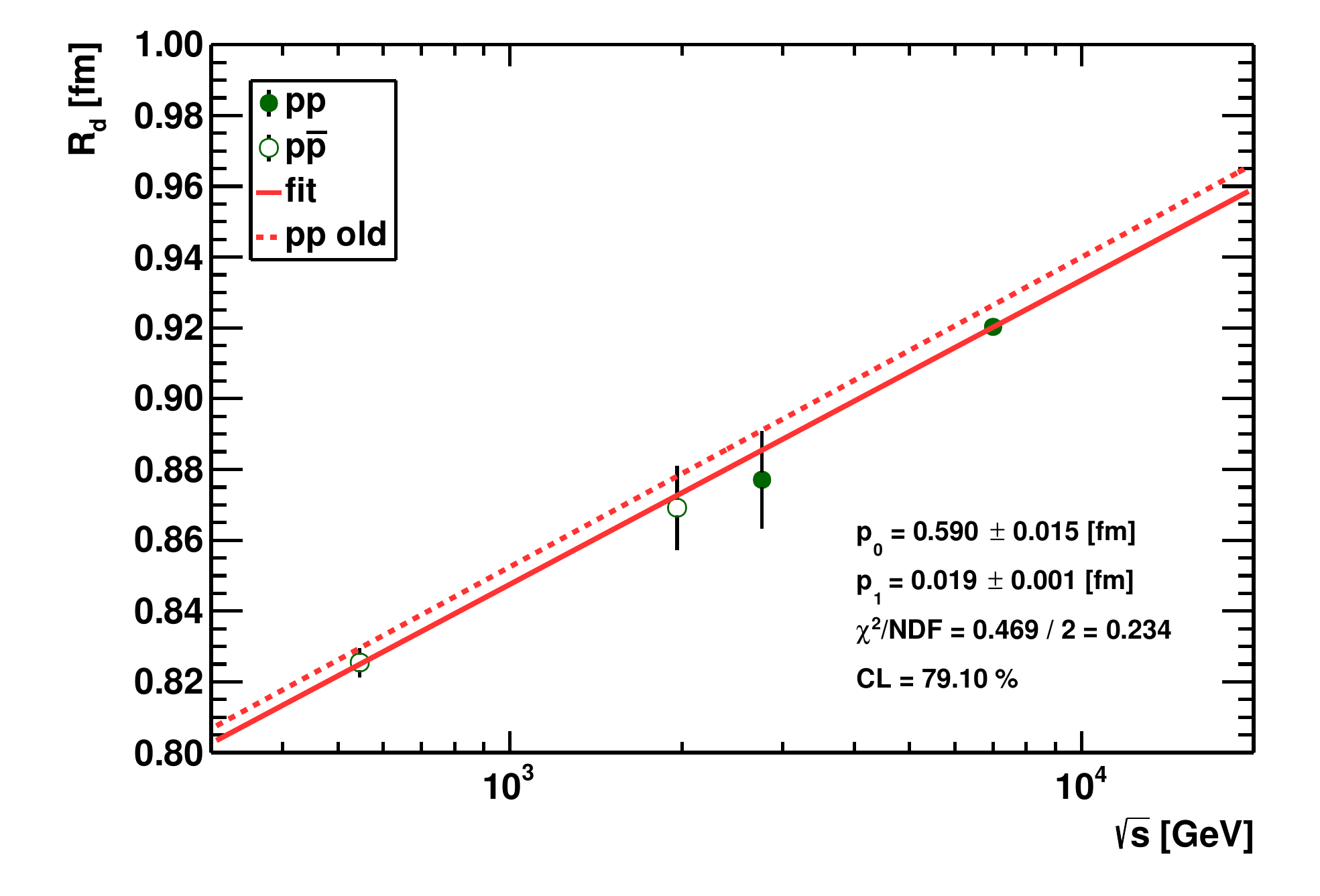}%
	}\hfill
	\subfloat[Parameter $R_{qd}$\label{fig:par_Rqd_lin}]{%
		\includegraphics[width=0.5\linewidth]{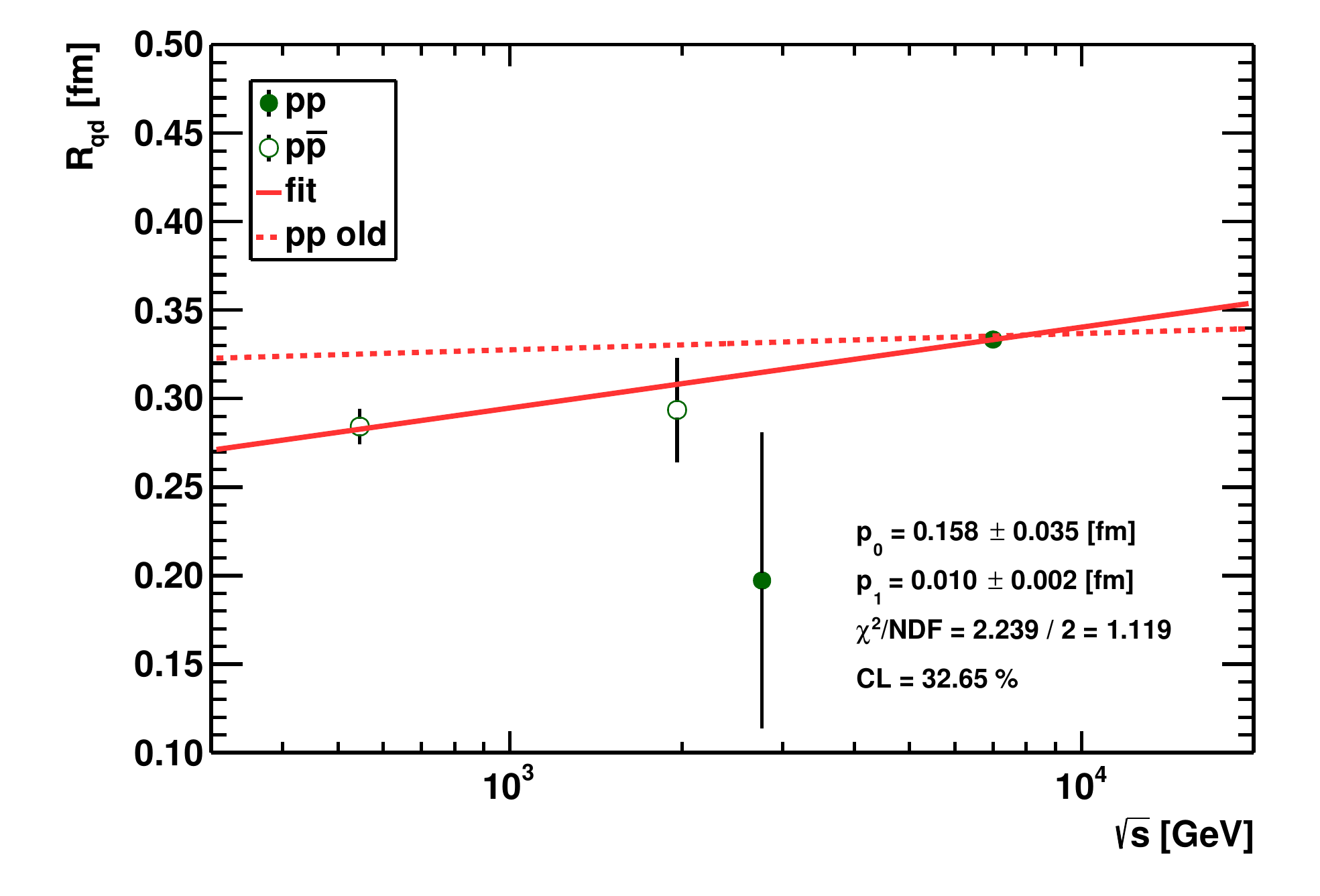}%
	}\hfill
	\subfloat[Parameter $\alpha$\label{fig:par_alpha_lin}]{%
		\includegraphics[width=0.5\linewidth]{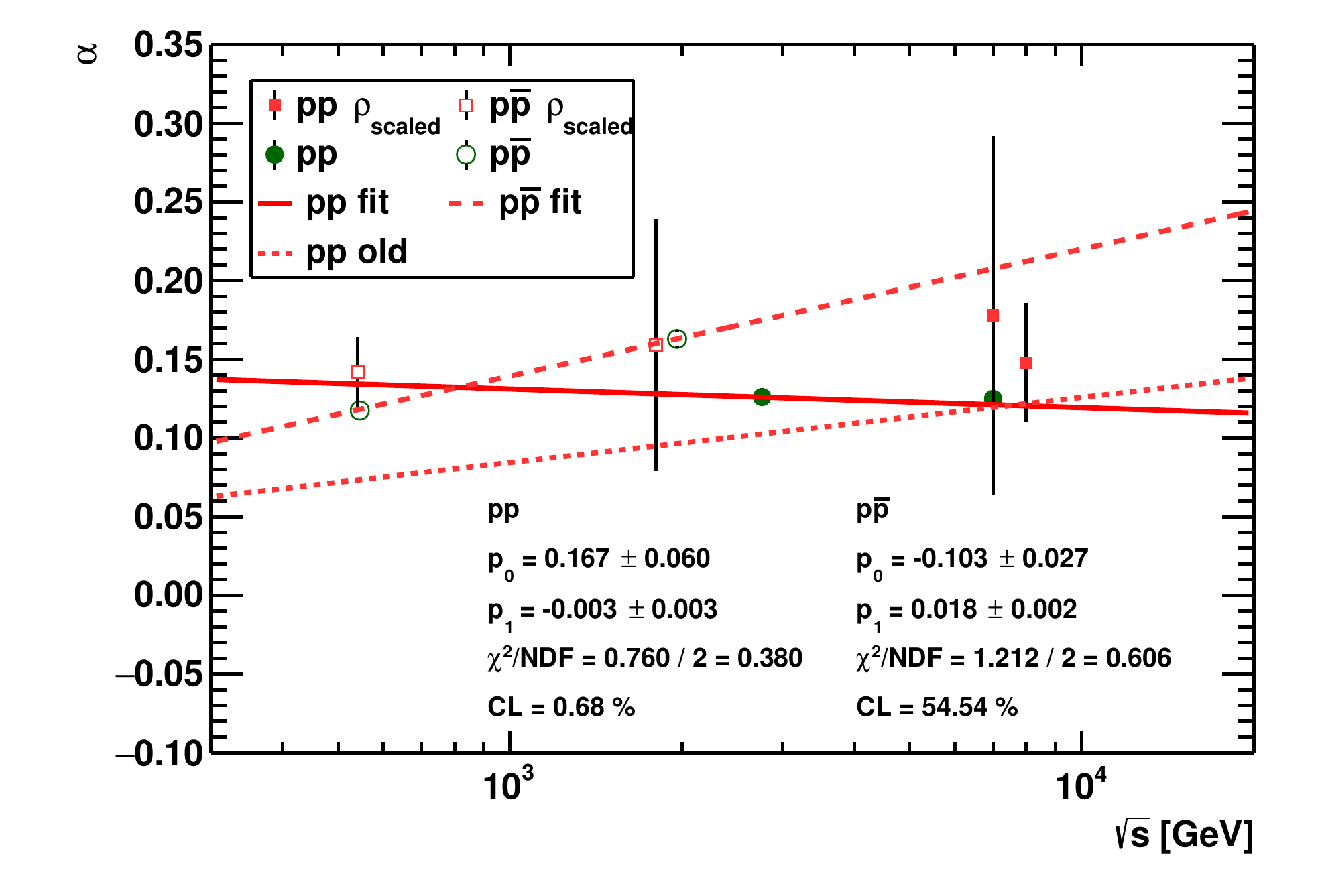}%
	}
	\caption{The energy dependence of the parameters of the ReBB model, $R_{q}$, $R_{d}$, $R_{qd}$ and $\alpha$, collected in  Table~\ref{tab:fit_parameters} and determined by fitting a linear logarithmic model, Eq.~(\ref{eq:parametrization_of_extrapolation_lin}), to each of them one by one.}
	\label{fig:reBB_model_log_lin_extrapolation_fits}
\end{figure}

The energy dependencies of the scale parameters, $R_q$, $R_d$, and $R_{qd}$ are graphically shown in Figs.~\ref{fig:par_Rq_lin}-\ref{fig:par_Rqd_lin}.
These figures clearly indicate that the energy dependence of the geometrical scale parameters consistent with the same evolution, namely the same linear rise in $\ln(s)$ for both $pp$ and $p\bar p$ scattering: when we fitted these parameters together, with a linear logarithmic function, we have obtained a statistically acceptable fit in each
of these three cases. This result extends and improves the earlier results published in ref.~\cite{Nemes:2015iia} for
elasic $pp$ scattering to the case of both $pp$ and $p\bar p$ collisions in a natural manner. For a comparision, these earlier results are also shown with a dotted red line on the panels of 
Fig. ~\ref{fig:reBB_model_log_lin_extrapolation_fits}, indicating the improved precision of the current analysis, due to more data points are included in the TeV energy range.

\begin{figure}[H]
	\centering
	\includegraphics[width=0.7\linewidth]{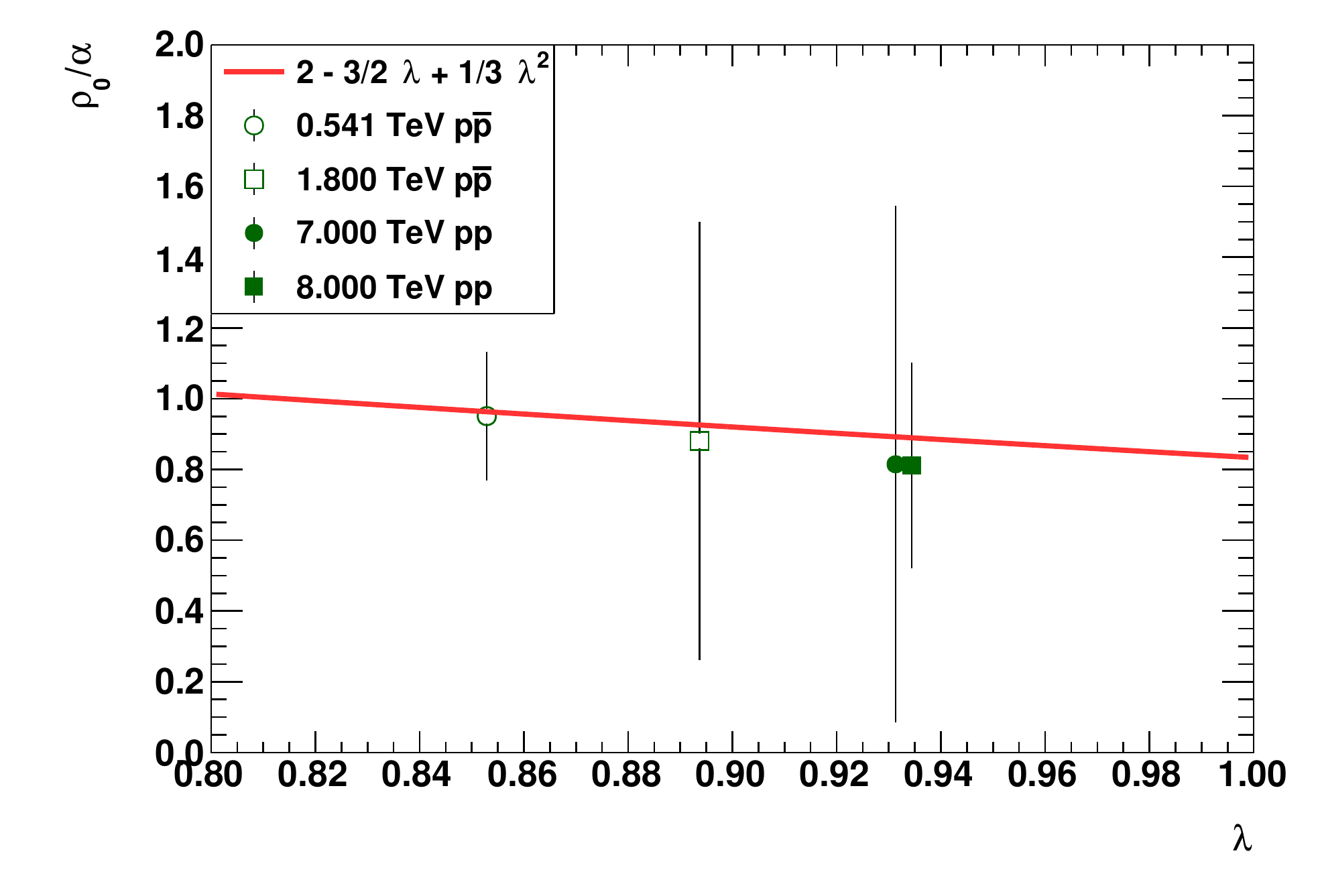}
	\caption{The dependence of $\rho_0/\alpha$ on $\lambda$ in the TeV energy range. The data points are generated numerically by using the trends of the ReBB model parameters, $R_q$, $R_d$, $R_{qd}$, shown in Figs.~\ref{fig:par_Rq_lin}-\ref{fig:par_Rqd_lin} and the experimentally measured ratio $\rho_0$ values. The red curve represents the result of the analytical calculation showing a good agreement with the numerical calculations.}
	\label{fig:rho0-per-alpha-vs-P0-LHC}
\end{figure}

For the opacity parameter $\alpha$, seen on panel (d) of Fig.~\ref{fig:reBB_model_log_lin_extrapolation_fits}, the situation is different: the $pp$ and $p\bar p$ points are not on the same trend, because the $\alpha$ parameters that characterize the dip in the ReBB model, are obtained with great precision both in the $pp$ and in the $p\bar p$ cases. The difference between the excitation functions of
$\alpha^{pp}(s)$ and $\alpha^{p\bar p}(s)$ corresponds to the qualitative difference between the differential cross-section
of elastic $pp$ and $p\bar p$ collisions in the few TeV energy range: 
the presence of a persistent dip and bump structure in the differential cross-section of elastic   $pp$ collisions,
and the lack of a similar feature in elastic $p\bar p$ collisions. Thus in the case of parameter $\alpha$ we have to consider, that there are only two, rather precisely determined  data points in both $pp$ and $p\bar p$ collisions from the presented
ReBB model studied so far. We can already conclude that they cannot be described by a single line as 
an affine linear fit with eq.~(\ref{eq:parametrization_of_extrapolation_lin}) would fail. Without additional information,
we cannot determine the trends and its uncertainties as two points can always connected with a straight line, so an affine
linear description of both the two $pp$ and the two $p\bar p$ data points would have a vanishing $\chi^2$ and an indeterminable
confidence level. This problem, however, is solved by utilizing the results of \ref{sec:appendix-b} on the proportionality between the model parameter $\alpha$ and the experimentally measurable real-to-imaginary ratio $\rho_0$. This proportionality is shown graphically in Fig.~\ref{fig:rho0-per-alpha-vs-P0-LHC}. The constant of proportionality in the few TeV region is an almost energy independent constant value, $\rho_0/\alpha = 0.85 \pm 0.01$, well within the errors of the $\rho_0$ measurements, in agreement with a theoretically obtained function, showed with a red solid line on Fig.~\ref{fig:rho0-per-alpha-vs-P0-LHC} and derived in \ref{sec:appendix-b}. This proportionality allows one to add new datapoints to the trends of $\alpha(s)$ both for the $pp$ and for the $p\bar p$ cases by simply rescaling the mesured $\rho_0$ values.

We found three additional published experimental data of $\rho_0$ for $p\bar p$ collisions, $\rho_0 =  0.135 \pm 0.015$
at $\sqrt{s} = 0.541 $ by the UA4/2 Collaboration in ref.~\cite{Haguenauer:1993kan} and $1.8$ TeV by the E-710 and the E811 collaborations in refs.~\cite{Amos:1991bp,Avila:2002bp}, respectively.
At $\sqrt{s} = 1.8$ TeV, we have utilized the combined value of these E-710 and E811 measurements~\cite{Avila:2002bp}, corresponding to $\rho_0(p\bar p) = 0.135 \pm 0.044$. The constancy of these $\rho_0(s)$ values in the few TeV energy range, when converted with the help of Fig.  ~\ref{fig:rho0-per-alpha-vs-P0-LHC}
to the opacity parameter $\alpha(p\bar p)$ of the Bialas-Bzdak model, leads to
the lack of diffractive minima hence an Odderon signal in elastic $p\bar p$ collisions, leading to an $\alpha(p\bar p) \approx 0.16 \pm 0.06$ which is within its large errors the same as the $\alpha = 0.163 \pm 0.005$ value
obtained from the ReBB model fit to D0 data at $\sqrt{s} = 1.96 $ TeV, summarized on Fig.~\ref{fig:reBB_model_fit_1_96_TeV}. Similarly the $\alpha$ parameter extracted from  $\rho_0$ at $\sqrt{s} = 0.541$ TeV is 
$\alpha \approx 0.16 \pm 0.02$ which is within twice the relatively large errors of the $\rho_0$ analysis the same as the value of $\alpha(p\bar p) = 0.117 \pm 0.002 $ obtained
from the analysis of the differential cross-section, shown  on Fig.~\ref{fig:reBB_model_fit_0_546_TeV}. These indicate a slowly rising value for $\alpha(p\bar p)$ or correspondingly, $\rho_0(p\bar p)$ in the TeV energy range. The final values of these datapoints together with the corresponding errors are connected with a long-dashed line in Panel (d) of Fig.~\ref{fig:reBB_model_log_lin_extrapolation_fits}. Table~\ref{tab:excitation_pars} indicates that for $\alpha(p\bar p)$ the coefficient  $p_1(p\bar p) = 0.018 \pm 0.002$ is a significantly positive number.

For the opacity coefficient in elastic $pp$ collisions, $\alpha(pp)$ on the other hand an oppisite effect is seen, when the $\rho_0$ measurements 
at $\sqrt{s} = 7$ and $8$ TeV are also taken into account, based on 
the data of the TOTEM Collaboration published in refs.~\cite{Antchev:2013iaa,Antchev:2016vpy}. As by now it is very well known,
these values indicate a nearly constant, actually decreasing trend, and based
on the fits of the extracted four data points of $\alpha(pp)$ we find that in the few TeV energy range, this trend is nearly constant, indicated by the solid red line of panel (d) of Fig. ~\ref{fig:reBB_model_log_lin_extrapolation_fits} . Table~\ref{tab:excitation_pars} indicates that for $\alpha(pp)$ the coefficient of increase with $\ln(s)$ is consistent with zero in this energy range, $p_1(pp) 
= - 0.003 \pm 0.003$, which is  significantly less from
the above quoted positive number for $p_1(p\bar p) = 0.018 \pm 0.002$.
Thus it is easy to see, that the Odderon signal in this analysis can be an estimated $6-7\sigma$ effect,
as a consequence of the inequality  $p_1(pp) \neq p_1(p\bar p)$ alone.

In the subsequent sections we first test if the excitation functions,
determined with the help of the  $p_0$ and $p_1$ parameters of Table~\ref{tab:excitation_pars} indeed reproduce the data at all the measured energies in the relevant kinematic range, then we proceed carefully to determine the significance of a model dependent Odderon signal.
We perform these cross-checks against all kind of available data,
including those data that were not utilized
in the determination of the trends for example because their acceptance was too limited to determine all the fit parameters of the ReBB model.

\begin{table*}[tbh]
\caption { Summary of the parameter values which determine the energy dependence by fitting a linear logarithmic model according to Eq.~(\ref{eq:parametrization_of_extrapolation_lin}). The values of the parameters are rounded up to three valuable decimal digits.
For $R_q$, $R_d$ and $R_{qd}$, the values  of the parameters $p_0$ and $p_1$ are given in units of femtometers (fm). For the parameters 
 $\alpha(pp)$ and $\alpha(p\bar p)$, the parameters $p_0$ and $p_1$ are dimensionless.}
\begin{center}
	{\begin{tabular}{|c|c|c|c|c|c|} \hline
	Parameter      & $R_{q}$ [$fm$]  & $R_{d}$ [$fm$]  & $R_{qd}$ [$fm$]  & $\alpha$ ($pp$)      &$\alpha$ ($p\bar p$)  \\ \hline
	$\chi^{2}/NDF$ & $1.596/2$       & $0.469/2$       & $2.239/2$ 	      & $0.760/2$             &$1.212/2$        \\	
	CL [\%]		   & 45.03	         & 79.10           & 32.65 	          & 0.68                 &54.54            \\	\hline
	$p_{0}$         & $0.131\pm0.010$ & $0.590\pm0.015$ & $0.158\pm0.035$  & $0.167\pm0.060$      &$-0.103\pm0.027$  \\ 
	$p_{1}$         & $0.017\pm0.001$ & $0.019\pm0.001$ & $0.010\pm0.002$  & $-0.003\pm0.003$     &$0.018\pm0.002$ 	\\   \hline
	\end{tabular}}
	\end{center}
\label{tab:excitation_pars}
\end{table*}

\section{Sanity tests\label{sec:sanity_tests}}

In this section we show that the determined energy dependence trends are reliable in the kinematic range of $0.546 \le \sqrt{s} \le 8 $ TeV and $0.37\leq-t\leq1.2 $ GeV$^2$. For this purpose we performed the so-called sanity tests: we have cross-checked if the trends summarized in Table~\ref{tab:excitation_pars} indeed represent all the available differential cross-section data on both $pp$ and $p\bar p$ elastic scattering in the mentioned kinematic range. We used both those data which were and which were not utilized in the determination of the energy dependence trends for example because their acceptance was too limited to determine all the fit parameters of the ReBB model. 

To perform these cross-checks, the differential cross sections are fitted with  all the four physical parameters of the ReBB model,
$\alpha(s)$, $R_q(s)$, $R_d(s)$ and $R_{qd}(s)$, fixed to their extrapolated value obtained with the help of the results summarized in Table~\ref{tab:excitation_pars}, while the correlation coefficients of the type B and C errors,
or the $\epsilon$ parameters in the $\chi^2$ definition of eq.~(\ref{eq:chi2-final}) are fitted to the data as free parameters.

\begin{figure}[H]
	\centering
	\includegraphics[width=0.8\linewidth]{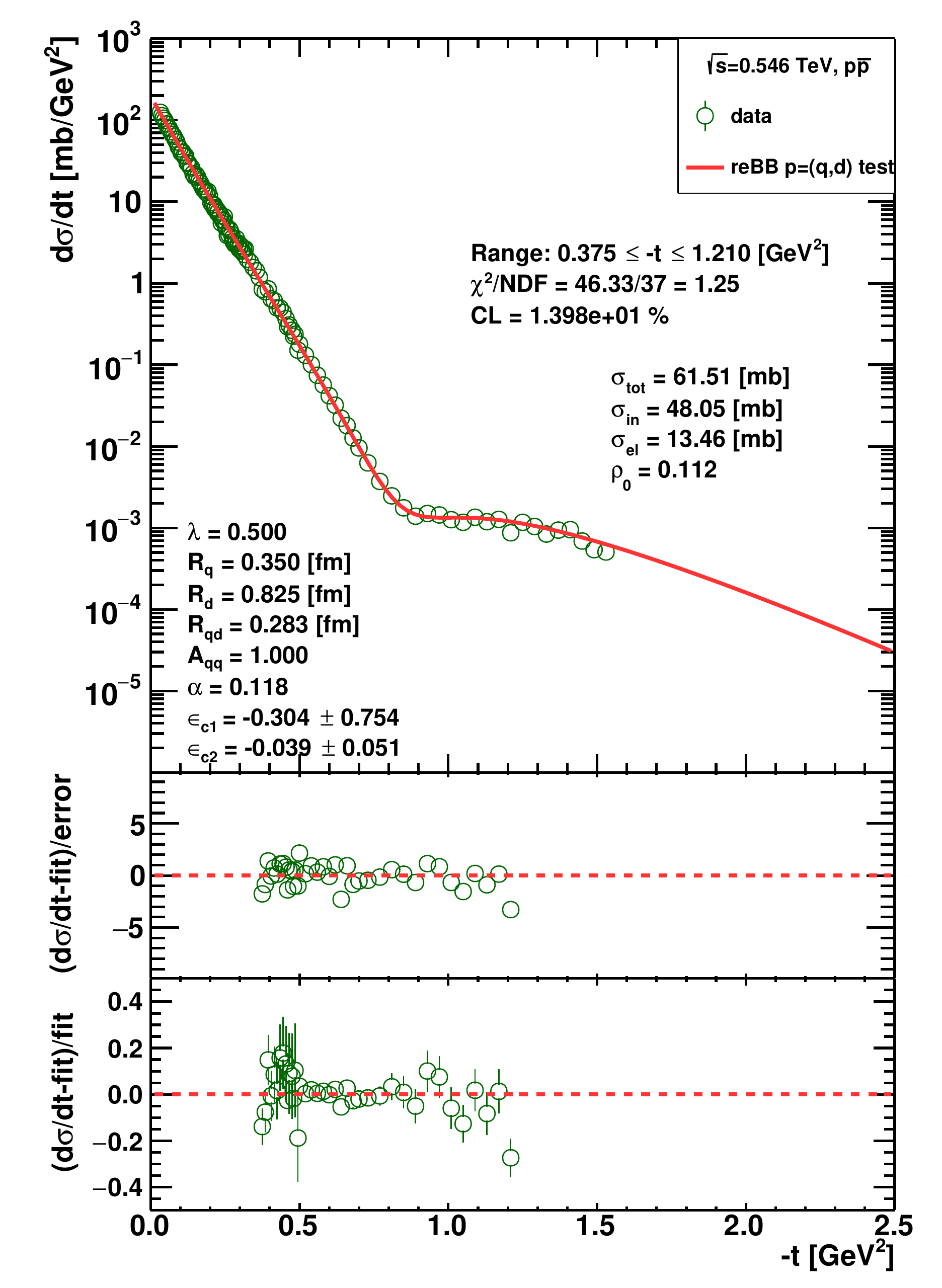}
	\caption{Result of the sanity test for the 0.546 TeV $p\bar p$ elastic differential cross section data \cite{Battiston:1983gp,Bozzo:1985th} in the range of $0.37\leq-t\leq1.2$ GeV$^2$. This sanity test was performed as a fit during which the model parameters $R_q$, $R_d$, $R_{qd}$ and $\alpha$ were fixed to their $s$-dependent value based on Table~\ref{tab:excitation_pars}, while correlation coefficients $\epsilon$-s  in the $\chi^2$ definition, Eq.~(\ref{eq:chi2-final}), were fitted as free parameters. Thus the physical parameters $R_q$, $R_d$, $R_{qd}$ and $\alpha$ are printed on the plot without error bars while the fitted correlation coefficients are given with their errors. The best parameter values are rounded up to three valuable decimal digits.}
	\label{fig:reBB_model_test_0_546_GeV}
\end{figure}

\begin{figure}[H]
	\centering
	\includegraphics[width=0.8\linewidth]{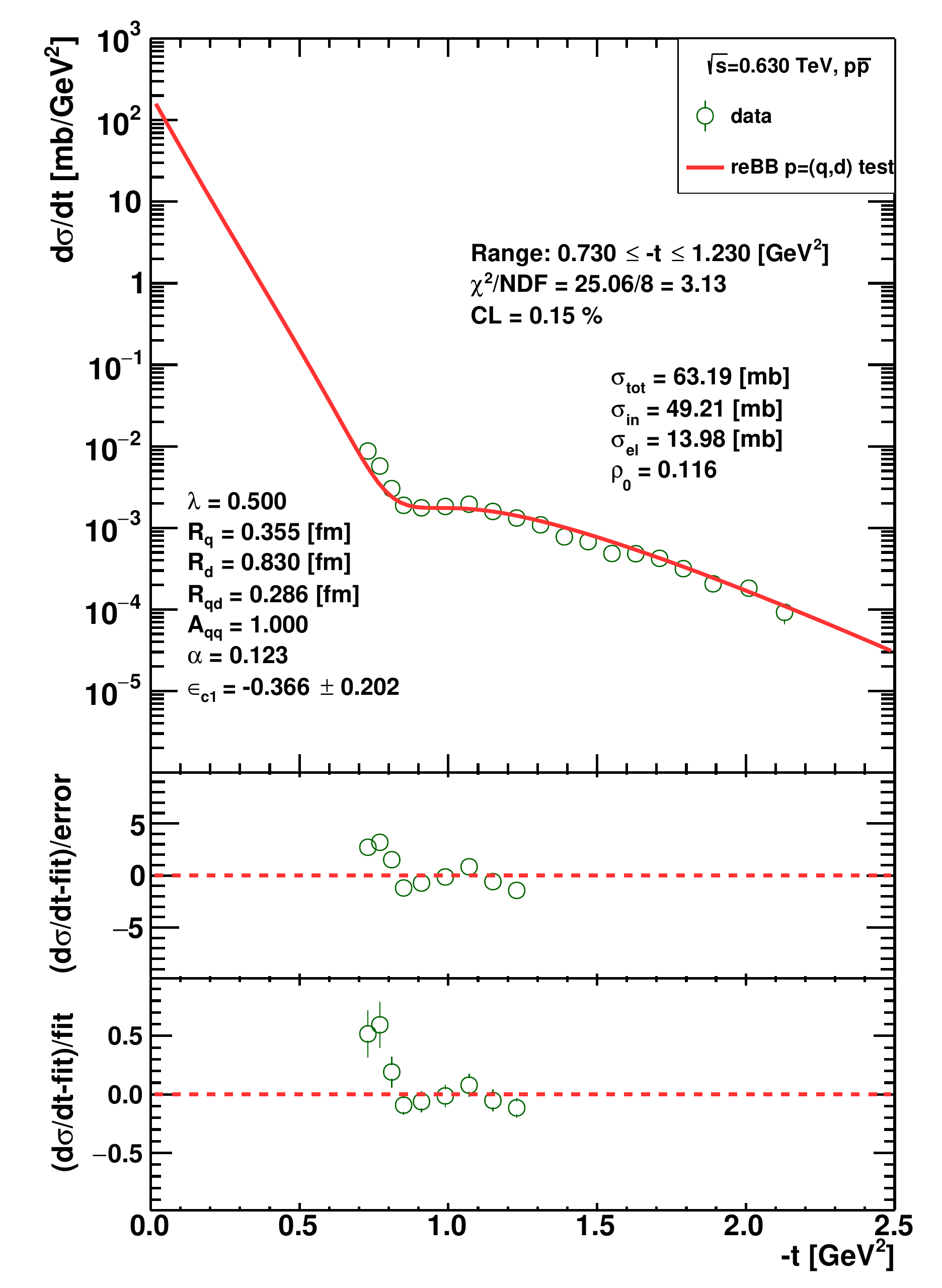}
	\caption{Result of a sanity test, similar to Fig.~\ref{fig:reBB_model_test_0_546_GeV}, but for the $\sqrt{s} = 0.63$ TeV $p\bar p$ elastic differential cross section data of ref.~\cite{Bernard:1986ye}, fitted in the range $0.7\leq-t\leq1.2$ GeV$^2$. }
	\label{fig:reBB_model_test_0_630_GeV}
\end{figure}

\begin{figure}[H]
	\centering
	\includegraphics[width=0.8\linewidth]{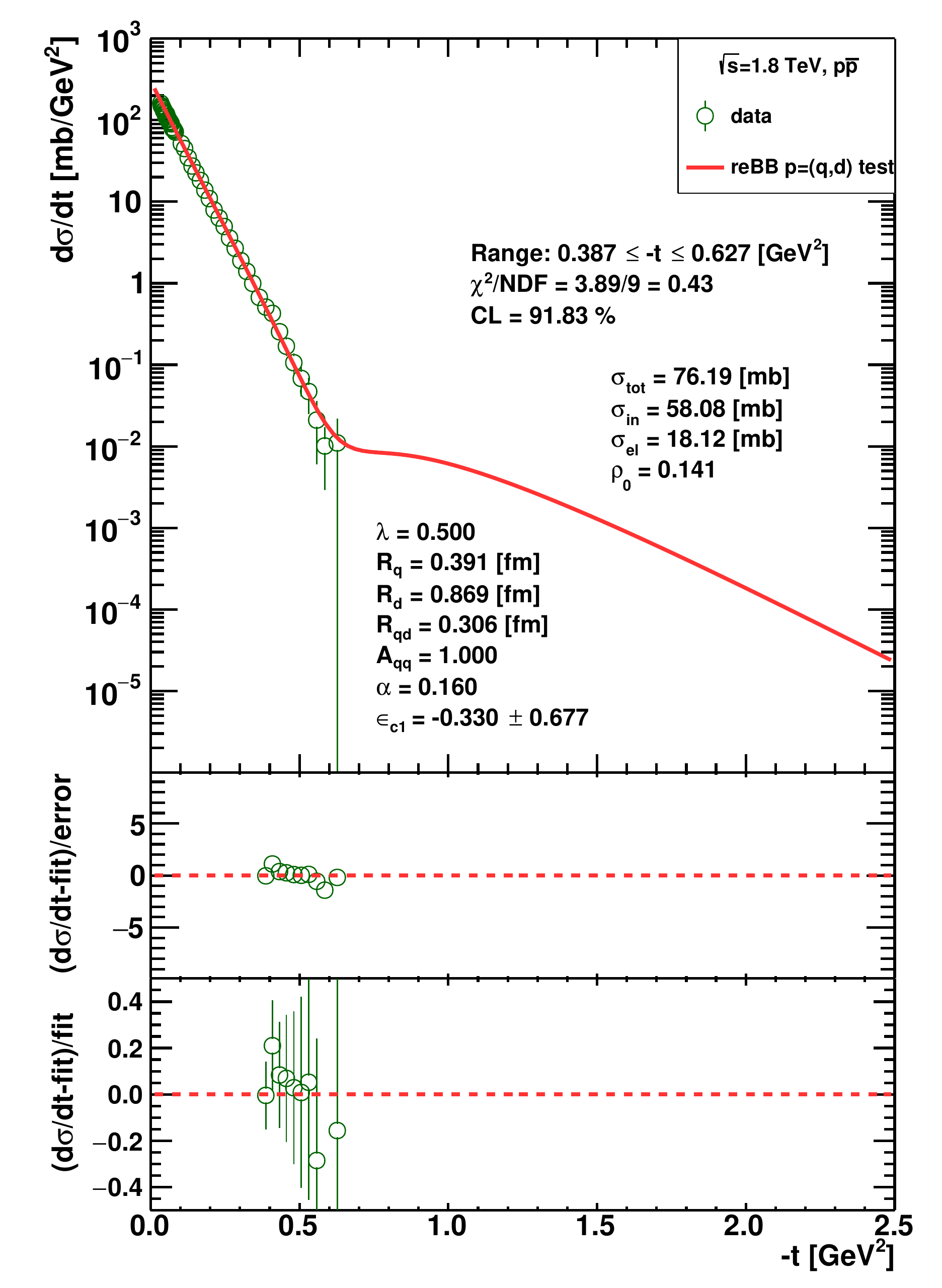}
	\caption{
	Result of a sanity test, same as Fig.~\ref{fig:reBB_model_test_0_546_GeV}, but for the 1.8 TeV $p\bar p$ elastic differential cross section data \cite{Amos:1990fw} in the range of $0.37\leq-t\leq0.6 $ GeV$^2$. 
	}
	\label{fig:reBB_model_test_1_8_TeV}
\end{figure}

\begin{figure}[H]
	\centering
	\includegraphics[width=0.8\linewidth]{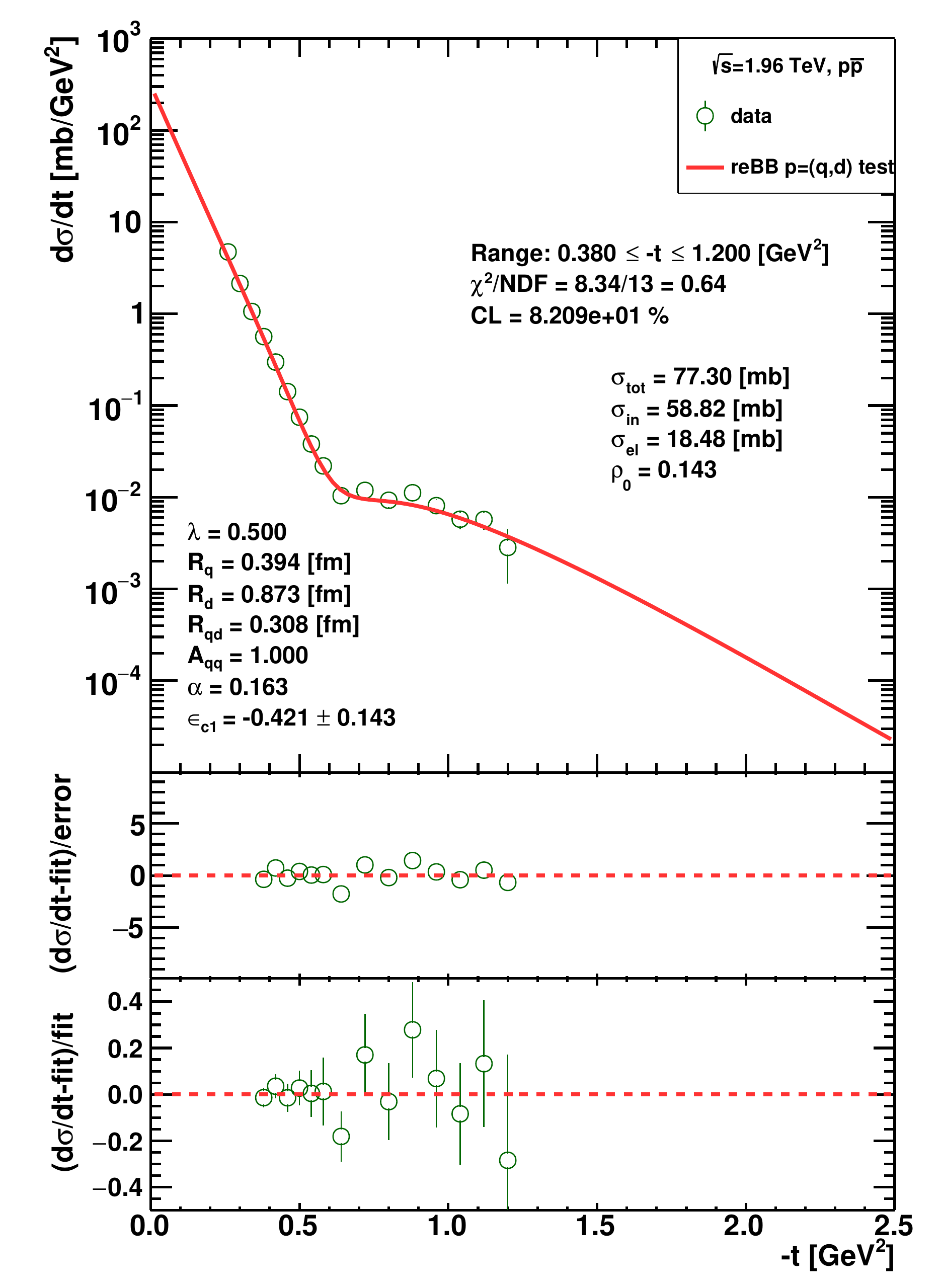}
	\caption{Result of a sanity test, same as Fig.~\ref{fig:reBB_model_test_0_546_GeV}, but 
	for the $\sqrt{s} = 1.96$ TeV $p\bar p$ elastic differential cross section data \cite{Abazov:2012qb} in the range of $0.37\leq-t\leq1.2 $ GeV$^2$. }
	\label{fig:reBB_model_test_1_96_TeV}
\end{figure}

\begin{figure}[H]
	\centering
	\includegraphics[width=0.8\linewidth]{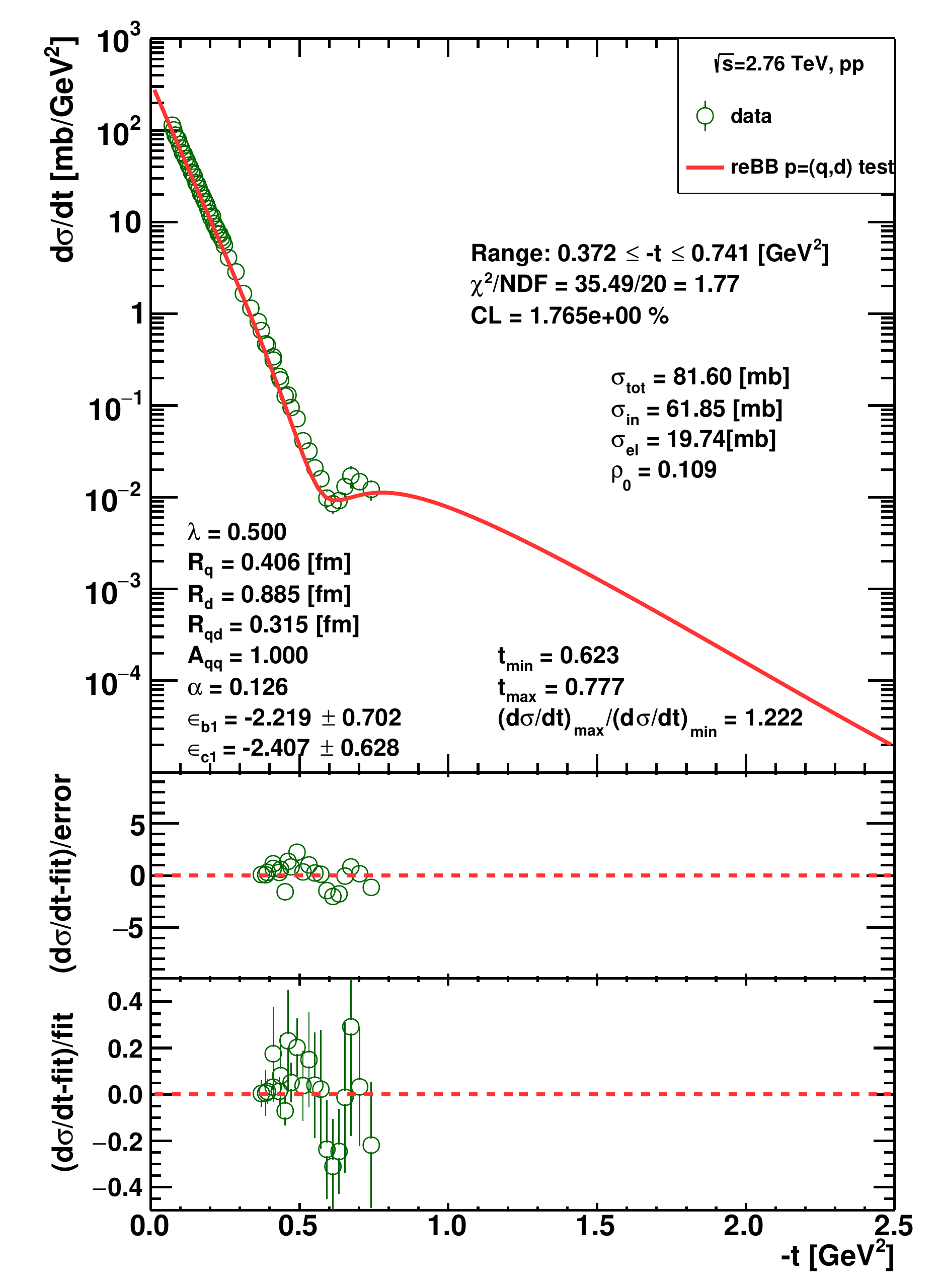}
	\caption{Result of a sanity test, same as Fig.~\ref{fig:reBB_model_test_0_546_GeV}, but 
	for the $\sqrt{s} = 2.76$ TeV $pp$ elastic differential cross section data \cite{Antchev:2018rec} in the range of $0.37\leq-t\leq0.7$ GeV$^2$. 
	}
	\label{fig:reBB_model_test_2_76_TeV}
\end{figure}

\begin{figure}[H]
	\centering
	\includegraphics[width=0.8\linewidth]{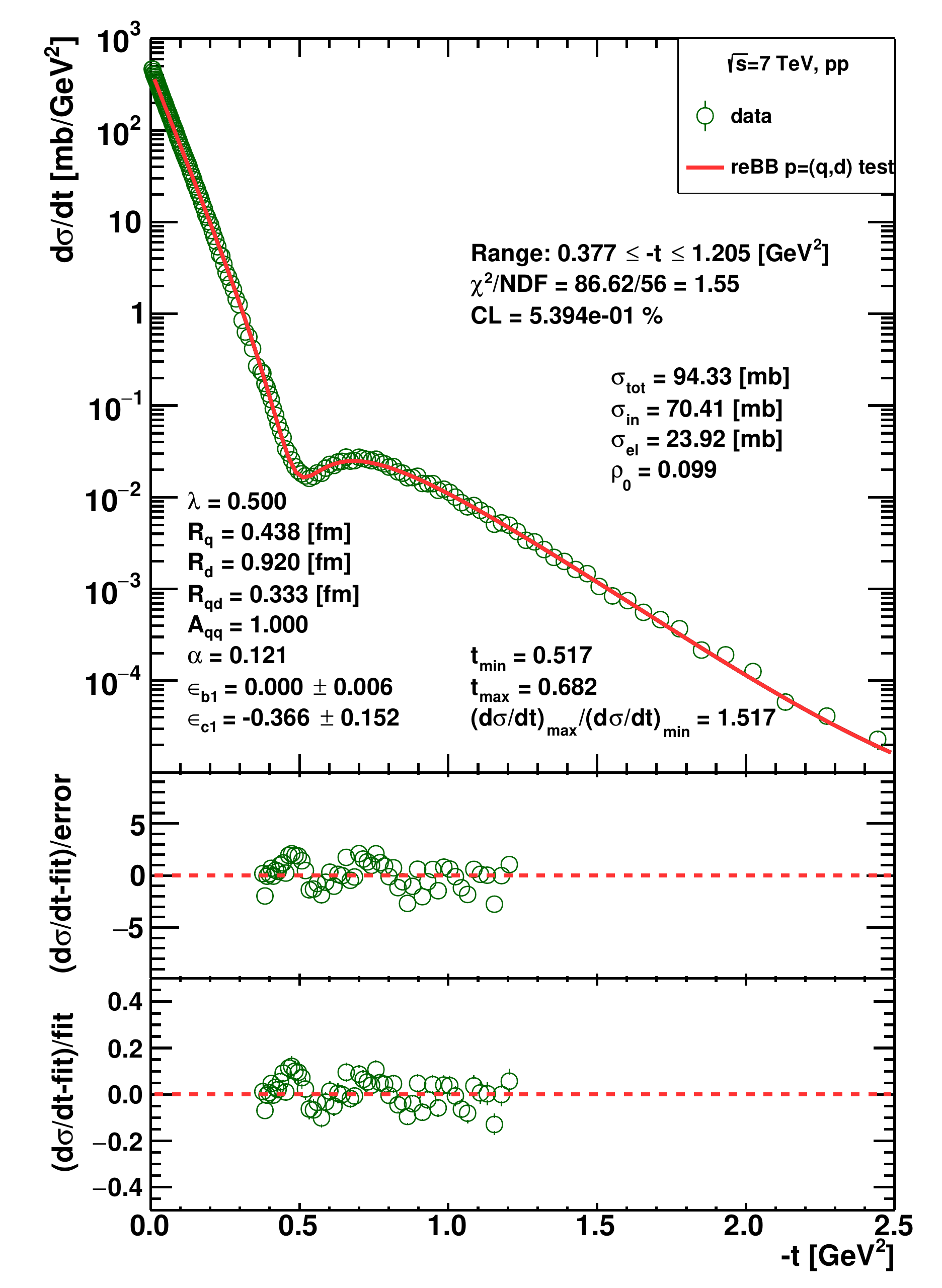}
	\caption{Result of a sanity test, same as Fig.~\ref{fig:reBB_model_test_0_546_GeV}, but 
	for the  $pp$ elastic differential cross section data at $\sqrt{s} = 7$ TeV from ref.~\cite{Antchev:2013gaa}, in the fitted range of $0.37\leq -t \leq 1.2~$GeV$^2$. 
	}
	\label{fig:reBB_model_test_7_TeV}
\end{figure}

The results for the data at $\sqrt{s} = 0.546$, $0.63$, $1.8$, $1.96$, $2.76$ and $7$ TeV are shown in Figs.~\ref{fig:reBB_model_test_0_546_GeV}-\ref{fig:reBB_model_test_7_TeV}. All of these sanity tests resulted in the description of 
these data with a statistically acceptable confidence level of CL $ \geq $ 0.1 \%.

\begin{figure}[H]
	\centering
	\includegraphics[width=0.8\linewidth]{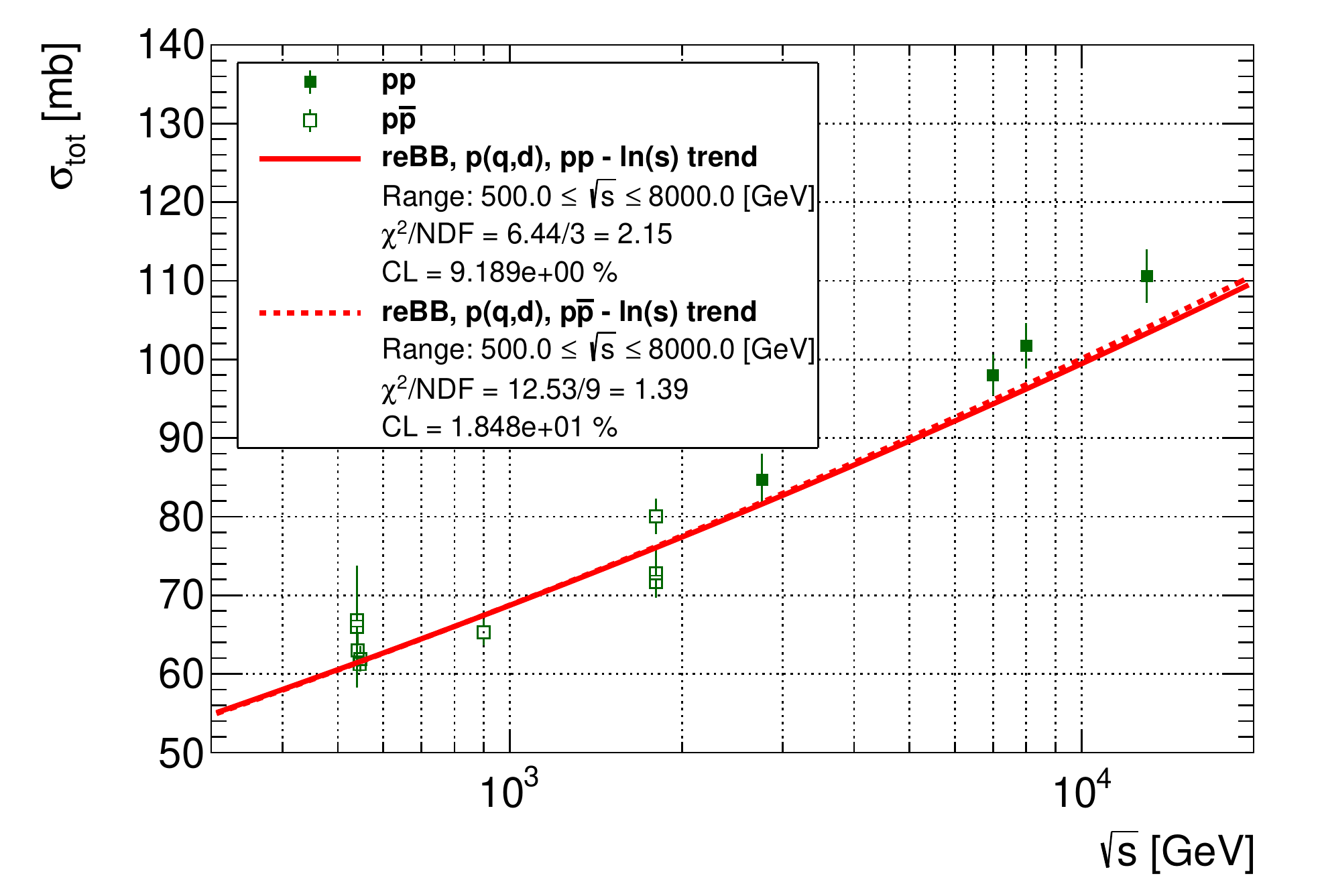}
	\caption{Result of the sanity test for $pp$ \cite{Antchev:2013iaa,Antchev:2013paa,Antchev:2017dia} and $p\bar p$ \cite{Tanabashi:2018oca} total cross section data. It was calculated from the model when the values of the parameters $R_q$, $R_d$, $R_{qd}$ and $\alpha$ were taken from eq.~\ref{eq:parametrization_of_extrapolation_lin} and Table~\ref{tab:excitation_pars}, corresponding to the linear curves shown on panels (a)-(d) of Fig.\ref{fig:reBB_model_log_lin_extrapolation_fits}. }
	\label{fig:reBB_model_test_sig_tot}
\end{figure}

\begin{figure}[H]
	\centering
	\includegraphics[width=0.8\linewidth]{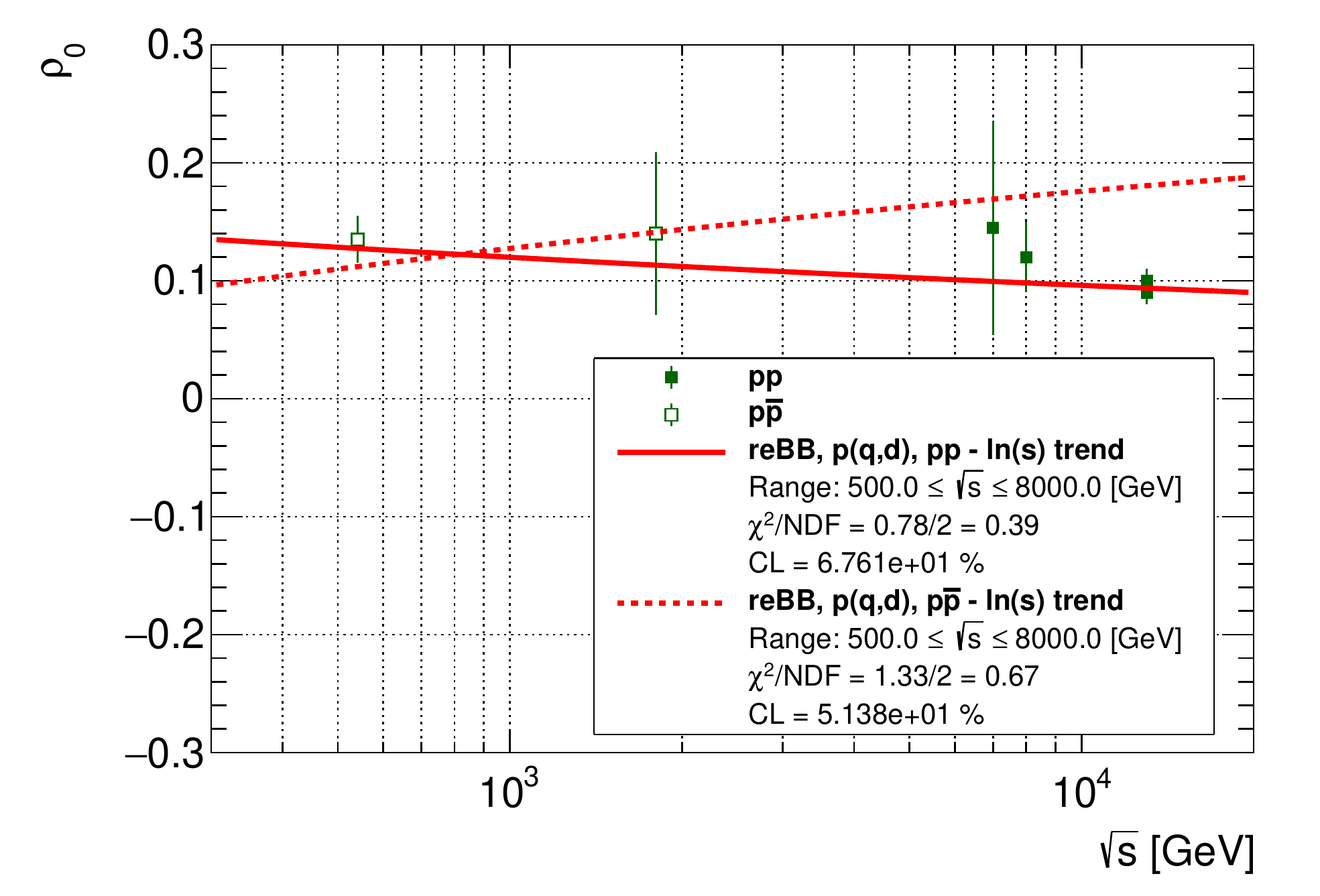}
	\caption{Sanity test result for $pp$ \cite{Antchev:2013iaa,Antchev:2016vpy,Antchev:2017yns} and $p\bar p$ \cite{Tanabashi:2018oca} parameter $\rho_0$ data, as calculated from the model when the values of the parameters $R_q$, $R_d$, $R_{qd}$ and $\alpha$ were taken from eq.~\ref{eq:parametrization_of_extrapolation_lin} and Table~\ref{tab:excitation_pars}, corresponding to the linear curves shown on
    panels (a)-(d) of Fig.\ref{fig:reBB_model_log_lin_extrapolation_fits}.
	On this plot, a model dependent Odderon effect is clearly identified: it corresponds to
	 $\rho_0^{pp}(s)\neq\rho_0^{p\bar p}(s)$, the non-vanishing difference between the excitation functions of $\rho_0$ for $pp$ and $\rho_0$  for  $p\bar p$ collisions, as detailed in \ref{sec:appendix-c}.
}
	\label{fig:reBB_model_test_rho}
\end{figure}

As an additional sanity test, we have  also cross-checked if this ReBB model describes the  $pp$ and $p\bar p$ total cross section $\sigma_{\rm tot}(s)$ and  real to imaginary ratio $\rho_0(s)$ data in a statistically acceptable manner, or not.
These results are presented in  Fig.~\ref{fig:reBB_model_test_sig_tot} and Fig.~\ref{fig:reBB_model_test_rho}, respectively.
As the calculated confidence levels are higher than 0.1  \% in all of these cases, we  can happily conclude that the 
energy dependent  trends of the ReBB model are really reasonable and  reliable in the investigated $0.541 \le \sqrt{s} \le 8$ TeV
energy and in the $0.377 \le -t \le 1.2$ GeV$^2$ squared four-momentum transfer range. Thus this model can be used reliably to 
extrapolate both the $pp$ and the $p\bar p$ differential cross-sections in this limited kinematic range of $(s,t)$, based only on 10
physical model parameters, summarized in Table~\ref{tab:excitation_pars}.

\clearpage

\clearpage

\section{Extrapolations\label{sec:extrapolations}}

According to our findings in Section~\ref{sec:excitation_functions} the energy dependencies of the scale parameters $R_{q}$, $R_{d}$ and $R_{qd}$ are identical for $pp$ and $p\bar p$ scattering, only the energy dependence of the opacity parameter $\alpha$ differs. The statistically acceptable quality of the fits shown in Fig.~\ref{fig:reBB_model_log_lin_extrapolation_fits} 
and the success of the sanity tests performed in the previous section allow for a reliable extrapolation  of the
differential cross-sections of elastic $pp$ and $p\bar p$ collisions with the help of the ReBB model~\cite{Nemes:2015iia},
limited to the investigated $0.541 \le \sqrt{s} \le 8$ TeV
center of mass energy and in the $0.377 \le -t \le 1.2$ GeV$^2$ four-momentum transfer range.

We extrapolate, in the TeV energy range, the $pp$ differential cross sections to energies where measured $p\bar p$ data exist and the other way round, the $p\bar p$ differential cross sections to energies where measured $pp$ data exist. Thus three of such extrapolations were performed: $pp$ extrapolation to $\sqrt{s} = 1.96$ TeV,
to compare it to the 1.96 TeV D0 $p\bar p$ $d\sigma/dt$ data, and $p\bar p$ extrapolations 
to $\sqrt{s} = 2.76$ and $7$ TeV, to compare them to the  $d\sigma/dt$ $pp$ data measured by TOTEM at these energies.

Since the energy dependencies of the scale parameters $R_{q}$, $R_{d}$ and $R_{qd}$ are identical for $pp$ and $p\bar p$ scattering, as discussed in Sec.~\ref{sec:excitation_functions}, in the course of the extrapolations their values are fixed at their fitted values given in Tab.~\ref{tab:fit_parameters}, furthermore, since the energy dependence of the $\alpha$ parameter differs for $pp$ and $p\bar p$ scattering, the $\alpha(pp)$ and  $\alpha(p\bar p)$ values are fixed from their energy dependence trend seen in Fig.~\ref{fig:par_alpha_lin}. In addition, during the extrapolations, the $\epsilon$ parameters in the $\chi^2$ definition, Eq.~(\ref{eq:chi2-final}), were optimized, furthermore the last two terms in Eq.~(\ref{eq:chi2-final}), i.e., the total cross section and $\rho_0$-parameter term, were not included. This way we handled the type B and type C errors of the published $pp$ differential cross-section  to match these data as much as possible to the differential cross-section of elastic  $p\bar p$ collisions within the allowed systematic errors, and vice versa. 

\begin{figure}[H]
	\centering
	\includegraphics[width=0.8\linewidth]{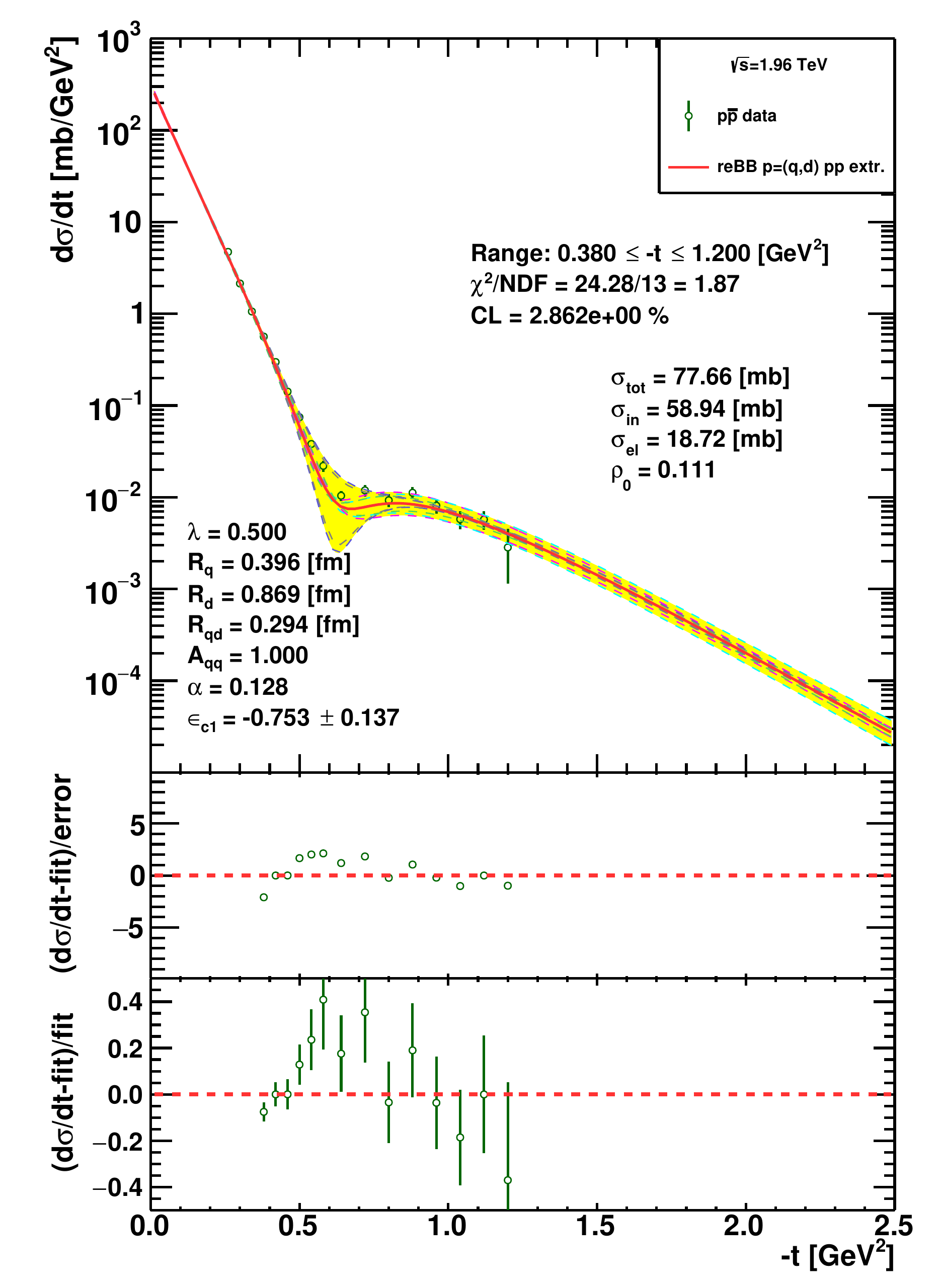}
	\caption{The ReBB model extrapolation for the $pp$ $d\sigma/dt$ at $\sqrt{s}=1.96$~TeV compared to the $p\bar p$ D0 $d\sigma/dt$ data \cite{Abazov:2012qb} measured at the same energy. The yellow band is the uncertainty of the extrapolation. The calculated CL value between the extrapolated model and the measured data does not indicate a significant difference between the $pp$ and $p\bar p$ differential cross sections.}
	\label{fig:reBB_model_extr_1_96_TeV}
\end{figure}

\begin{figure}[H]
	\centering
	\includegraphics[width=0.8\linewidth]{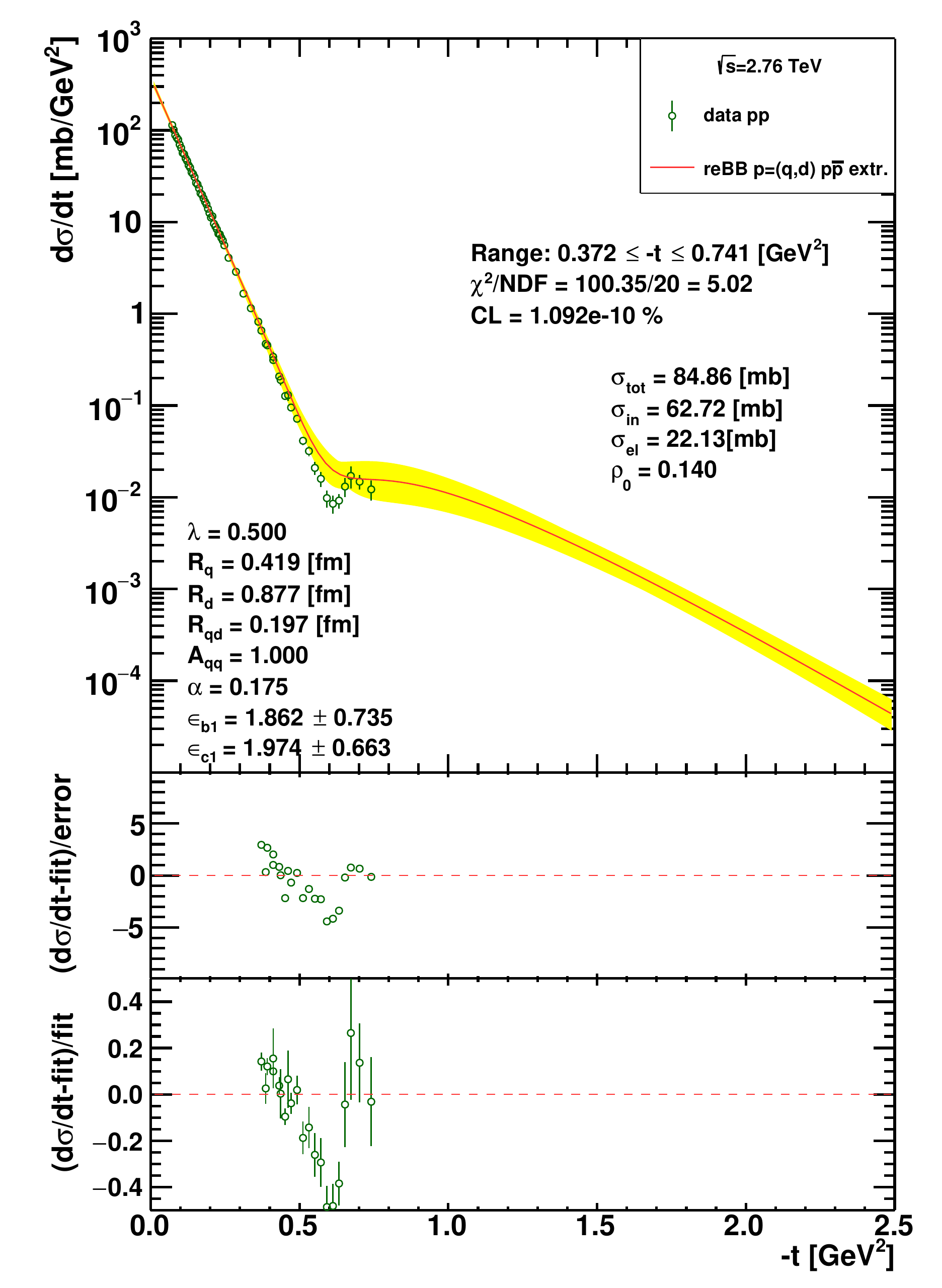}
	\caption{The ReBB model extrapolation for the $p\bar p$ $d\sigma/dt$ at $\sqrt{s}=2.76$~TeV compared to the $pp$ TOTEM $d\sigma/dt$ data\cite{Antchev:2018rec} measured at the same energy. The yellow band is the uncertainty of the extrapolation. The calculated CL value between the extrapolated model and the measured data  indicates a significant difference between the $pp$ and $p\bar p$ differential cross sections, corresponding to a 7.1 $\sigma$  significance for the $t$-channel Odderon exchange.}
	\label{fig:reBB_model_extr_2_76_TeV}
\end{figure}

\begin{figure}[H]
	\centering
    \includegraphics[width=0.8\linewidth]{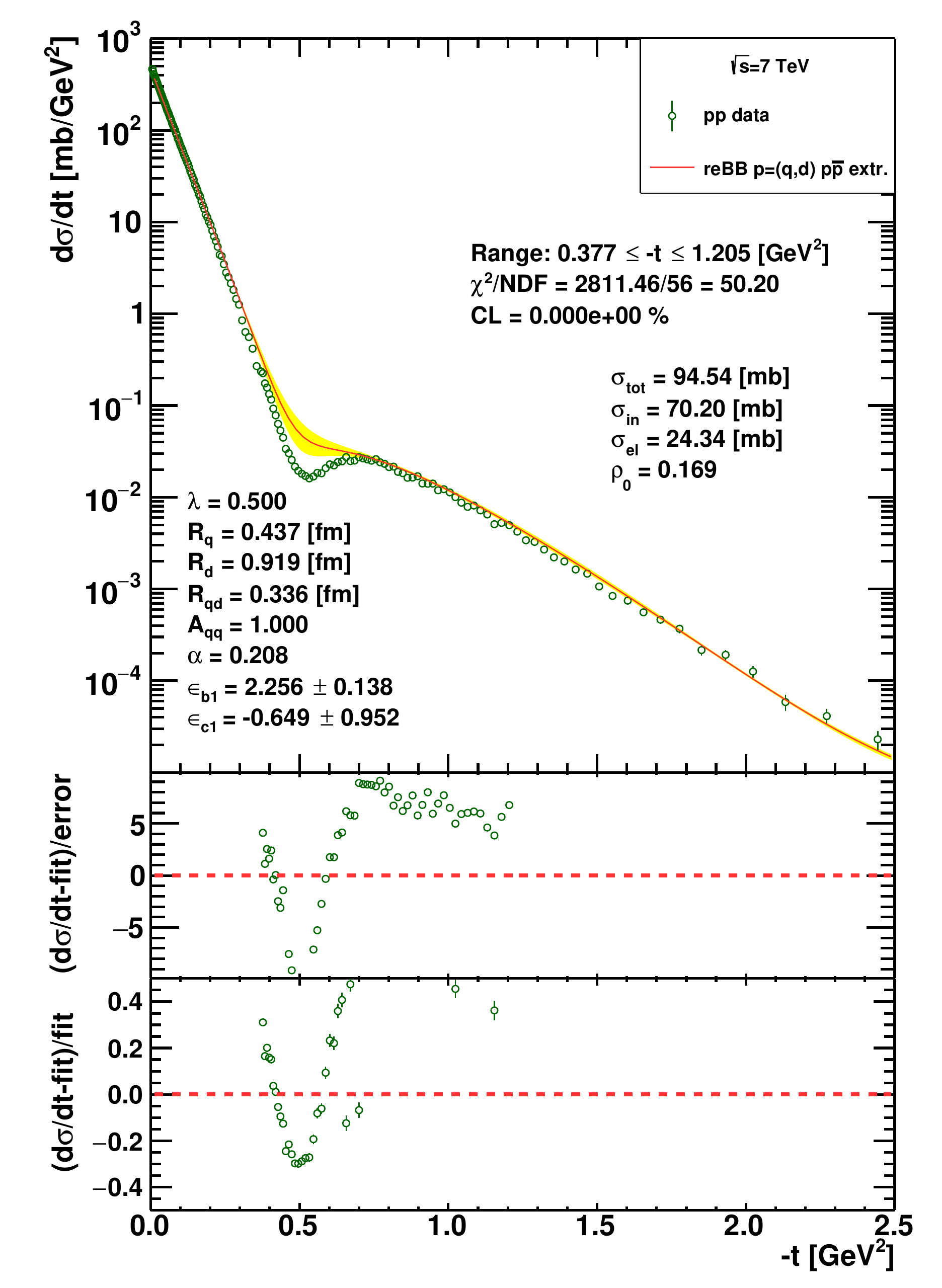}
	\caption{The ReBB model extrapolation for the $p\bar p$ $d\sigma/dt$ at $\sqrt{s}=7$~TeV compared to the $pp$ TOTEM $d\sigma/dt$ data \cite{Antchev:2013gaa} measured at the same energy. The yellow band is the uncertainty of the extrapolation. The calculated CL value between the extrapolated model and the measured data indicates a significant difference between the $pp$ and $p\bar p$ differential cross sections, hence a significant Odderon effect, that is dominant around the dip region.}
	\label{fig:reBB_model_extr_7_TeV}
\end{figure}

The results of the extrapolations are shown in Fig.~\ref{fig:reBB_model_extr_1_96_TeV}, Fig.~\ref{fig:reBB_model_extr_2_76_TeV} and Fig.~\ref{fig:reBB_model_extr_7_TeV}. The error band around these extrapolations is also evaluated, based on the envelope of one standard deviation errors
of the $R_q(s)$, $R_d(s)$, $R_{qd}(s)$ model parameters and the $p_0$ and $p_1$ parameters of $\alpha(s)$. As an example, the resulting ten curves -- considering that the values of the scale parameters are taken from the original fit while the value $\alpha$ is taken from the trend -- are explicitly shown for 1.96 TeV in Fig.~\ref{fig:reBB_model_extr_1_96_TeV}.

While at $\sqrt{s} = 1.96$ TeV no statistically significant difference is observed between the extrapolated $pp$ and measured $p\bar p$ differential cross sections, at $\sqrt{s} = 2.76$ and $7$ TeV,  remarkable 
and statistically significant differences can be observed. In Figs.~\ref{fig:reBB_model_extr_2_76_TeV} and \ref{fig:reBB_model_extr_7_TeV}, even an untrained  eye can see, 
that the dip is filled in case of elastic $p\bar p$ scattering, while it is not filled in elastic
$pp$ scattering. Thus we confirm the prediction of ref.~\cite{Donnachie:1983hf}, that predicted, based on a
three-gluon exchange picture that dominates at larger values of $-t$, that the dip will be filled in high energy $p\bar p$
elastic collisions.

In this work, the  differences between elastic $pp$ and $p\bar p$ collisions are quantified 
by  the confidence levels obtained from the comparision of the  extrapolated curves
to the measured data: at 2.76 TeV, the hypothesis that these extrapolations agree with the data is  characterized by a $CL = 1.092 \times 10^{-10}$ \%, while at 7 TeV, CL = 0 \%. Theoretically the observed difference can be attributed only to the effect of a C-odd exchange, as detailed recently in refs.~\cite{Csorgo:2019ewn,Csorgo:2020msw,Csorgo:2020rlb}. At the TeV energy scale, the secondary Reggeon exchanges are generally known to be negligible. This effect has been also specifically cross-checked and confirmed recently in ref.~\cite{Szanyi:2018ain}. Thus in the few TeV energy range of the LHC,
the only source of a difference between the differential cross-sections of elastic $pp$ and $p\bar p$ collisions can be a $t$-channel Odderon exchange. In the modern language of QCD, the Odderon exchange corresponds to the exchange of C-odd colorless bound states consisting of odd number of gluons~\cite{Bartels:1999yt,Donnachie:1990wd,Donnachie:1983hf}. 

Thus the CL, calculated for the 2.76 TeV $p\bar p$ extrapolation, 
corresponds to an Odderon observation with a probability of $ P $ $ = $ $1 - CL$ $ = $ $ 1 - 1.092 \times 10^{-12}$.
This corresponds to a $\chi^2/NDF = 100.35/20$ and to a 7.12  $\sigma$ model dependent significance for the observation of a $t$-channel Odderon exchange, and the existence of the colorless  bound states containing odd number of gluons.
When extrapolating the $pp$ differential cross-sections from 2.76 down to 1.96 TeV,
however, significance is lost, corresponding to a $\chi^2/NDF = 24.28/13$ and to a 2.19  $\sigma$ effect, less than a 3 $\sigma$ effect in this comparison. However, these two significances at 1.96 and 2.76 TeV can be combined, providing a combined $\chi^2/NDF = 124.63/33$,
that corresponds to a statistically significant, 7.08 $\sigma$ effect.

This 7.08 $\sigma$ combined significance increases to 
an even larger significance of an Odderon observation,
when we extrapolate the differential cross-section of elastic proton - anti\-proton collisions
to $\sqrt{s} = 7.0$ TeV,
where the probability of Odderon observation becomes practically unity. Given that a 7.08  $\sigma$ effect is already well above the usual 5 $\sigma$, statistically
significant discovery level, we
quote this as the possibly lowest level of the significance of our model-dependent Odderon observation.

As already mentioned in the introduction we have also been recently involved in a truly  model-independent search for Odderon effects in the comparision of the scaling properties of the differential cross-sections of elastic $pp$ and $p\bar p$ collisions
in a similar $s$ but in the complete available $t$ range. As compared to the model-dependent studies
summarized in this manuscript, the advantage of the model-independent scaling studies of refs.~\cite{Csorgo:2019ewn,Csorgo:2020msw,Csorgo:2020rlb} is that they scale out all the effects from the differences between
$pp$ and $p\bar p$ elastic collisions due to possible differences in their 
$\sigma_{\rm el}(s)$,
$B(s)$ and their product, the $\sigma_{\rm el}(s)
B(s)$ = $\sigma_{\rm tot}^2(s) (1 +  \rho^2_0(s))$ functions. As part of the Odderon signal in the ReBB model is apparently in the difference between
the $\rho_0(s)$ excitation functions for $pp$ and $p\bar p$ collisions, the significance of the Odderon signal is reduced
in this model independent analysis. When considering the interpolations as theoretical curves, the significance is reduced to
a 6.55 $\sigma$ effect~\cite{Csorgo:2019ewn}, but when considering that the interpolations between experimental data
have also horizontal and vertical, type A and type B errors, the signicance of the Odderon signal is further reduced to
a 6.26 $\sigma$ effect~\cite{Csorgo:2020msw,Csorgo:2020rlb}. Thus we conclude that the Odderon is now discovered, both in a model-dependent and in a model-independent manner, with a statistical significance that is well above the 5 $\sigma$ discovery limit
of high energy particle physics.

\begin{figure}[H]
	\centering
    \includegraphics[width=0.85\linewidth]{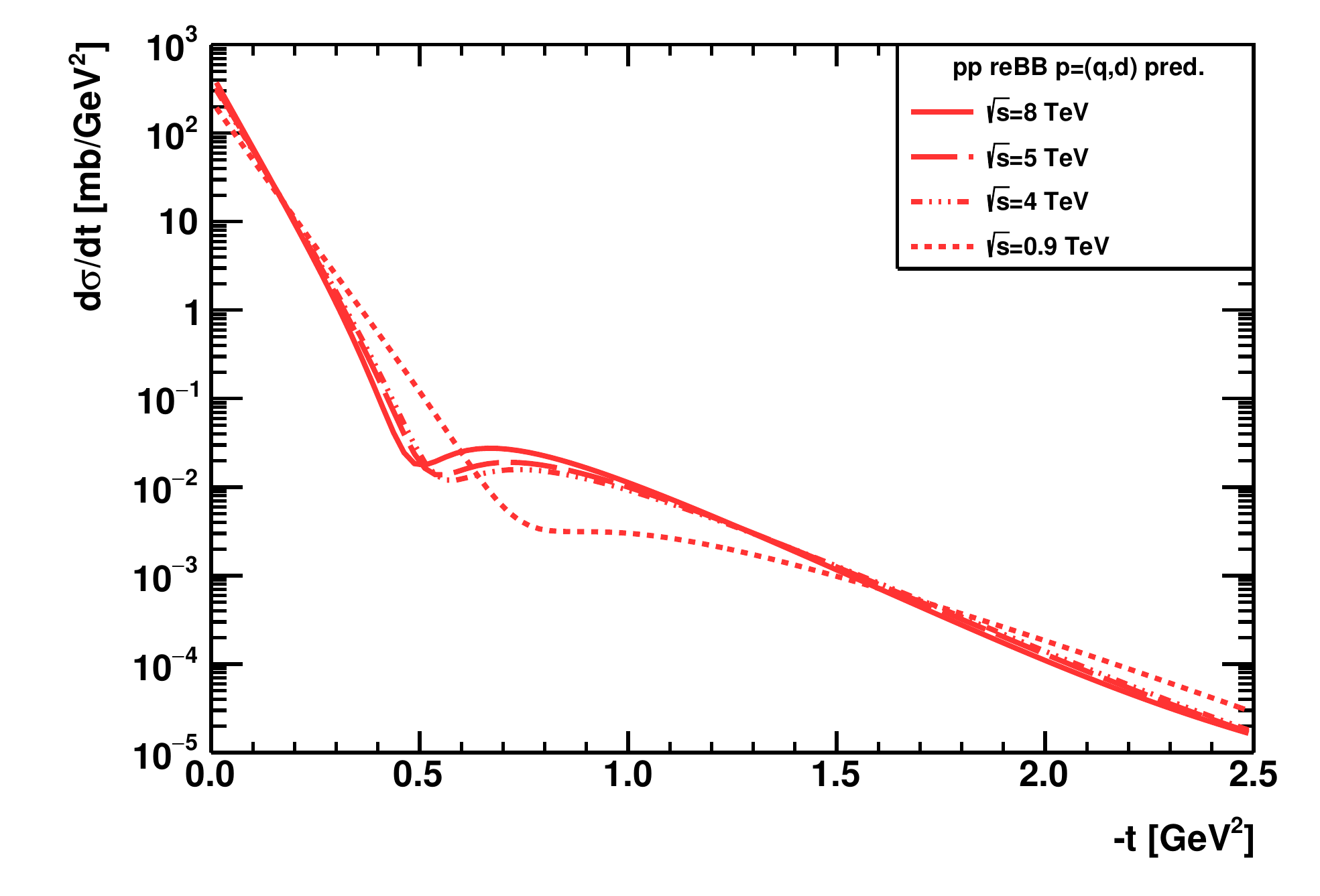}
	\caption{Predictions from the ReBB model, for the $d\sigma/dt$ of elastic $pp$ collisions at $\sqrt{s}=8$, 5, 4, and 0.9 TeV.}
	\label{fig:reBB_model_extr_8_TeV}
\end{figure}

Finally we close this section with the predictions to the experimentally not yet available large-$t$ differential cross-section of $pp$ collisions at $\sqrt{s} = 0.9$, $4$, $5$ and $ 8$ TeV shown in Fig.~\ref{fig:reBB_model_extr_8_TeV}.

\clearpage

\section{Discussion \label{sec:discussion}}

In the previous sections, we have investigated what happens if we interpret the data in terms of a particular model,
the Real Extended Bialas-Bzdak Model. This allows also to consider the Odderon signal in the excitation function of
the model parameter $\alpha$. We have shown in ~\ref{sec:appendix-b} that this model parameter is proportional to the experimentally
measured parameter $\rho_0$, the ratio of the real to the imaginary part of the scattering amplitude at the optical point,
and related the coefficient of proportionality to the value of the imaginary part of the scattering amplitude at vanishing impact parameter, $\lambda(s)=Imt_{el}(s,b=0)$,
for the $\sqrt{s} \leq 8$ TeV elastic proton-proton collisions, and we have shown that within the framework of this ReBB model, the very different trend of $\rho_0(s)$ in proton-proton and in proton-antiproton collisions enhances the model-independent
Odderon signal, from a 6.26 $\sigma$ and 6.55 $\sigma $ effect to a combined, at least $7.08 $ $\sigma$ effect.

Recently, the TOTEM Collaboration concluded, that only one condition is yet to be satisfied to see a statistically significant
Odderon signal: namely, the logarithmically small energy gap between the lowest TOTEM energy of $\sqrt{s} = 2.76$ TeV at LHC and the highest D0 energy of 1.96 TeV at Tevatron needs to be closed. This energy gap has been closed in a model-independent way in refs.~\cite{Csorgo:2020msw,Csorgo:2020rlb,Csorgo:2019ewn}, using the scaling properties of elastic scattering, and by comparing the $H(x) = \frac{1}{B \sigma_{el}} \frac{d\sigma}{dt}$
scaling functions of elastic proton-proton and proton-antiproton collisions, as a function of $x = - t B$
at $\sqrt{s} = 1.96$, $2.76$ and $7.0$ TeV. The advantages of that method, 
with respect to comparing the cross sections directly include the scaling out of
the $s$-dependencies of 
$\sigma_{\rm el}(s)$,
$B(s)$ and their product, $\sigma_{\rm el}(s)
B(s)$ = $\sigma_{\rm tot}^2(s) (1 +  \rho^2_0(s))$,
as well as
the normalization of the $H(x)$ scaling function that cancels the point-to-point correlated and $t$-independent normalization errors. The validity of the $H(x)$ scaling 
for $pp$ collisions and its violation in $p \bar p$ collisions in the few TeV energy range
resulted in a discovery level statistical significance of an Odderon signal, 
characterized in refs.~\cite{Csorgo:2020msw,Csorgo:2020rlb,Csorgo:2019ewn} to be at least $6.26$ $\sigma$,
model independently, based on a careful interpolation of the experimental data-points, their point-to-point fluctuating,
point-to-point correlated and data point dependent as well as point-to-point correlated and data point independent errors.
If these errors are considered as errors on a theory curve, then the significance goes up to at least $6.55$ $\sigma$~\cite{Csorgo:2019ewn}. 

In high energy particle physics, the standard accepted discovery threshold corresponds to a 5$\sigma$ effect.
In the previous section, we have shown, that the statistical significance of an Odderon observation in the limited
 $0.541 \le \sqrt{s} \le 8$ TeV center of mass energy and in the $0.377 \le -t \le 1.2$ GeV$^2$ four-momentum transfer range is at least
 a combined 7.08  $\sigma$ effect, corresponding to a statistically significant and model dependent Odderon observation.

The $\sqrt{s} = 7 $ TeV $pp$ differential cross-sections are measured with asymmetric type B errors. In order to make sure that our results are reliable and reproducible, we have performed several cross-checks to test the reliability of our fit at $\sqrt{s} = 7 $ TeV. One of these tests related to the handling of the asymmetric type B, $t$-dependent systematic errors. We have performed cross-checks for taking at every point either the smaller or the
larger of the up and down type B errors to have a lower or an upper limit on their effects.
We found that the parameters of the ReBB model remained stable for such a symmetrization of 
the type B systematic errors, as the modification of the fit parameters due to such a symmetrization was within the quoted errors on the fit parameters. Our final fits, presented before, were done with asymmetric type B errors, as detailed in Section~\ref{sec:fit_results}.
So we conclude that our  fit  at $\sqrt{s} = 7 $ TeV is stable even for the symmetrization of the type B systematic errors.

We have also investigated the stability of our result for the case, when the energy range is extended towards lower values of $\sqrt{s}$, in the ISR energy range, detailed in \ref{sec:appendix-d}. When the $\sqrt{s} = 23.5$ GeV energy data are included to those summarized in Table~\ref{tab:fit_parameters}, the energy dependence of the model parameters becomes quadratic in $\ln(s)$. This provides  3x5 = 15 model parameters for this broader energy range, as summarized in
Table~\ref{tab:excitation_pars_quadratic} and detailed in \ref{sec:appendix-d}. 
This way, the non-linear terms are confirmed to be negligibly small in the TeV energy range, where
we find the significant Odderon effects, with the help of as little as only 10 model parameters. These 10 parameters are given in Table ~\ref{tab:excitation_pars}.

It turns out in Sec.~\ref{sec:fit_results}, that the ReBB model as presented in ref.~\cite{Nemes:2015iia} does  not yet provide a statistically acceptable fit quality  to the differential cross-section of $\sqrt{s} = 13$ TeV elastic $pp$ scattering. This might be due to the emergence of the black-ring limit of elastic proton-proton scattering instead of the expected
black-disc limit. In what follows we shortly discuss the earlier and more recent results on the black ring shaped interaction region of the colliding protons.

A complementary way of studying the high-energy scattering processes is by passing from the momentum transfer $t$ to the impact parameter $b$. In 1963 van Hove introduced the inelasticity profile or the overlap function~\cite{van1963phenomenological,VanHove:1964rp}, which corresponds to the impact parameter distribution of the inelastic cross section characterizing the shape of the interaction region of two colliding particles. The natural expectation is that the most inelastic collisions are central, i.e., the inelasticity profiles have a maximum at $b=0$ consistently with the black disc terminology. The possibility of a minimum at $b = 0$, i.e., the peripheral form of the inelastic function was first considered in Ref.~\cite{TROSHIN1993175} which implies the shape of a black ring rather than that of a black disc. 

In Ref.~\cite{Dremin:2014dea}, it was shown that the inelasticity profile of protons is governed by the ratio of the slope of the diffraction cone to the total cross section through the variable $Z=4\pi B/\sigma_{tot}$ and the evolution to values of $Z<1$ at LHC energies implies a transition from the black disk picture of the interaction region to a black ring (or torus-like) shape. These results were reviewed in Ref.~\cite{Dremin:2014spa} using the unitarity relation in combination with experimental data on elastic scattering in the diffraction cone. Ref.~\cite{Dremin:2014dea} concludes that the shape of the interaction region of colliding protons could be reliably determined if the behavior of the elastic scattering amplitude at all transferred momenta was known.

The black ring shape of the interaction region can be interpreted as the presence of a hollow at small impact parameter values. 

In Refs.~\cite{Arriola:2016bxa,RuizArriola:2016ihz,Broniowski:2017aaf,Broniowski:2017rhz} the authors study the hollowness phenomenon within an inverse scattering approach based on empirical parameterizations. Ref.~\cite{RuizArriola:2016ihz} concludes that the very existence of the hollowness phenomenon is quantum-mechanical in nature. Hollowness has also been reported to emerge from a gluonic hot-spot picture of the $pp$ collision at the LHC energies~\cite{Albacete:2016pmp}.  It is shown in Ref.~\cite{Broniowski:2017rhz} that the emergence of such a hollow strongly depends on the phase of the scattering amplitude. In Ref.~\cite{Broniowski:2018xbg} the authors demonstrated the occurrence of the hollowness phenomenon in a Regge model above $\sqrt{s} \sim 3$~TeV. 

Ref.~\cite{Troshin:2017ucy} discusses the absorptive (saturation of the black disk limit) 
and reflective (saturation of the unitarity limit)
scattering modes of proton-proton collisions concluding that a distinctive feature of the transition to the reflective scattering mode is the developing peripheral form of the inelastic overlap function. Reflective scattering is detailed also in Refs.~\cite{Troshin:2007fq,Troshin:2020dvd,Troshin:2020bcn}.

The authors of Ref.~\cite{Petrov:2018rbi} argue that the presence of nonzero real
part of the elastic scattering amplitude in the unitarity condition enables
to conserve the traditional black disk picture refuting the existence of the hollowness effect. However, as noted in Ref.~\cite{Broniowski:2018xbg}, the criticism that has been raised in Ref.~\cite{Petrov:2018rbi} is based on an incorrect perception of the approximations involved and does
not address the arbitrariness of the $t$-dependence of the ratio $\rho$ which is crucial for hollowness.

In Refs.~\cite{Campos:2018tsb,Campos:2020edy} the hollowness effect is interpreted as a consequence of fundamental thermodynamic processes.

Ref.~\cite{Csorgo:2019fbf} notes that the onset of the hollowness effect is possibly connected to the opening of a new channel between $\sqrt{s}$ = 2.76 and 7 TeV as indicated by the measured $\sigma_{el}/\sigma_{tot}$ ratio and the slope parameter $B_0$ data.

In Ref.~\cite{Csorgo:2019egs} the model independent L\'evy imaging method is employed  to reconstruct the   proton inelasticity profile function and its error band at different energies. This method established a statistically significant proton hollowness effect, well beyond the 5$\sigma$ discovery limit. This conclusion is based on a  model independent description of the TOTEM proton-proton differential cross-section data at $\sqrt{s} = 13$ TeV with the help of the L\'evy imaging method, that represents the TOTEM data in a statistically acceptable manner, corresponding to  a confidence level of CL = 2 \%.


\section{Summary \label{sec:summary}}

Currently, the statistically significant observation of the elusive Odderon is a hot research topic, with several interesting 
and important results and contributions. In the context of this manuscript, Odderon exchange 
corresponds to a crossing-odd component of the scattering amplitude of elastic  proton-proton and proton-antiproton collisions, that does not vanish at asymptotically high energies, as probed experimentally by the D0 Collaboration for proton-antiproton and by the TOTEM Collaboration for proton-proton elastic collisions in the TeV energy range. Theoretically, the observed differences can be attributed only to the effect of a C-odd exchange, as detailed recently in refs.~\cite{Csorgo:2019ewn,Csorgo:2020msw,Csorgo:2020rlb}. Those model independent studies resulted in an at least 6.26 $\sigma$ statistical significance of the Odderon exchange~\cite{Csorgo:2019ewn,Csorgo:2020msw,Csorgo:2020rlb}.
The goal of the research summarized in this manuscript was to cross-check, in a model-dependent way, the persistence  of these Odderon-effects, and to provide a physical picture to interpret these results. 
Using the ReBB model of ref.~\cite{Nemes:2015iia}, developed originally to describe precisely the
differential cross-section of elastic proton-proton collisions, we were able to describe also  the proton-antiproton differential cross section at $\sqrt{s} = 0.546$ and $1.96$ TeV without any modification of formalism. We have shown also that this model 
describes   the proton-proton differential cross section at $\sqrt{s} = 2.76$ and $7$ TeV, also in a statistically acceptable manner, with a CL $> $ 0.1 \%. 

Using our good quality, statistically acceptable fits for the $0.5 \le\sqrt{s} \le 8$ TeV energy region, we have determined the energy dependence of the model parameters to be an affine linear function of $\ln(s/s_0)$. We have verified this energy dependence by demonstrating that 
the exctitation functions of the physical parameters of the Real Extended Bialas-Bzdak model satisfy the so-called sanity tests: they describe in a statistically acceptable manner not only those four datasets that formed the basis of the determination of the excitation function, but all other published datasets in the $\sqrt{s} = 0.541$ - $8.0$ TeV energy domain. We have also demonstrated that the excitation functions for the total cross-sections and the $\rho_0$ ratios correspond
to the experimentally estabished trends. 

Remarkably, we have observed that the energy dependence of the geometrical scale parameters for $pp$ and $p\bar p$ scattering are identical  in elastic proton-proton and proton-antiproton collisions: only the energy dependence of the shape or opacity parameter $\alpha(s)$ differs significantly between $pp$ and $p\bar p$ collisions. After determining the energy dependence of the model parameters we made extrapolations in order to compare the $pp$ and $p\bar p$ differential cross sections in the few TeV energy range, corresponding to the energy of D0 measurement at $\sqrt{s} = 1.96$ TeV in ref.~\cite{Abazov:2012qb}
and the TOTEM measurements at $\sqrt{s} = 2.76$ and $7.0$ TeV.  
Doing this, we found evidence for the Odderon exchange with a high statistical significance. 
We have cross-checked, 
that this evidence withstands several reasonable cross-checks, for example the possible presence of small quadratic
terms of $\ln(s/s_0)$ in the excitation functions of the parameters of this model. Subsequently, we have also predicted the details of the diffractive interference (dip and bump) region at $\sqrt{s} =0.9$, $4$, $5$ and $8$ TeV\footnote{Currently, TOTEM preliminary  experimental data are publicly  presented from an on-going analysis 
at $\sqrt{s} = 8 $ TeV, see ref.~\cite{Kaspar:2018ISMD} for further details.}

We have shown that within the framework of this ReBB model, the very different trend of $\rho_0(s)$ in proton-proton and in proton-antiproton collisions enhances the model-independent
Odderon signal, from a 6.26 $\sigma$ and 6.55 $\sigma $ effect of refs.~\cite{Csorgo:2019ewn,Csorgo:2020msw,Csorgo:2020rlb} 
to an at least $7.08 $ $\sigma$ effect. This gain of significance is due to the possibility of extrapolating the differential cross-sections of elastic $p\bar p$ scattering
from $\sqrt{s }$ $=$ 1.96 TeV to 2.76 TeV. It is important to note that in the evaluation of the
7.08 $\sigma$ Odderon effect, only $p\bar p$ data at $\sqrt{s} = 1.96$ TeV and $pp$ data at $\sqrt{s} = 2.76$ TeV were utilized,
amounting to a model dependent but successful closing of the energy gap between D0 and TOTEM measurements. Let us also emphasize that our Odderon observation is valid in the limited kinematic range of 
 $0.541 \le \sqrt{s} \le 8$ TeV center of mass energy and in the $0.377 \le -t \le 1.2$ GeV$^2$ four-momentum transfer range.

When extrapolating the $pp$ differential cross-sections from 2.76 down to 1.96 TeV,
however, significance is lost, corresponding to a $\chi^2/NDF = 24.28/13$ and to a 2.19  $\sigma$ effect, which is less than a 3 $\sigma$ effect at 1.96 TeV.  However, these two significances at 1.96 and 2.76 TeV can be combined, providing a  $\chi^2/NDF = 124.63 /33$,
that corresponds to a statistically significant, combined 7.08 $\sigma$ effect. 

This 7.08 $\sigma$ combined significance increases to  an even larger significance of an Odderon observation,
when we extrapolate the differential cross-section of elastic proton-antiproton collisions
to $\sqrt{s} = 7.0$ TeV. Given that a 7.08  $\sigma$ effect is already well above the usual 5 $\sigma$, statistically
significant discovery level, we
quote this as the possibly lowest level of the significance of our model-dependent Odderon observation.

Concerning the direction of future research: Odderon is now discovered both in a model-independent way, described in refs.~\cite{Csorgo:2019ewn,Csorgo:2020msw,Csorgo:2020rlb}, and in a model-dependent way, described in this manuscript;
so the obvious next step is to extract its detailed properties, both in a model-independent and in a model-dependent manner. The main properties of the Odderon as well as the Pomeron, based on the ReBB model, are already summarized in ~\ref{sec:appendix-c}.

Let us also note, that the ReBB model as presented in ref.~\cite{Nemes:2015iia} does  not yet provide a statistically acceptable fit quality  to the differential cross-section of $\sqrt{s} = 13$ TeV elastic $pp$ scattering.
This might be due to the emergence of the black-ring limit of elastic proton-proton scattering instead of the expected
black-disc limit, as detailed in Sec.~\ref{sec:discussion}, or due to the very strong non-exponential features of the 
differential cross-sections in these collisions at low $-t$\footnote{
We see that the ReBB model has a leading order exponential feature. If we want to describe the significantly non-exponential features of  differential cross-section in the low-$|t|$ range~\cite{Antchev:2015zza,Antchev:2017yns}, the model has to be generalized for a possible non-exponential behaviour at low $|t|$.}, as shown in ref.~\cite{Antchev:2017dia,Antchev:2017yns}.

So we conclude that the Real Extended Bialas-Bzdak model needs to be further generalized for the top LHC energies and above.
This work is in progress, but it goes clearly well beyond the scope of the current, already rather detailed manuscript. 
Importantly, any possible outcome of these follow-up studies is not expected to modify the model behavior at the presently
investigated energy range, and hence our work is apparently completed,
refinements are not necessary from the point of view of the task solved in this manuscript. 

In short, we determined the model-dependent statistical significance of the Odderon observation to be an at least 7.08  $\sigma$ effect
in the $0.5 \le \sqrt{s} \le 8$ TeV center of mass energy and
$0.377 \le -t \le 1.2$ GeV$^2$ four-momentum transfer range. Our analysis is based on the analysis of published D0 and 
TOTEM data of refs.~\cite{Abazov:2012qb,Antchev:2017dia,Antchev:2018rec}
and uses as a tool the Real Extended Bialas-Bzdak model of ref.~\cite{Nemes:2015iia}.
We have cross-checked that this unitary model works  in a statistically acceptable, carefully tested and verified manner in this particular kinematic range. Our main results are illustrated on Figs.~\ref{fig:reBB_model_extr_2_76_TeV} and \ref{fig:reBB_model_extr_7_TeV}.

\section*{Acknowledgments}
First of all, we would like to thank F. Nemes,  who initiated this project, and started to test the Real Extended Bialas-Bzdak model agains the proton-antiproton data. He also provided several valuable technical help  for us in the initial phases of these studies.

We greatfully acknowledge inspiring  discussions with C. Avila, S. Giani, P. Grannis, G. Gustaf\-son, L. Jenkovszky, E. Levin, B. Nicolescu, T. Nov\'ak, K. \"Osterberg, R. Pasechnik, C.  Royon, A. Ster and M. Strikman.  Our research has been partially supported by the \'UNKP-18-2
New National Excellence Program of the Ministry of Human Capacities, and by the NKFIH grants No. FK-123842, FK-123959 and K133046 as well as by the EFOP 3.6.1-16-2016-00001 grant (Hungary), as well as by the framework of the COST Action CA15213  ``Theory of hot matter and relativistic heavy-ion collisions (THOR)" of the European Union.

\appendix 

\section{Basics of the Bialas-Bzdak model and its unitary extension  \label{sec:appendix-a}}
\setcounter{equation}{0}

In the followings, let us shortly introduce the details of the $p=(q,d)$ ReBB model based on Refs.~\cite{Bialas:2006qf,Nemes:2015iia}.

The elastic scattering amplitude in the $b$ impact parameter space can be written in the so called eikonal form as:
\begin{equation}\label{eq:uBB_ansatz}
t_{el}(s,b)=i\left[1-e^{-\Omega(s,b)}\right], 
\end{equation}
where $\Omega(s,b)$ is the opacity or eikonal function and $b=|{\vec b}|$.
In general this opacity  is a complex valued function \cite{Glauber:1984su,Glauber:1970jm}. The shadow profile function is given as
\begin{equation}
    P(s,b) = 1 - |\exp(-\Omega)|^2 = \tilde \sigma_{in}(s,b), 
\end{equation}
and this is the reason why the shadow profile function is also frequently called as the inelastic profile function,
as it describes the probability distribution of inelastic collisions in the impact parameter space. This way the inelastic $pp$ scattering may be characterized by a probability distribution. However, let us stress that 
elastic scattering is an inherently quantum process, as evidenced by a diffractive interference that results in
diffractive minima and maxima of the differential cross-sections. Probabilistic interpretation can be given only to the inelastic
scattering, or to the sum of elastic scattering plus propagating without interactions.

If the real part of the scattering amplitude can be neglected, then  the $\Omega(s,b)$ has only a real part given as
\begin{equation}
\text{Re}\,\Omega(s,b)=-\frac{1}{2}\ln\left[1-\tilde\sigma_{in}(s,b)\right]\,.
\label{eq:re_omega}
\end{equation}

The inelastic profile function was evaluated with the help of Glauber's multiple diffraction theory~\cite{Glauber:1970jm}
for the colliding protons consisting  a constituent quark and diquark or $p = (q,d)$ picture in Section 2.2 of ref.
~\cite{Nemes:2015iia} and the results were visualized in Figs. 5 and 9 of that paper.

The imaginary part of the opacity function in Ref.~\cite{Nemes:2015iia}, which generates the real part of the scattering amplitude, is defined to be proportional to the inelastic scattering probability,
\begin{equation}
\text{Im}\,\Omega(s,b)=-\alpha\cdot\tilde\sigma_{in}(s,b)\,,
\label{eq:im_omega}
\end{equation}
were $\alpha$, mentioned earlier, is a free parameter and proportional to $\rho_0$ (see \ref{sec:appendix-b}). This ansatz assumes that the inelastic collisions at low four-momentum transfers correspond to the cases when the parts of proton suffer elastic scattering but these parts are scattered to different directions, not parallel to one another. Other models were also tested on TOTEM data in ref.~\cite{Nemes:2015iia}, but this physically motivated assumption worked well and was shown to be consistent with the experimental data at $\sqrt{s} = 7$ TeV in
ref.~\cite{Nemes:2015iia}.

\begin{figure}[hbt] 
	\centering
	\includegraphics[scale=0.35]{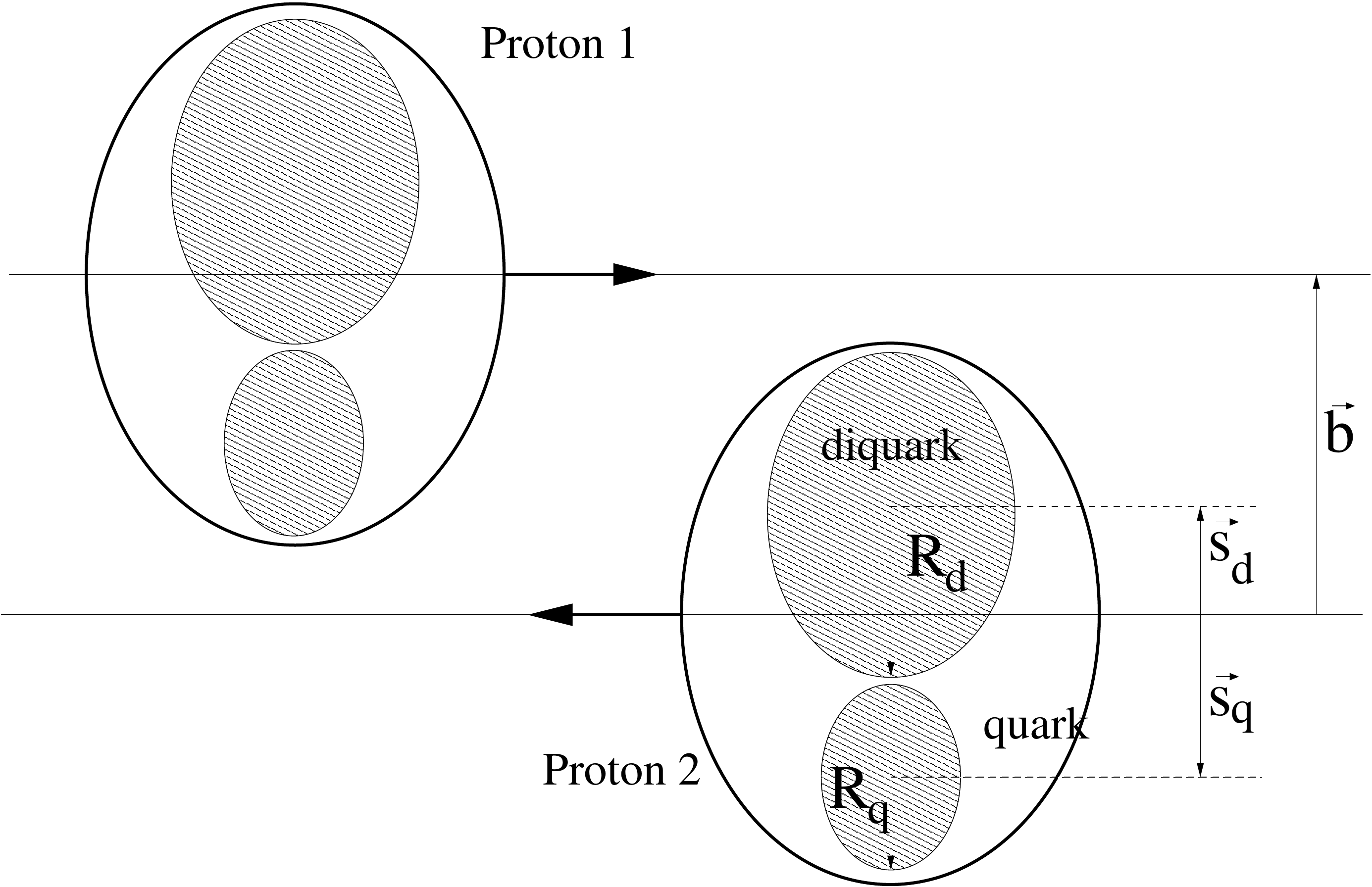}
	\caption{Visualization of the proton-proton scattering in the quark-diquark BB model. The figure is taken from Ref.~\cite{Nemes:2012cp}.}
	\label{fig:BB_qd}
\end{figure}

The inelastic scattering probability in the BB model \cite{Bialas:2006qf} for a fixed impact parameter $\vec b$ as
a probability distribution, given  as
\begin{equation}
\tilde\sigma_{in}(b)=\int_{-\infty}^{+\infty}...\int_{-\infty}^{+\infty} d^{2}\vec s_{q}d^{2}\vec s_{q}^{\,\prime}d^{2}\vec s_{d}d^{2}\vec s_{d}^{\,\prime
}D(\vec s_{q},\vec s_{d})D(\vec s_{q}^{\,\prime},\vec s_{d}^{\,\prime})\sigma(\vec s_{q},\vec s_{d}%
;\vec s_{q}^{\,\prime},\vec s_{d}^{\,\prime};b),
\label{eq:tilde_sigma_inel}
\end{equation}
where $\vec s_{q}$, $\vec s_{q}^{\,\prime}$, $\vec s_{d}$ and $\vec s_{d}^{\,\prime}$ are the transverse positions of the quarks and diquarks  in the two colliding protons (see Fig.~\ref{fig:BB_qd}). $D({\vec s}_q,{\vec s}_d)$ denotes the distribution of quark and diquark inside the proton which is considered to be Gaussian: 
\begin{equation}
D\left({\vec s}_q,{\vec s}_d\right)=\frac{1+\lambda^2}{R_{qd}^2\,\pi}e^{-(s_q^2+s_d^2)/R_{qd}^2}\delta^2({\vec s}_d+\lambda{\vec s}_q),
\label{eq:quark_diquark_distribution}
\end{equation}
where $\lambda=m_q/m_d$ is the ratio of the quark and diquark masses, furthermore $R_{qd}$ is the standard deviation of the quark and diquark distance emerging as a free parameter. The two-dimensional Dirac $\delta$ function fixes the center-of-mass of the proton and reduces the dimension of the integral in Eq.~(\ref{eq:tilde_sigma_inel}) from eight to four.
The diquark positions can be expressed by that of the quarks:
\begin{equation}
{\vec s}_d=-\lambda\,{\vec s}_q,\,\; {\vec s}^{\,\prime}_{d}=-\lambda\,{\vec s}^{\,\prime}_{q}\,.
\label{Dirac_deltas}
\end{equation}

The term $\sigma(\vec s_{q},\vec s_{d};\vec s_{q}^{\,\prime},\vec s_{d}^{\,\prime};b)$ is the probability of inelastic interactions at a fixed impact parameter and transverse positions of all constituents and given by a Glauber expansion as follows: 
\begin{align}
\sigma(\vec s_{q},\vec s_{d}
;\vec s_{q}^{\,\prime},\vec s_{d}^{\,\prime};b)=1-\prod_{a}\prod_{b}\left[1-\sigma_{ab}({\vec b} + {\vec s}^{\,\prime}_{a} - {\vec s}_b )\right]\,
\label{Glauber_expansion}
\end{align}
where $a,b\in\{q,d\}$. The terms $\sigma_{ab}\left({\vec s}\right)$ are the inelastic differential cross-sections of the binary collisions of the constituents having Gaussian shapes: 
\begin{equation}
\sigma_{ab}\left({\vec s}\right) = A_{ab}e^{-s^2/S_{ab}^2},\;S_{ab}^2=R_a^2+R_b^2,\quad a,b \in \{q,d\}\,.
\label{eq:inelastic_cross_sections}
\end{equation}
where $R_{q}$, $R_{d}$ and $A_{ab}$ are free parameters. The physical meaning of the $R_{q}$ and $R_{d}$ parameters 
as well as the impact parameter $b$ and the coordinates $s_q$ and $s_d$ is illustrated  on Fig.~\ref{fig:BB_qd} and detailed in
ref.~\cite{Nemes:2015iia}.

The inelastic quark-quark, quark-diquark and diquark-diquark cross sections are obtained by integrating Eq.~(\ref{eq:inelastic_cross_sections}):
\begin{equation}
\sigma_{ab,\text{inel}}=\int^{+\infty}_{-\infty}\int^{+\infty}_{-\infty}\sigma_{ab}\left(\vec s\right)\text{\rm d}^2s=\pi A_{ab}S_{ab}^2\,.
\label{eq:totalinelastic}    
\end{equation}

The number of the free parameters of the model can be reduced demanding that the ratios of the cross sections are
\begin{equation}
\sigma_{qq,\text{inel}}:\sigma_{qd,\text{inel}}:\sigma_{dd,\text{inel}}=1:2:4\,, 
\label{eq:ratiosforsigma}
\end{equation}
expressing the idea that the constituent diquark contains twice as many partons than the constituent quark and also that the colliding constituents do not "shadow" each other.

Using Eq.~(\ref{eq:totalinelastic}) and the assumptions given by Eq.~(\ref{eq:ratiosforsigma}), the $A_{qd}$ and $A_{dd}$ parameters can be expressed through $A_{qq}$: 
\begin{equation}
A_{qd}=A_{qq}\frac{4R_q^2}{R_q^2+R_d^2}\,,\;A_{dd}=A_{qq}\frac{4R_q^2}{R_d^2}\,.
\end{equation}
Counting the number of free parameters one finds that the model now contains six of them: $R_{q}$, $R_{d}$, $R_{qd}$, $\alpha$, $\lambda$ and $A_{qq}$. However, it was shown in Ref.~\cite{Nemes:2015iia} that the latter two parameters can be fixed. $\lambda = 0.5$ if the diquark is very weakly bound, so that its mass is twice as large as that of the valence quark. The Real Extended Bialas Bzdak model describes the 
experimental data in the $\sqrt{s} \leq 8$ TeV region with $A_{qq} = 1$ fixed, assuming that head-on $qq$ collisions are inelastic with a probability of 1, corresponding to Eq.~(\ref{eq:inelastic_cross_sections}).

Substituting Eq.~(\ref{eq:re_omega}) and Eq.~(\ref{eq:im_omega}) to Eq.~(\ref{eq:uBB_ansatz}) one obtains for the scattering amplitude:
\begin{equation}\label{eq:uBB_ansatz_sub}
t_{el}(s,b)=i\left(1-e^{i\, \alpha\, \tilde\sigma_{in}(s,b)}\sqrt{1-\tilde\sigma_{in}(s,b)}\right).
\end{equation}
This equation is, in fact, a special solution of the unitarity relation, obtained from the optical theorem. 
The integral for $\tilde\sigma_{in}(s,b)$ defined by Eq.~(\ref{eq:tilde_sigma_inel}) can be calculated analytically with the methods described in Refs.~\cite{Bialas:2006qf,Nemes:2015iia}.

In order to compare the theoretical model to the experimental data, the amplitude in impact parameter space, given by Eq.~(\ref{eq:uBB_ansatz_sub}), has to be transformed into momentum space by a Fourier-Bessel transformation:
\begin{equation}
T(s,t)= 2\pi \int\limits_0^{\infty}{J_{0}\left(\Delta\cdot b\right)t_{el}(s,b)b\, {\rm d}b}\,.
\label{eq:elastic_amplitude_app}
\end{equation}
In the above formula $\Delta=|{\vec\Delta}|$ is the transverse momentum and $J_{0}$ is the zeroth order
Bessel-function of the first kind. Here the high energy limit is considered, \textit{i.e.}, $\sqrt{s}\rightarrow\infty$ and then $\Delta(t) \simeq \sqrt{-t}$. 

Substituting the expression for the elastic scattering amplitude given by Eq.~(\ref{eq:elastic_amplitude_app}) into Eqs.~(\ref{eq:differential_cross_section})-(\ref{eq:rho_parameter}) the model for fitting the scattering data is complete. The Fourier-Bessel integral in the amplitude can be calculated numerically during the fitting procedure.

\section{\bf On the proportionality between $\rho_0(s)$ and $\alpha(s)$ in the ReBB model}

 \label{sec:appendix-b}
\setcounter{equation}{0}
Let us first of all note that the detailed description of the Real Extended Bialas-Bzdak (ReBB) model is given in
Section 2.2 of ref.~\cite{Nemes:2015iia} and also summarized in \ref{sec:appendix-a}. We have utilized this formalism throughout the fits described in the body of the manuscript, however in this Appendix, we need to develop this formalism a bit further,
as in the earlier publications the details of the relations
between the $\rho_0$ parameter (the ratio of the real to the imaginary part of the scattering amplitude at $t=0$) and the
parameter $\alpha$ of the ReBB model (that is responsible for filling up the singular dip of the original Bialas-Bzdak model of
refs.~\cite{Bialas:2006kw,Bialas:2006qf,Bialas:2007eg,Bzdak:2007qq}) has not yet been detailed before.

Let us stress, that ReBB model is unitary, by definition. Thus 
the elastic scattering amplitude in the ReBB model too has unitary form given by Eq.~\ref{eq:uBB_ansatz},
where the opacity function $\Omega(s,b)$ is, in general, a complex valued function.

In the ReBB model, the impact parameter dependent scattering amplitude is given by Eq.~\ref{eq:uBB_ansatz_sub}. Now we develop two small set of approximations
that are based on the physical domain of the ReBB model parameters. From the fits performed so far, we always find $\alpha \lessapprox 0.165$, corresponding to
Table 1 of ref.~\cite{Nemes:2015iia} and Table ~\ref{tab:fit_parameters} of the current manuscript.

In the physical case, when $\alpha \, \tilde\sigma_{in}(s,b) \ll 1$ is small, 
one obtains for the real and imaginary parts of the scattering amplitude, respectively,
\begin{equation}\label{ReTelsb}
\text{Re} \, t_{el}(s,b) \simeq \alpha\, \tilde\sigma_{in}(s,b)\, \sqrt{1-\tilde\sigma_{in}(s,b)}
\end{equation}
and
\begin{equation}\label{ImTelsb}
\text{Im} \, t_{el}(s,b) \simeq 1-\sqrt{1-\tilde\sigma_{in}(s,b)}.
\end{equation}

Given that the real part of the scattering amplitude is thus proportional to $\alpha$ while the imaginary part 
is independent of $\alpha$, we indeed find that 
\begin{equation}
    \rho_0 \propto \alpha, \qquad \mbox{\rm if} \quad \alpha \ll 1.
\end{equation}

Based on Figs. 5 and 9. of ref.
~\cite{Nemes:2015iia} and the model-independent results  of the Levy series method detailed in ref.~\cite{Csorgo:2019egs},
if the colliding energy is in the $\sqrt{s} \le 8$ TeV domain, corresponding to the domain of our extrapolations,
the shadow profile function is nearly Gaussian. Such a behaviour can be obtained easily as follows.

Let us approximate the imaginary part of the scattering amplitude with a Gaussian, \textit{i.e.}, 
\begin{equation}
\text{Im} \, t_{el}(s,b) \simeq \lambda(s) \, \exp\left(-\frac{b^2}{2R^2(s)}\right), 
\end{equation}
where $\lambda(s)\simeq\text{Im} \, t_{el}(s,b=0)$.
Then the inelastic profile or shadow profile function takes the form of 
\begin{equation}
\tilde\sigma_{in}(s,b)= 2 \lambda(s) \exp\left(-\frac{b^2}{2R^2(s)}\right) - \lambda(s)^2 \exp\left(-\frac{ b^2}{R^2(s)}\right) .
\end{equation}
This expression, up to second order terms, starts as a Gaussian, but it actually corresponds to the subtraction of a broader and smaller Gaussian from a  narrower and larger Gaussian in the physical domain of $\lambda(s) \leq 1$.

As $P_0 \equiv P(s,0) = \tilde \sigma_{inel}(s, b= 0)$ is the value of the  profile or inelastic profile  function at $b=0$,
we find  the following relation between $P_0$ and $\lambda(s)$:
\begin{equation}
    P_0(s) = 2 \lambda(s) - \lambda^2(s)  \le 1 . \label{eq:P0-lambda}
\end{equation}

When performing the transformation from the impact parameter space to momentum space, the result for the real to imaginary part ratio of the forward scattering amplitude, defined by Eq.~(\ref{eq:rho_parameter}), is 
\begin{equation}
\rho_0(s)=\alpha(s)\,\left(2-\frac{3}{2}\lambda(s)+\frac{1}{3}\lambda^2(s)\right) .
\label{eq:rho-vs-alpha-appendix}
\end{equation}
In the above equation, we may consider that $\lambda \equiv \lambda(s)$ is a function of $P_0(s)$ based on Eq.~(\ref{eq:P0-lambda}).
Based on the formalism of Section 2.2 of ref.~\cite{Nemes:2015iia}, $P_0 \equiv P_0(s)$ is a function of $R_q(s)$, $R_d(s)$
and $R_{qd}(s)$ only, 
but otherwise it is independent of  the fourth physical parameter of the Real Extended Bialas-Bzdak model,
$\alpha(s)$. Hence the excitation function of $P_0(s)$ is determined completely by the parameters $p_1$ and $p_0$ of the excitation functions of the scale parameters ($R_q$, $R_d$, $R_{qd}$), as summarized in Table~\ref{tab:excitation_pars}. 
This way, the $P_0 = P_0\left(R_q(s),R_d(s),R_{qd}(s)\right)$ function is uniquely given by with the help 
of eq.~(\ref{eq:uBB_ansatz_sub}), corresponding to eq. (29) of 
ref.~\cite{Nemes:2015iia}.

We have cross-checked the result of these analytic considerations compared to the fit results on $\alpha(s)$ and the measured
values of $\rho_0(s)$ at the ISR energies and we find an excellent agreement between the analytic approximations and the numerical results at ISR, corresponding to  the $\lambda(s)$ range of 0.73 - 0.78, as illustrated in Fig.~\ref{fig:alpha-rho0-ISR}.
The linear relationship between $\rho_0$ and the ReBB model parameter $\alpha$ is also indicated at the ISR energy range, in 
Fig.~\ref{fig:alpha-per-rho-ISR}. Similarly, we find an excellent agreement between the analytic calculations of eq.~(\ref{eq:rho-vs-alpha-appendix}) and the numerical and experimental results at the energy scale of
$0.5 \le \sqrt{s} \le 8$ TeV, as demonstrated on Fig.~\ref{fig:rho0-per-alpha-vs-P0-LHC}, presented in the body of this manuscript. 


\section{Pomeron and Odderon from the Real Extended Bialas-Bzdak model  \label{sec:appendix-c}}
\setcounter{equation}{0}

In this Appendix we summarize, for the sake of clarity, how we can determine the crossing-even and crossing-odd components of the scattering
amplitude, based on the ReBB model. In the TeV energy range, we indentify these components with the Pomeron and the Odderon amplitude,
given that the Reggeon contributions in this energy range are generally expected to be negligibly small, as confirmed also by explicit calculations for example in ref.~\cite{Broniowski:2018xbg}.

In this energy range, the proton-proton ($pp$) as well as the proton-antiproton ($p\bar p$) elastic scattering amplitudes can be written as
\begin{eqnarray}
     T_{el}^{pp} & = & T_{el}^+ - T_{el}^- , \\
          T_{el}^{p\bar p} & = & T_{el}^+ + T_{el}^- ,
\end{eqnarray}
where we have suppressed the dependence of these amplitudes on the Mandelstam variables: $T_{el}^{pp} \equiv T_{el}^{pp}(s,t)$ etc.

If the $pp$ and the $p\bar p$ scattering amplitudes are known, then the crossing even and the crossing odd components of the elastic scattering amplitude can be reconstructed as 
\begin{eqnarray}
     T_{el}^{+} & = & \frac{1}{2} \left(T_{el}^{p\bar p} + T_{el}^{p p}\right) , \label{e:pomeron-amplitude} \\
          T_{el}^{-} & = & \frac{1}{2}  \left(T_{el}^{p\bar p} - T_{el}^{pp} \right). \label{e:odderon-amplitude}
\end{eqnarray}

In this manuscript, we have utilized the Real Extended Bialas-Bzdak or ReBB model of ref.~\cite{Nemes:2015iia}, to determine the 
elastic scattering amplitude for elastic $pp$ and $p\bar p$ scattering. 
This model is based on R. J. Glauber's theory of multiple diffractive scattering~\cite{Glauber:1955qq,Glauber:1970jm,Glauber:2006gd}, and assumes that the elastic proton-proton scattering is based on multiple diffractive scattering of the constituents of the protons. Hence this ReBB model
has two main variants: the case when the proton is assumed to have a constituent quark and a diquark component is referred to as $p = (q,d)$
model, while the case when the diquark is assumed to be further resolved as a weakly bound state of two constituent quarks is the $p = (q, (q,q))$
model. It was shown before that  this $p = (q,(q,q))$ variant predicts too many diffractive minima for the differential cross-section, hence in this paper we utilize the $p = (q,d)$ variant as formulated in ref.~\cite{Nemes:2015iia}, without any change.

With the help of the ReBB model of ref.~\cite{Nemes:2015iia}, we have described in a statistically acceptable manner the $pp$ and $p\bar p$ differential cross-sections. In this ReBB model the $pp$ elastic scattering amplitude depends on $s$ only through four energy dependent parameters, that we denote here, for the sake of clarity, as $R_q^{pp}(s)$, $R_d^{pp}(s)$, $R_{qd}^{pp}(s)$ and $\alpha^{pp}(s)$:
\begin{equation}
    T_{el}^{pp}(s,t) = F(R_q^{pp}(s), R_d^{pp}(s), R_{qd}^{pp}(s),\alpha^{pp}(s); t). \label{e:Telpp-parameters}
\end{equation}
Similarly, we described the amplitude of the elastic  $p\bar p$ scattering with 4 energy dependent parameters, 
that we denote here for the sake of clarity as 
$R_q^{p\bar p}(s)$, $R_d^{p\bar p}(s)$, $R_{qd}^{p\bar p}(s)$ and $\alpha^{p\bar p}(s)$:
\begin{equation}
    T_{el}^{p\bar p}(s,t) = F(R_q^{p\bar p}(s), R_d^{p\bar p}(s), R_{qd}^{p\bar p}(s),\alpha^{p\bar p}(s); t).  \label{e:Telpbarp-parameters}
\end{equation}
Here $F$ stands for  a symbolic short-hand notation for a function, that indicates how the left hand side of the $pp$ and $p\bar p$ 
scattering amplitude depend on $s$ through their $s$-dependent parameters.
The scale parameters $R_q$, $R_d$, and $R_{qd}$ correspond to the Gaussian sizes of the constituent quarks, diquarks and their separation
in the scattering (anti)protons. Each of these parameters is $s$-dependent. 
Since the trends of $R_q(s)$, $R_d(s)$ and $R_{qd}(s)$ follow, within errors,  the same excitation functions in both $pp$ and 
$p\bar p$ collisions, as indicated on panels a, b and c of Fig.~\ref{fig:reBB_model_log_lin_extrapolation_fits}, we have denoted these in principle different scale parameters with the same symbols in the body of the manuscript:
\begin{eqnarray}
    R_q(s) & \equiv & R_q^{pp}(s) \, = \, R_q^{p\bar p}(s), \label{e:Rqs} \\
    R_d(s) & \equiv & R_d^{pp}(s) \, = \, R_d^{p\bar p}(s), \label{e:Rds}\\
    R_{qd}(s) & \equiv & R_{qd}^{pp}(s) \, = \, R_{qd}^{p\bar p}(s). \label{e:Rqds}
\end{eqnarray}
On the other hand, the opacity or dip parameters $\alpha(s)$ are different in elastic $pp$ and $p\bar p$ reactions: if they too were the same, then the
scattering amplitude for $pp$ and $p\bar p$ reactions were the same, correspodingly the differential cross-sections were the same in these reactions, while experimental results indicate that they are qualitatively different. Hence
\begin{eqnarray}
     \alpha^{pp}(s) & \neq & \alpha^{p\bar p}(s), 
\end{eqnarray}
corresponding to panel d of Fig.~\ref{fig:reBB_model_log_lin_extrapolation_fits} and to Table~\ref{tab:excitation_pars}.

In this form, the ReBB model of ref.~\cite{Nemes:2015iia} provides a statistically acceptable description of the elastic scattering amplitude, both for $pp$ and  $p\bar p$ elastic scattering, in the  kinematic range that extends to at least $ 0.372 \leq -t \leq 1.2$ GeV$^2$ and $0.546 \leq \sqrt{s} \leq 8$ TeV. Now, for the sake of clarity, let us note that the $s$-dependence of the Pomeron and Odderon (crossing-even and crossing-odd) components
of the scattering amplitude thus happens through the $s$-dependence of five parameters only:
Based on eqs.~(\ref{e:pomeron-amplitude},\ref{e:odderon-amplitude}),  eqs.~(\ref{e:Telpp-parameters},\ref{e:Telpbarp-parameters}) 
and eqs. (\ref{e:Rqs},\ref{e:Rds},\ref{e:Rqds}) we find that
\begin{eqnarray}
     T_{el}^{\mathbb P}(s,t)  & = &T_{el}^{+}(s,t) \, =
          G(R_q^{pp}(s), R_d^{pp}(s), R_{qd}^{pp}(s),\alpha^{pp}(s), \alpha^{p\bar p}(s);t),  
     \label{e:pomeron-amplitude-in-terms-of-5}  \\ 
     T_{el}^{\mathbb O}(s,t)  & = &     T_{el}^{-}(s,t) \, =
          H(R_q^{pp}(s), R_d^{pp}(s), R_{qd}^{pp}(s),\alpha^{pp}(s), \alpha^{p\bar p}(s);t). 
     \label{e:odderon-amplitude-in-terms-of-H} 
\end{eqnarray}
Here $G$ and $H$ are just sympolic short-hand notations that summarize how the left hand side of the above equations depend on $s$ through
their $s$-dependent parameters.




The differential cross section, Eq.~(\ref{eq:differential_cross_section}), the total, elastic and inelastic cross sections, Eqs.(\ref{eq:total_cross_section})-(\ref{eq:inelastic_cross_section}), as well as the real to imaginary ratio, Eq.~(\ref{eq:rho_parameter}),
and the nuclear slope parameter,
\begin{equation}
     B(s,t) =  \frac{d}{dt} \ln \frac{d\sigma(s,t)}{dt}, \label{e:B-st}
\end{equation}
 characterize experimentally the 
 $(s,t)$ dependent elastic scattering amplitudes, $T_{el}(s,t)$ discussed above. These quantities can be evaluated for a specific process like
 the elastic $pp$ or $p\bar p$ scattering. Given that we evaluate the elastic scattering amplitude for both of them in the TeV energy range, that yields also the $(s,t)$ dependent elastic scattering amplitude also for the Pomeron and the Odderon exchange, we have the possibility to evaluate
 these quantities for the crossing-even Pomeron ($\mathbb P$) and for the crossing-odd Odderon ($\mathbb O$) exchange.
 
The momentum space dependent scattering amplitude $T_{el}(s,t)$, for spin independent processes, is related to a  Fourier-Bessel transform of the impact parameter dependent elastic scattering amplitudes $t_{el}(s,b)$ as given by Eq.~\ref{eq:elastic_amplitude_app}.

This impact parameter dependent amplitude is constrained by the unitarity of the scattering matrix $S$,
\begin{equation}
   S S^{\dagger} = I 
\end{equation}
where $I$ is the identity matrix. Its  decomposition is $S = I + iT$ , where the matrix  $T$ is the
transition matrix.  In terms of $T$, unitarity leads  to the relation
\begin{equation}
   T - T^{\dagger} =  i T T^{\dagger}
\end{equation}
which can be rewritten in terms of the  impact parameter or $b$ dependent amplitude $t_{el}(s,b)$ as
\begin{equation}
2 \, {\mathcal Im} \, t_{el}(s, b) = |t_{el}(s, b)|^2 + \tilde\sigma_{inel}(s,b),
\end{equation}
where $\tilde\sigma_{inel}(s,b)$ is the impact parameter dependent probability of inelastic scattering.
It can be equivalently expressed from the above unitarity relation as
\begin{equation}
\tilde\sigma_{inel}(s,b)  = 1 - ({\mathcal Re} \, t_{el}(s, b))^2 - ({\mathcal Im} \, t_{el}(s, b) -1)^2.
\end{equation}
It follows that 
\begin{eqnarray}
\tilde\sigma_{inel}(s,b) & \leq &  1,
\end{eqnarray}
as a consequence of unitarity.

Given  that the ReBB model of ref. ~\cite{Nemes:2015iia} is unitary, those dispersion relations that are consequences of the unitarity of the scattering amplitude are automatically satisfied. For example, the dispersion relations discussed in refs.~\cite{Andreev:1969qg}
and \cite{Dremin:2018orv} are automatically satisfied by the unitary ReBB model. 

The impact parameter dependent elastic scattering amplitudes for elastic $pp$ and $p\bar p$ scatterings are given in terms of the complex
opacity or eikonal functions $\Omega(s,b)$. The defining relations are
\begin{eqnarray}
    t_{el}^{pp}(s,b)  & = & i \, (1 - \exp\left(-\Omega^{pp}(s,b) \right), \\  
    t_{el}^{p\bar p}(s,b)  & = & i \, (1 - \exp\left(-\Omega^{p\bar p}(s,b) \right).  
\end{eqnarray}
As another consequence of the unitary relations, we have 
\begin{eqnarray}
\sqrt{1 - \tilde\sigma_{inel}(s,b)} & = & \exp\left(- {\mathcal Re} \, \Omega(s,b)\right).
\end{eqnarray}

In ref.~\cite{Nemes:2015iia}, three different possibilities were considered for the solution of the unitarity relation, using various functions to model the imaginary part of the complex opacity function $\Omega$, that corresponds to the real part of the scattering amplitude.
Out of the considered three possible choices, the assumption that was found to be consisent with the experimental data on $pp$ elastic scattering at the ISR and LHC energies is defined by Eq.~(\ref{eq:uBB_ansatz_sub}).
At that time it was not yet clear that a similar relation works also for $p\bar p$ collisions. A very important advantage of this particular solution to the unitarity equation is that the multiple diffractive scattering theory of R. J. Glauber predicts $\tilde\sigma_{inel}(s,b)$ to depend only on the $s$-dependent geometrical scales $(R_q(s), R_d(s), R_{qd}(s))$. Given that the $R_q(s)$, $R_d(s)$, $R_{qd}(s)$ scales are found in panels a, b, and c of Fig.~\ref{fig:reBB_model_log_lin_extrapolation_fits} 
to be independent of the type of the elastic collisions \textit{i.e.} to be the same in elastic $pp$ and $p\bar p$ collisions in the body of this paper, the imaginary part of the complex opacity function in elastic  $pp$ and  $p\bar p$ collisions has the same $b$-dependent factor,
but has an $s$-dependent prefactor that is in principle a different function in the cases of elastic $pp$ and $p\bar p$ collisions:
\begin{eqnarray}
    {\mathcal Im} \Omega^{pp}(s,b) & = & - \alpha^{pp}(s) \tilde\sigma_{inel}(s,b), \\
    {\mathcal Im} \Omega^{p\bar p}(s,b) & = & - \alpha^{p\bar p}(s) \tilde\sigma_{inel}(s,b).
\end{eqnarray}
These relations yield the following simple expressions for the impact parameter dependent elastic $pp$ and $p\bar p$ scattering amplitudes
\begin{eqnarray}
    t_{el}^{pp}(s,b) = i\left(1-e^{i\, \alpha^{pp}(s) \, \tilde\sigma_{in}(s,b)} \, \sqrt{1-\tilde\sigma_{in}(s,b)}\right), \\
    t_{el}^{p\bar p}(s,b) = i\left(1-e^{i\, \alpha^{p\bar p}(s) \, \tilde\sigma_{in}(s,b)} \, \sqrt{1-\tilde\sigma_{in}(s,b)}\right).
\end{eqnarray}
It then clearly follows that in the ReBB model $t_{el}^{pp}(s,b) \equiv t_{el}^{p\bar p}(s,b)$ and $t_{el}^{\mathbb O}(s,b)\equiv0$ if, and only if 
$\alpha^{pp}(s)\equiv\alpha^{p\bar p}(s)$.

As detailed in~\ref{sec:appendix-b}, $\alpha(s)\sim\rho_0(s)$ both for $pp$ and $p\bar p$ elastic collisions. At the same time $\alpha(s)$ controls the value of the differential cross section in the region of the dip in these collisions. Thus, within  the ReBB model there is a deep connection between the $t=0$ and the dip region. This supports the findings that the recently observed decrease in $\rho_0(s)$ around $\sqrt{s}=$13 TeV, the dip-bump structure in $pp$ scattering and its absence in $p\bar p$ scattering are both the consequences of the Odderon contribution.
In the ReBB model, this Odderon contribution  is encoded in the difference between $\alpha^{pp}(s)$ and $\alpha^{p\bar p}(s)$. This conclusion
is supported also by the detailed calculations of the ratios of the modulus squared Odderon to Pomeron scattering amplitudes.

Thus if  $\rho_0^{pp}(s)\neq\rho_0^{p\bar p}(s)$, within the ReBB model it follows that $\alpha^{pp}(s)\neq\alpha^{p\bar p}(s)$ or equivalently
$t_{el}^{\mathbb O}(s,b)\neq0$ in the TeV region.

Within the framework of the ReBB model, we thus can significantly sharpen an Odderon theorem noted in ref.~\cite{Csorgo:2019ewn}.
The weaker, original form of this theorem was formulated in ref.~~\cite{Csorgo:2019ewn} as follows:

{\bf Theorem 1}
\begin{itemize} 
    \item If the $pp$ differential cross sections differ from that of $p\bar p$ scattering at the same value of $s$ in a TeV energy domain, then the Odderon contribution to the scattering amplitude cannot be equal to zero, i.e.
     \begin{equation}
        \frac{d\sigma^{pp}}{dt} \neq  \frac{d\sigma^{p\bar p}}{dt} \,\,\, \mbox{ \rm for}\,\, \sqrt{s}\ge 1 \,\, \mbox{\rm TeV}
        \implies 
                T_{\rm el}^O(s,t) \neq 0  \, .
    \end{equation}
\end{itemize}
This theorem is model-independenty true as it depends only on the general structure of the theory of elastic scattering.
The outline of the proof is that the differential cross-section, Eq.~(\ref{eq:differential_cross_section}), is proportional to the modulus squared elastic scattering amplitudes both for
$pp$ and $p\bar p$ scattering. If the modulus square of two complex functions is different, then the two complex functions, corresponding to the
elastic scattering amplitudes of $pp$ and $p\bar p$ collisions, cannot be identical. Hence their difference, proportional to the Odderon amplitude
in the TeV energy range, cannot be zero.

Within the ReBB model, this theorem can be significantly sharpened.
This sharpened, stronger version of the above theorem thus reads as follows:

{\bf Theorem 2:}
In the framework of the unitary Real Extended Bialas-Bzdak (ReBB) model, the elastic  $pp$ differential cross sections differ from that of elastic $p\bar p$ scattering at the same value of $s$ in a TeV energy domain, 
    if and only if the Odderon contribution to the scattering amplitude is not  equal to zero. This happens if and only if
    $\alpha^{pp}(s) \neq \alpha^{p\bar p}(s)$ and as a consequence, if and only if  $\rho_0^{pp} \neq \rho_0^{p\bar p}  $:
     \begin{eqnarray}
        \frac{d\sigma^{pp}}{dt}  & \neq  & \frac{d\sigma^{p\bar p}}{dt} 
        \iff 
                T_{\rm el}^O(s,t) \neq 0  \, \nonumber \\  &\iff&  \rho_0^{pp}(s) \neq \rho_0^{p\bar p}(s) \nonumber \\
                & \iff &  \alpha^{pp}(s) \neq \alpha^{p\bar p}(s) \nonumber \\
                & & \,\,\, \mbox{ \rm for}\,\, \sqrt{s}\ge 1 \,\, \mbox{\rm TeV}. \nonumber
    \end{eqnarray}

This theorem is proven by the explicit expressions for the impact parameter dependent elastic scattering amplitude for the C-even Pomeron 
and the C-odd Odderon exchange in the ReBB model as detailed below. These relations are consequences of the unitarity of the ReBB model.
\begin{eqnarray*}
    t_{el}^{\mathbb P}(s,b)  & = & i \, \left(1 - \frac{1}{2} \left(\exp\left(-\Omega^{pp}(s,b) \right) + \exp\left(-\Omega^{p\bar p}(s,b) \right)\right)\right), \\  
    t_{el}^{\mathbb O}(s,b)  & = & i \, \frac{1}{2}
                            \left(\exp\left(-\Omega^{pp}(s,b) \right) - \exp\left(-\Omega^{p\bar p}(s,b) \right)\right).  
\end{eqnarray*}
It is obvious to note that the Pomeron amplitude given above is crossing-even, while the Odderon amplitude is crossing-odd: 
$ C t_{el}^{\mathbb P}(s,b) = t_{el}^{\mathbb P}(s,b)  $  and $C t_{el}^{\mathbb O}(s,b) = - t_{el}^{\mathbb O}(s,b)$ .

These relations can be equivalently rewritten for the Pomeron amplitude, using the shorthand notation $\tilde\sigma_{in}\equiv \tilde\sigma_{in}(s,b)$ 
and suppressing the $s$ dependencies of $\alpha^{pp}(s)$ and $\alpha^{p\bar p}(s)$, as follows:
\begin{eqnarray*}
        {\mathcal Im }\, t_{el}^{\mathbb P}(s,b)  & = & 1 - \sqrt{1-\tilde\sigma_{in} }
        \cos\left( \frac{\alpha^{pp}+\alpha^{p\bar p}}{2}\tilde\sigma_{in}  \right)
            \cos\left( \frac{\alpha^{p\bar p}- \alpha^{pp}}{2}\tilde\sigma_{in}  \right), \\
        {\mathcal Re}\,  t_{el}^{\mathbb P}(s,b)  & = & \sqrt{1-\tilde\sigma_{in} } \, 
        \sin\left( \frac{\alpha^{pp}+\alpha^{p\bar p}}{2}\tilde\sigma_{in}  \right) 
        \cos\left( \frac{\alpha^{p\bar p}-\alpha^{pp}}{2}\tilde\sigma_{in}  \right).
\end{eqnarray*}
This form of the Pomeron amplitude is explicitely C-even, as corresponding to the Pomeron amplitude in the unitary, Real Extended Bialas-Bzdak model.
Thus if the difference between the opacity parameters $\alpha$ for $pp$ and $p\bar p$ elastic collisions is small, the Pomeron is 
predominantly imaginary, with a small real part that is proportional to $  \sin\left( \frac{\alpha^{pp}+\alpha^{p\bar p}}{2}\tilde\sigma_{in}  \right) $ .
Similary, for the Odderon, we have in the ReBB model the following amplitude
\begin{eqnarray}
        {\mathcal Re }\, t_{el}^{\mathbb O}(s,b)  & = & \sqrt{1-\tilde\sigma_{in} }
        \sin\left( \frac{\alpha^{p\bar p}-\alpha^{pp}}{2}\tilde\sigma_{in}  \right)
            \cos\left( \frac{\alpha^{p\bar p}+ \alpha^{pp}}{2}\tilde\sigma_{in}  \right), \label{e:Re-Odderon}\\
        {\mathcal Im}\,  t_{el}^{\mathbb O}(s,b)  & = & \sqrt{1-\tilde\sigma_{in} } \, 
        \sin\left( \frac{\alpha^{p\bar p}-\alpha^{pp}}{2}\tilde\sigma_{in}  \right) 
        \sin\left( \frac{\alpha^{pp}+\alpha^{p\bar p}}{2}\tilde\sigma_{in}  \right).\label{e:Im-Odderon}
\end{eqnarray}
This form of the Odderon amplitude is explicitely C-odd and satisfies unitarity, corresponding to the Real Extended Bialas-Bzdak model.
If the  difference between the opacity parameters $\alpha$ for $pp$ and $p\bar p$ elastic collisions becomes vanishingly  small, both the real and the imaginary part of the Odderon amplitude vanishes, as they are both proportional to
$        \sin\left( \frac{\alpha^{p\bar p}-\alpha^{pp}}{2}\tilde\sigma_{in}  \right) $.
If this term is non-vanishing, but  $(\alpha^{p\bar p}+ \alpha^{pp}) \sigma_{in}$ remains small,
the above Odderon amplitude remains predominantly real, with a small, leading order linear in 
$(\alpha^{p\bar p}+\alpha^{pp}) \tilde \sigma_{in}$ imaginary part. Given that $\alpha(s)  \propto \rho_0(s) $ in the ReBB model,
as detailed in~\ref{sec:appendix-b}, and experimentally $\rho_0(s)  \leq 0.15$ at LHC energies,
the ReBB model Odderon amplitude is predominantly real at small values of $t$.

Eqs. ~(\ref{e:Re-Odderon},\ref{e:Im-Odderon}) complete the proof, that the Odderon amplitude in the ReBB model vanishes if and only if
the opacity parameters $\alpha(s)$ for elastic $pp$ and $p\bar p$ scattering are equal, corresponding to  $\alpha^{p\bar p}(s) =\alpha^{pp}(s) $ .
Note that these proofs are independent of the detailed calculations of the inelastic scattering probability $\tilde \sigma_{in} = \sigma_{inel} (s,b)$, hence they are valid both in the $p = (q,d)$ and in the $p = (q, (q,q)) $ variant of the ReBB model. In fact they are valid for possible
further generalized ReBB models as well, where for example the distribution of the scattering by quarks or diquarks is not assumed to be a Gaussian anymore, or if further parton contributions get resolved in a future paper.
 
In the following plots, we have evaluated the differential and total cross-sections of the Pomeron and the Odderon exchange, as well as the ratios of these differential cross-sections, to determine the main properties of these processes with the help of the ReBB model of ref.~\cite{Nemes:2015iia}.

\begin{figure}[hbt]
	\centering
		\includegraphics[width=0.8\linewidth]{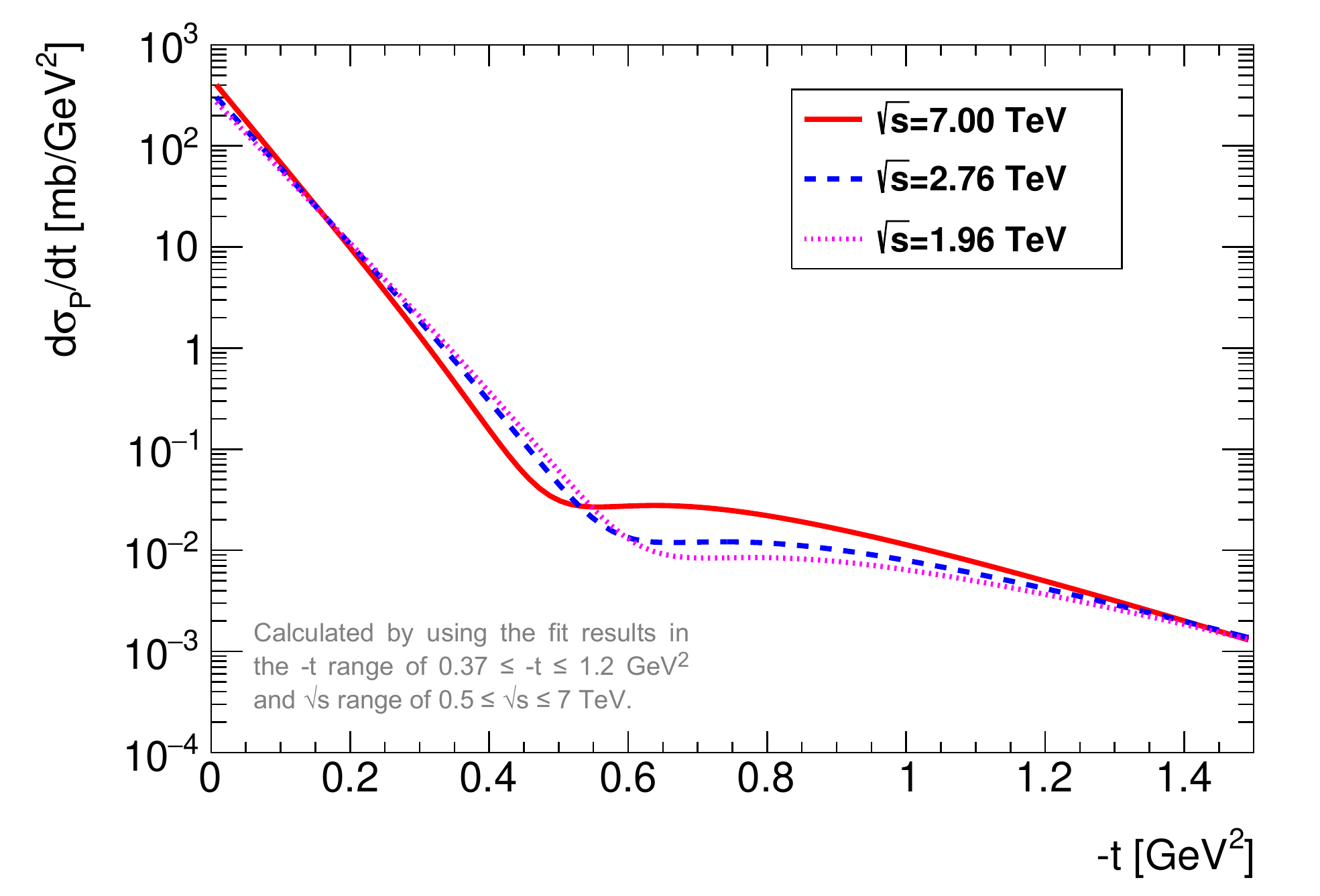}
	\includegraphics[width=0.8\linewidth]{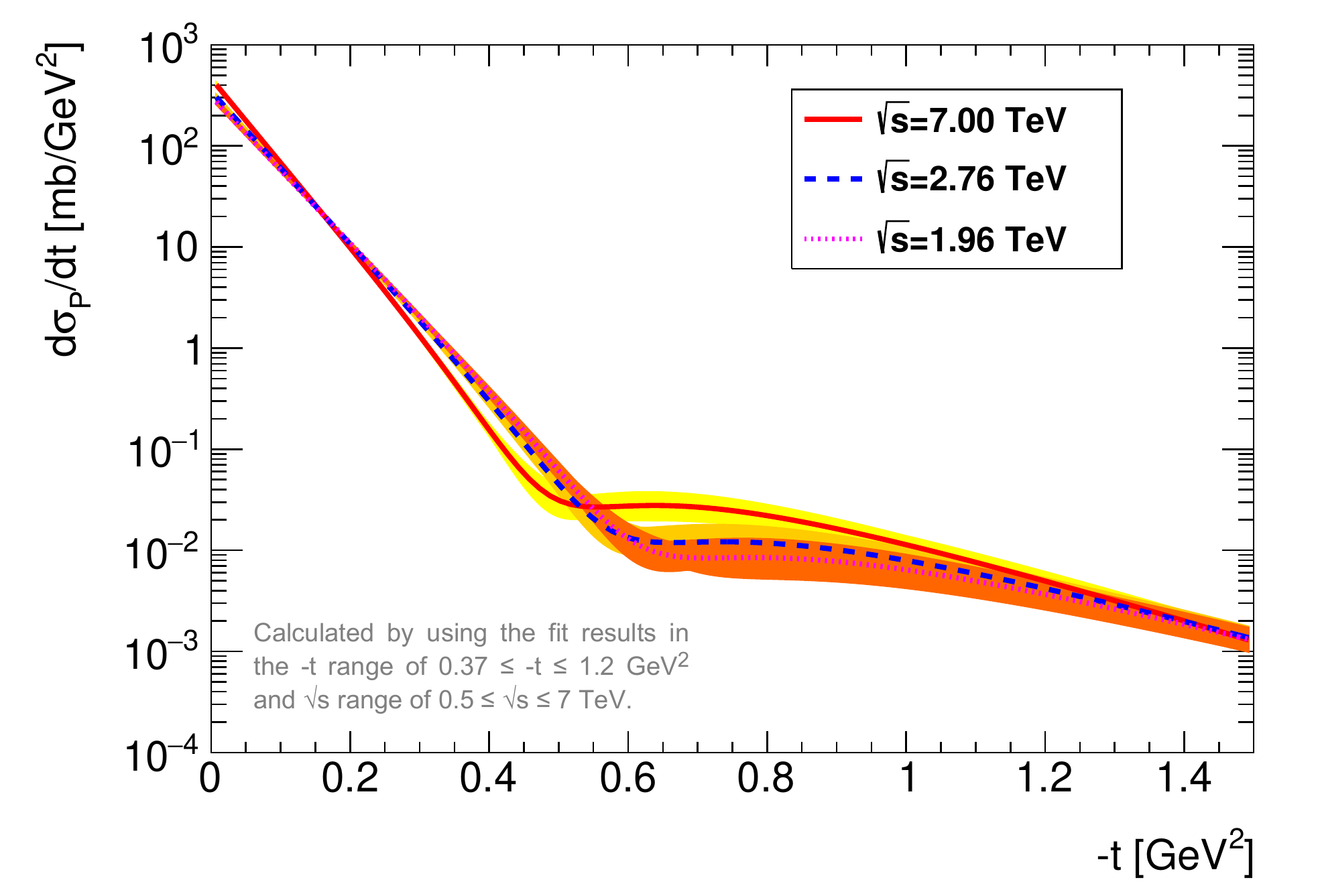}
	\caption{The differential cross-section of Pomeron exchange, as calculated from ReBB model of ref.~\cite{Nemes:2015iia},
	based on the log-linear excitation functions of $\alpha^{pp}(s)$, $\alpha^{p\bar p}(s)$ and  the scale parameters, $R_q(s)$, $R_d(s)$, $R_{qd}(s)$, corresponding to Fig.~\ref{fig:reBB_model_log_lin_extrapolation_fits} as summarized in 
	Table~\ref{tab:excitation_pars} in the body of this manuscript. 
	The top panel indicates the central values, while the lower panel includes our first estimates on the systematic errors of this reconstruction. The presented (over)estimates of the systematic  error bands were obtained by neglecting the possible correlations between the parameters $p_0$ and $p_1$ for each of the excitation functions given in  Table~\ref{tab:excitation_pars}.  }
	\label{fig:differential-cross-section-for-Pomeron-exchange}
\end{figure}

Fig.~\ref{fig:differential-cross-section-for-Pomeron-exchange} indicates the calculated differential cross-section for Pomeron exchange based
on the fits presented in the body of this manuscript, utilizing the ReBB model of ref.~\cite{Nemes:2015iia}. This result is based on
Table~\ref{tab:excitation_pars}, that summarizes the parameters of the excitation functions for the opacity parameters
$\alpha^{pp}(s)$, $\alpha^{p\bar p}(s)$ and  the scale parameters, $R_q(s)$, $R_d(s)$, $R_{qd}(s)$, corresponding to Figs.~\ref{fig:par_Rq_lin}-\ref{fig:par_alpha_lin} in the body of this manuscript. The top panel indicates the central values for the differential cross-section of Pomeron exchange at various colliding energies $\sqrt{s}$, while the lower panel includes our first estimates on the systematic errors of this reconstruction. These  first error band estimates were obtained by neglecting the possibly strong correlations between the parameters $p_0$ and $p_1$. These figures also indicate that Pomeron exchange does not lead to a pronounced diffractive minimum structure, in contrast to the experimental results for the diffractive minimum in elastic $pp$ collisions. This differential cross-section is more similar to the neck and shoulder type of structure, experimentally observed  in  elastic $p\bar p$ collisions, as discussed in the body of this manuscript.

\begin{figure}[hbt]
	\centering
		\includegraphics[width=0.8\linewidth]{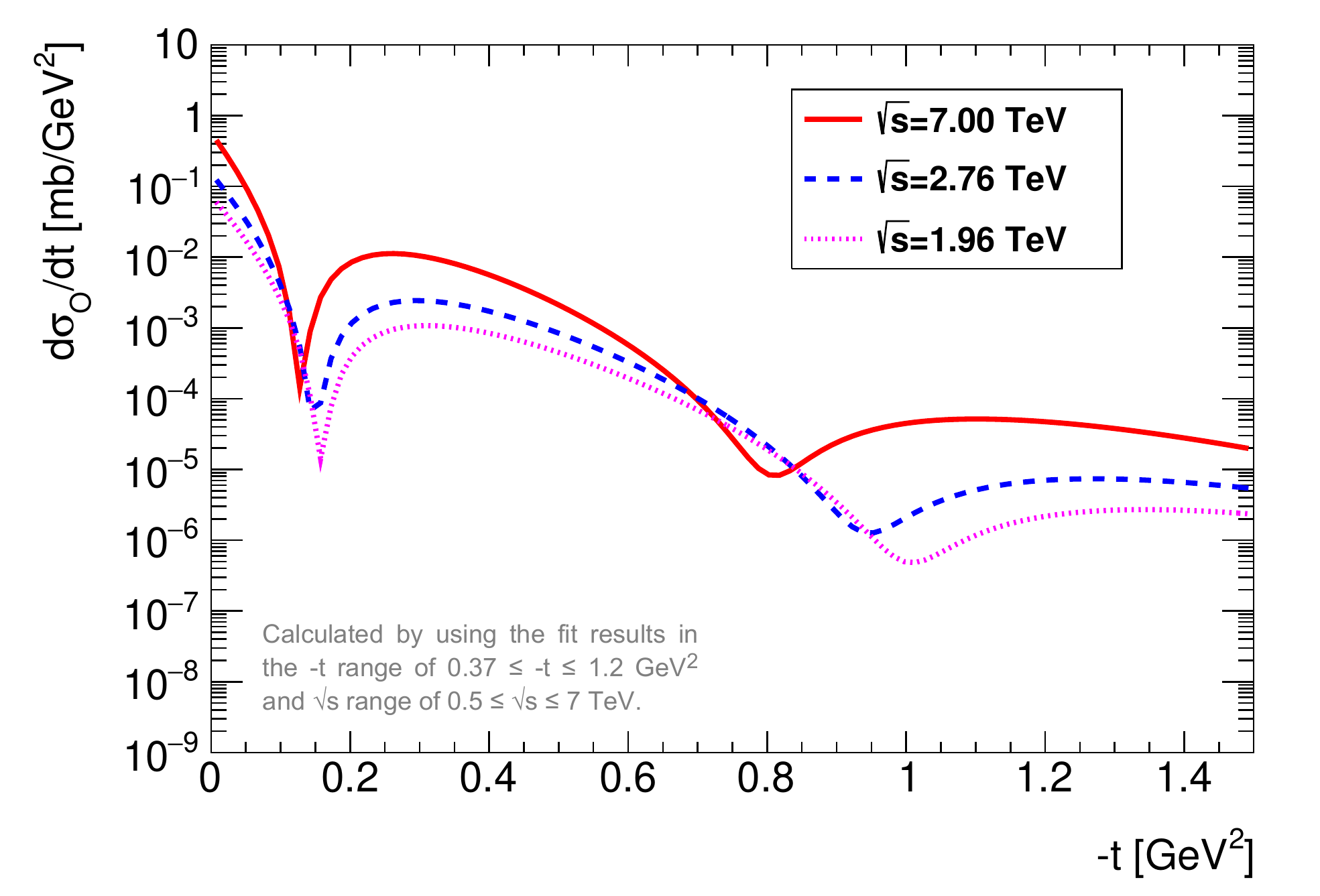}
	\includegraphics[width=0.8\linewidth]{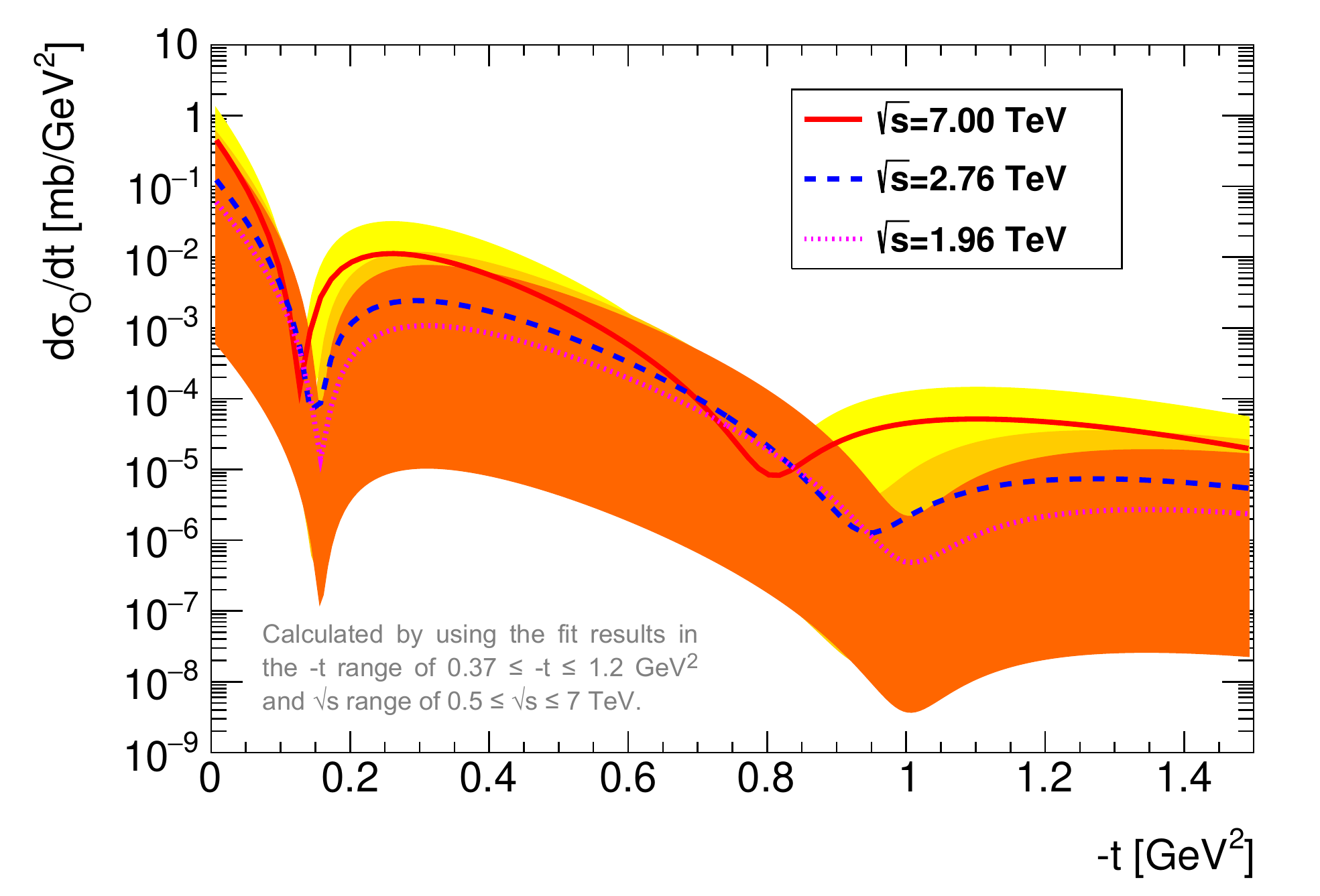}
	\caption{Same as Fig.~\ref{fig:differential-cross-section-for-Pomeron-exchange}, but for the differential cross-section of the Odderon exchange.
	}
	\label{fig:dsigmadt-for-Odderon}
\end{figure}

Fig.~\ref{fig:dsigmadt-for-Odderon} is the same as Fig.~\ref{fig:differential-cross-section-for-Pomeron-exchange}, but for the C-odd Odderon exchange as evaluated from the ReBB model of ref.~\cite{Nemes:2015iia}. The top panel indicates the central values for the differential cross-section of Odderon exchange at various colliding energies $\sqrt{s}$ in the TeV domain, while the lower panel includes our first estimates on the systematic errors of this reconstruction, obtained by neglecting the possibly strong correlations between the parameters $p_0$ and $p_1$. These figures also indicate that Odderon exchange may lead even to two pronounced diffractive minima, in contrast to the experimental results for the diffractive minimum in elastic $pp$ collisions. However, the interference between the Pomeron and the Odderon exchange leads to a single well defined and experimentally resolvable diffractive minimum in elastic $pp$ collisions at the TeV scale.

\begin{figure}[hbt]
	\centering
	\includegraphics[width=0.8\linewidth]{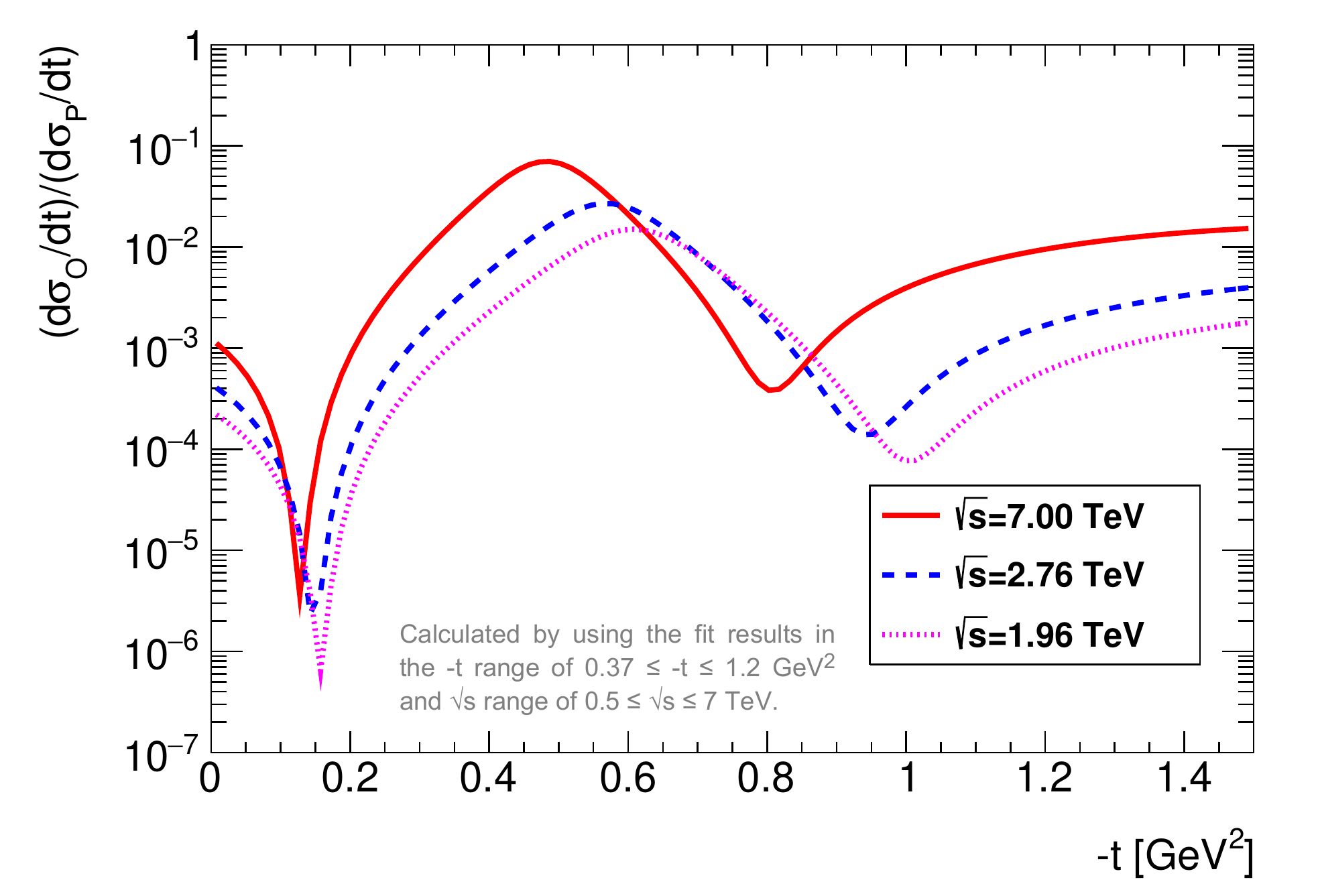}
	\includegraphics[width=0.8\linewidth]{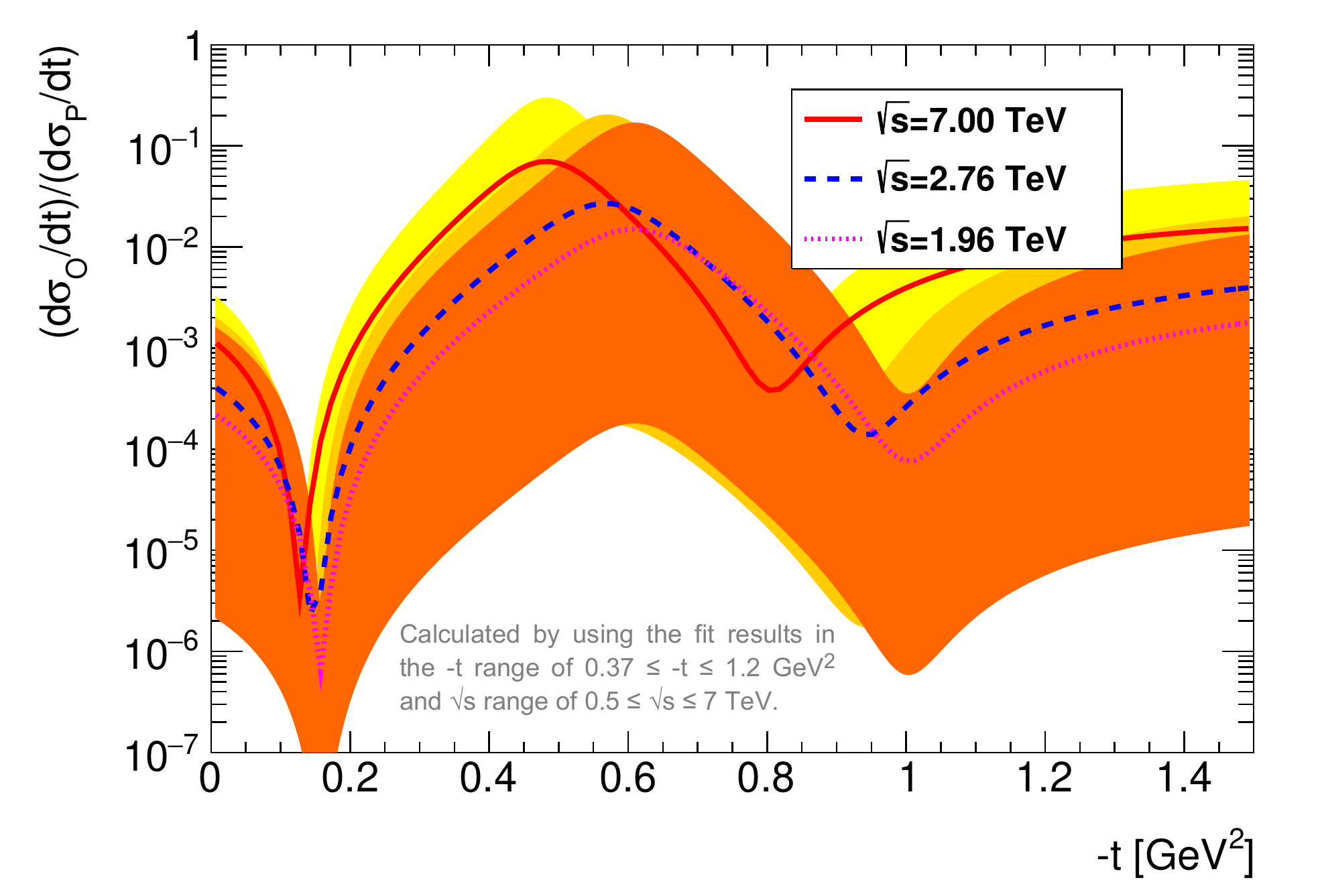}
	\caption{Same as Fig.~\ref{fig:differential-cross-section-for-Pomeron-exchange}, but for the ratio of the differential cross-sections of Odderon to Pomeron exchange. 
 }
	\label{fig:dsigma-dt-Odderon-to-Pomeron-ratio}
\end{figure}

Fig.~\ref{fig:dsigma-dt-Odderon-to-Pomeron-ratio} indicates the ratio of the differential cross-sections for Odderon to Pomeron exchange.
This figure indicates that the Odderon contribution is important and relatively large in three kinematic regions:
near to the $t=0$ optical point, near to the position of the diffractive minimum of elastic 
$pp$ collisions, $t_{ dip} \approx -0.5$ GeV$^2$, and then at higher squared momentum transfer values, $-t\gtrsim1$ GeV$^2$.
This figure also highlights with an explicit calculation, that the Odderon contribution to the dip region is correlated with the Odderon contribution at $ -t = 0$, thus the Odderon signals at the dip region appear simultaneously   with the Odderon signals at $-t = 0$ .

The last three figures characterize the modulus square of the amplitude for Pomeron and for Odderon exchanges in the ReBB model.
A very important information, however, is included to the phase of these amplitudes, that are shown on the subsequent two figures.

\begin{figure}[hbt]
	\centering
		\includegraphics[width=0.8\linewidth]{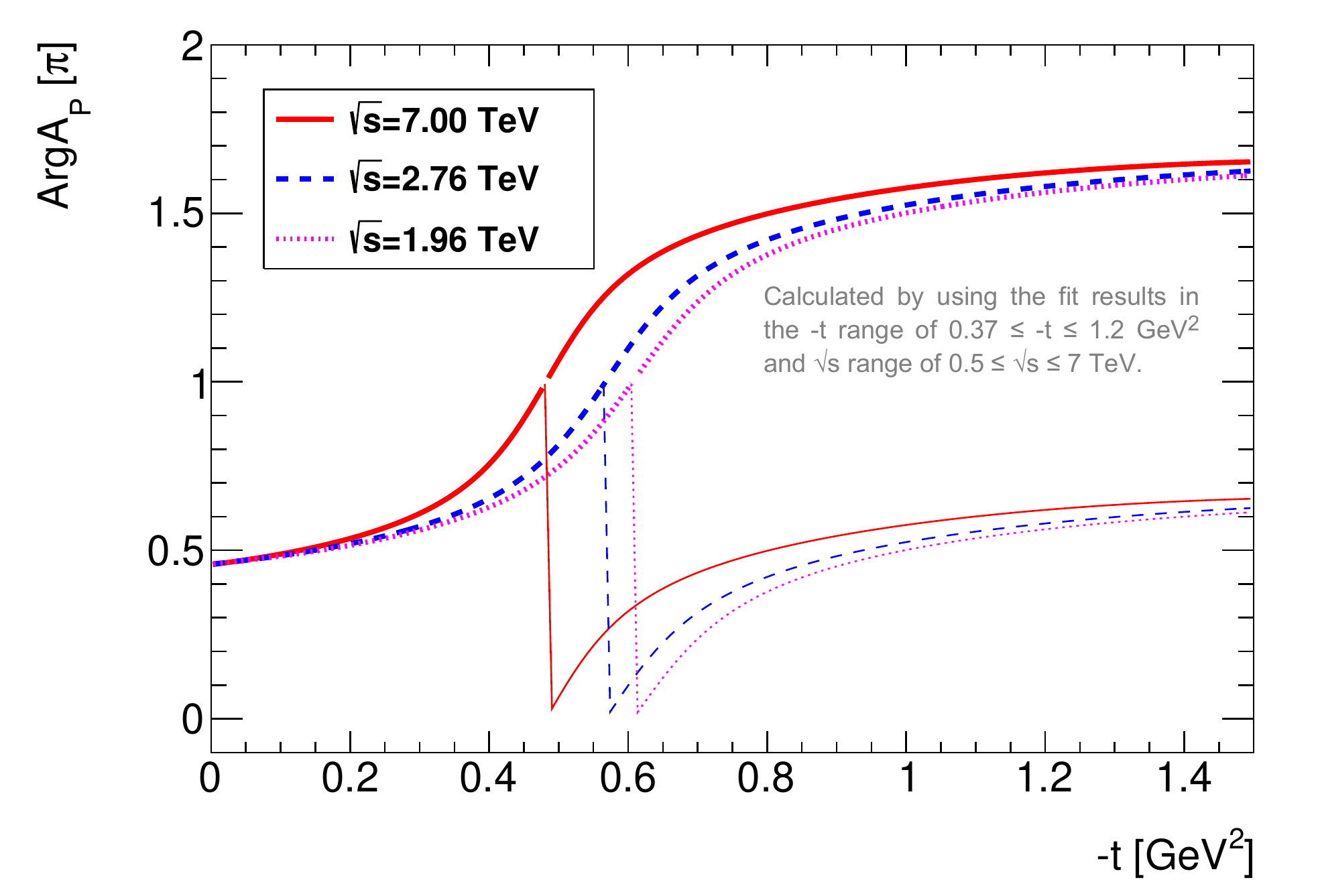}
	\includegraphics[width=0.8\linewidth]{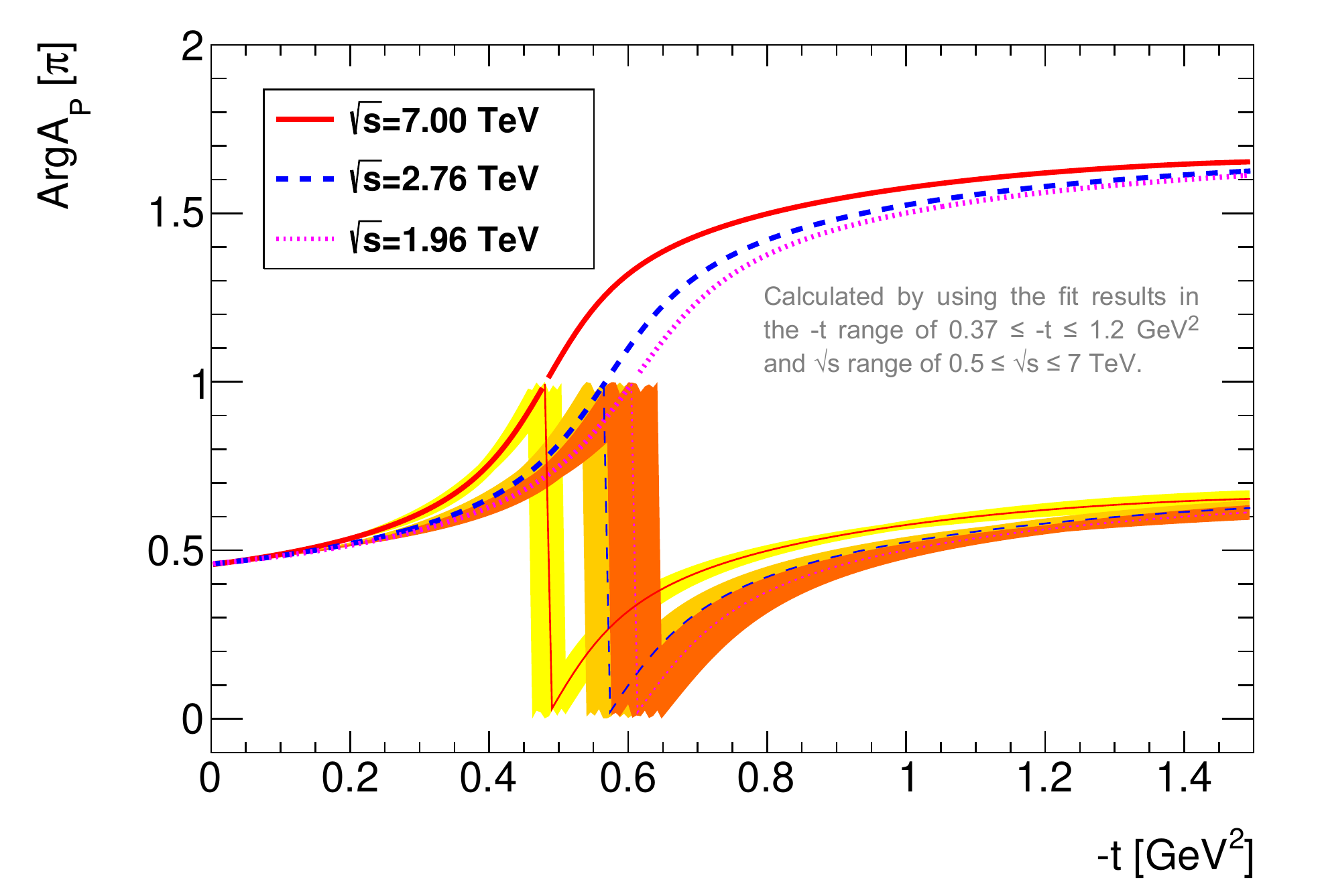}
	\caption{Same as Fig.~\ref{fig:differential-cross-section-for-Pomeron-exchange}, but for the phase of the amplitude of Pomeron exchange. 
	}
	\label{fig:phase-Pomeron-exchange}
\end{figure}

The phase of Pomeron exchange is indicated on Fig.~\ref{fig:phase-Pomeron-exchange}. This indicates that at low $-t$, the Pomeron contribution is predominantly imaginary, with a real component of the Pomeron exchange starting to be important near the diffractive minimum of elastic $pp$ collisions. On this plot, the principal value of the phase of the Pomeron (C-even) amplitude is indicated with a thin line, while the continuously varying phase
evaluated from the multi-valued inverse tangent function is shown with the thick line.

\begin{figure}[hbt]
	\centering
	\includegraphics[width=0.8\linewidth]{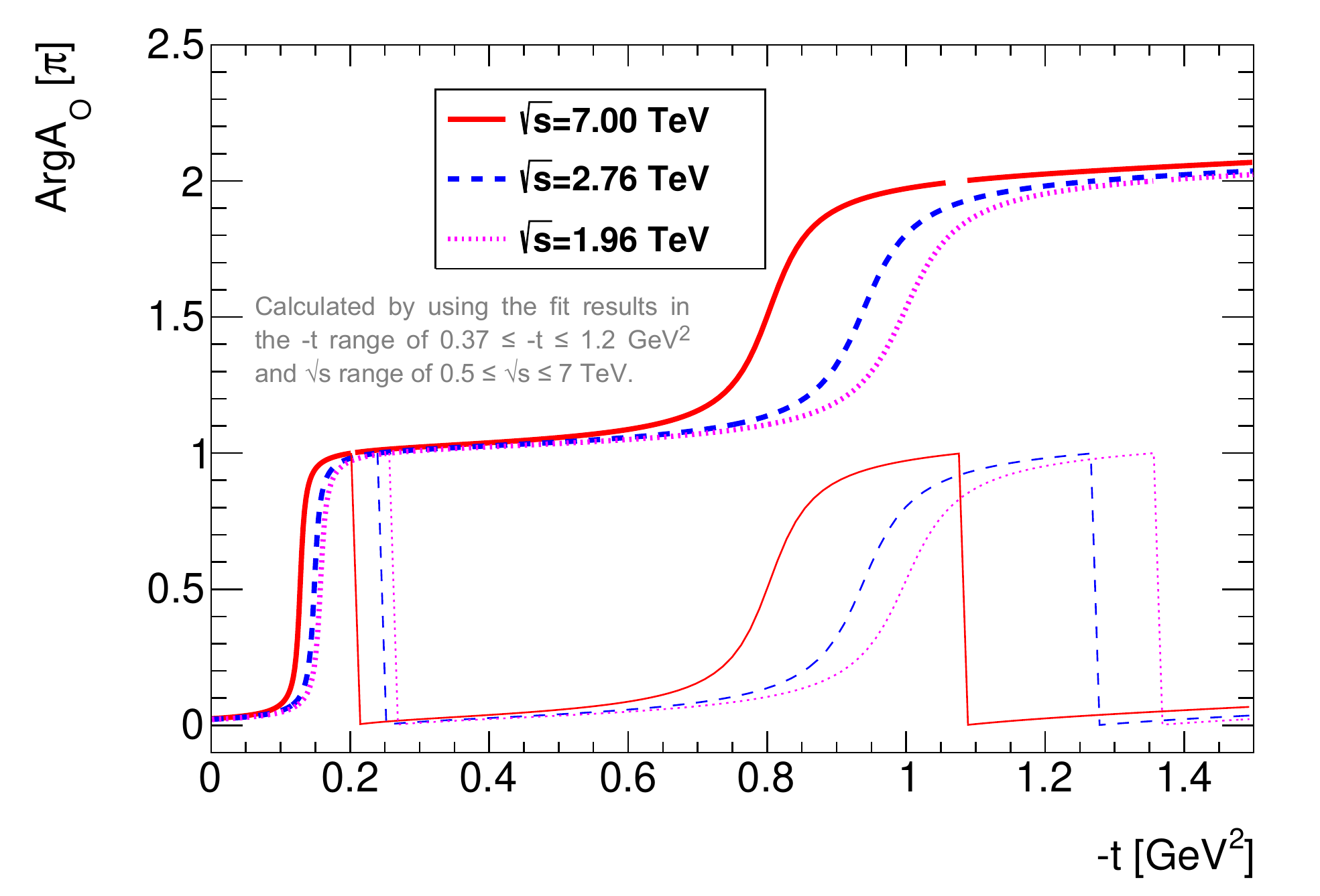}
	\includegraphics[width=0.8\linewidth]{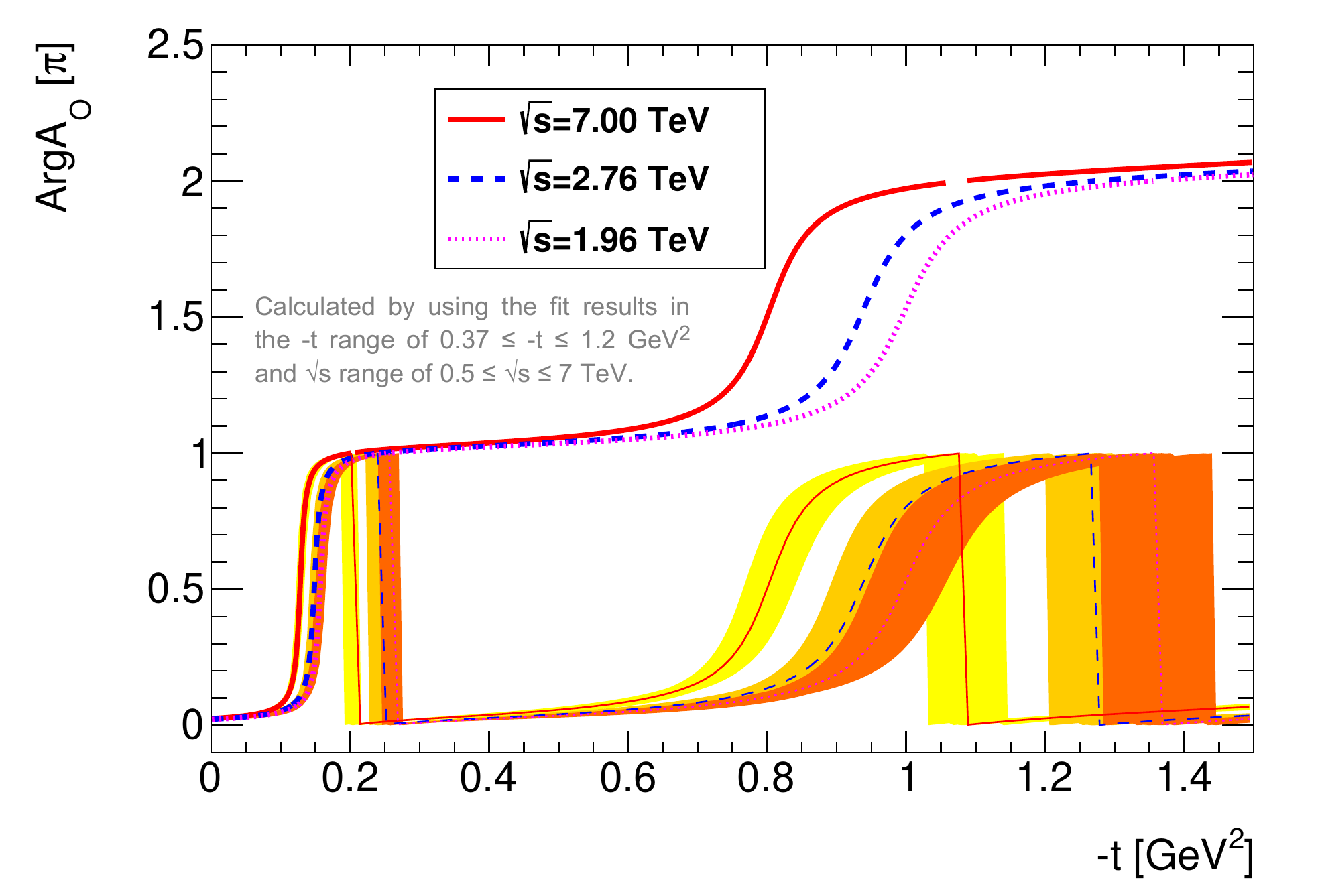}
	\caption{Same as Fig.~\ref{fig:differential-cross-section-for-Pomeron-exchange}
	but for the phase of the amplitude of  Odderon exchange.
	}
	\label{fig:phase-Odderon-exchange}
\end{figure}

The phase of Odderon exchange is indicated on Fig.~\ref{fig:phase-Odderon-exchange}. This indicates that at low $-t$, the Odderon contribution is predominantly real, with an imaginary component of the Odderon exchange starting to be important already at low $-t$ near to 0.1 GeV$^2$.
This phase starts to change quickly and the Odderon becomes predominantly real again near the diffractive minimum of elastic $pp$ collisions.
On this plot, the principal value of the phase of the Odderon (C-odd) amplitude is indicated with a thin line, while the continuously varying phase
evaluated from the multi-valued inverse tangent function is shown with the thick line.

\begin{figure}[hbt]
	\centering
	\includegraphics[width=0.8\linewidth]{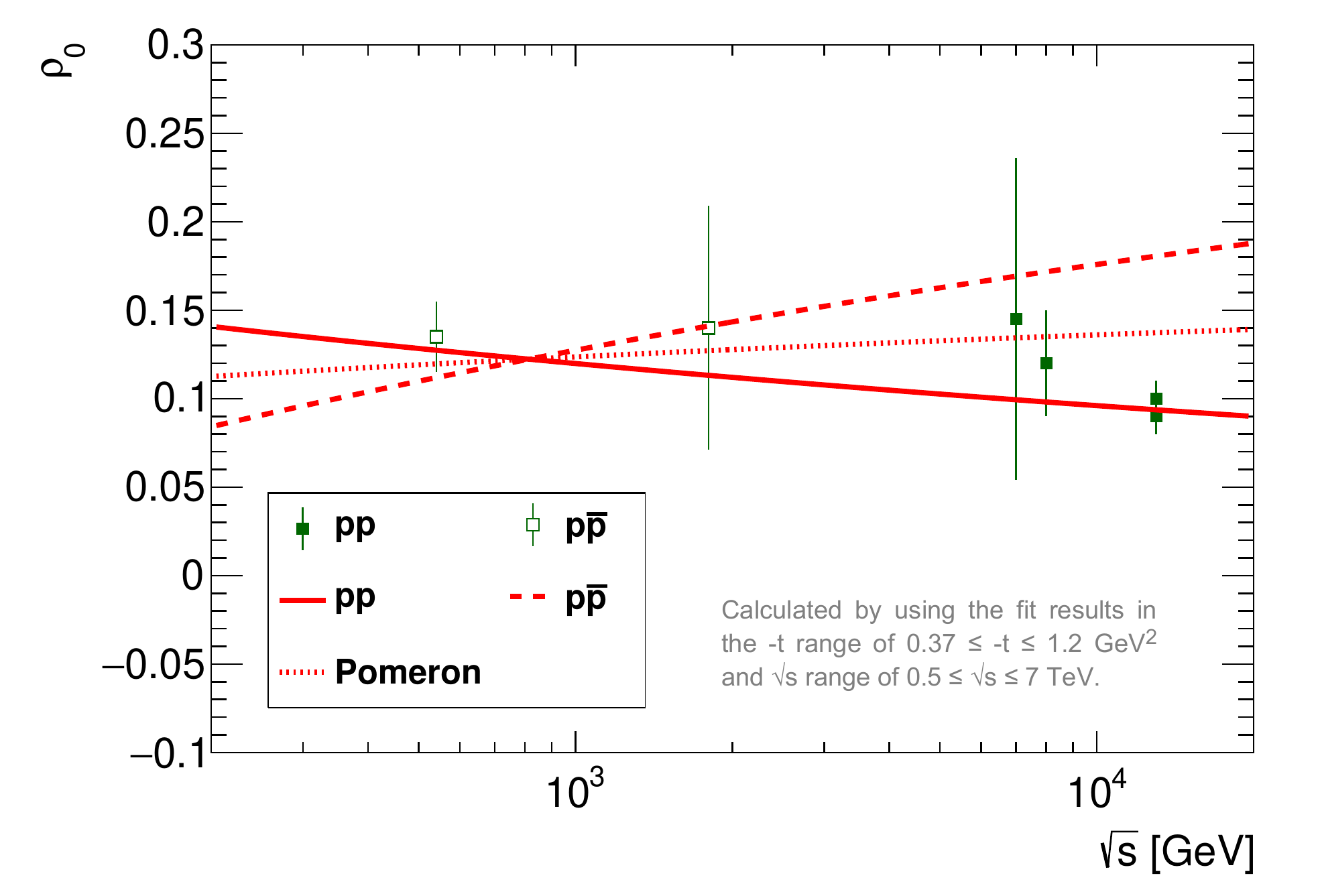}
	\caption{The 
	real to imaginary ratio $\rho_0(s)$ for elastic $pp$ and $p\bar p$ collisions and their Pomeron component 
	evaluated from  the log-linear excitation functions of the opacity parameters $\alpha^{pp}(s)$ and  $\alpha^{p\bar p}(s)$ as well as that of the scale parameters, $R_q(s)$, $R_d(s)$, $R_{qd}(s)$, corresponding to Table ~\ref{tab:excitation_pars}.
	}
		\label{fig:rho0-pp-pbarp-Pomeron}
\end{figure}

Fig. ~\ref{fig:rho0-pp-pbarp-Pomeron} indicates the value of the real to imaginary ratio of the scattering amplitude $\rho_0(s)$ for elastic 
proton-proton, proton-antiproton scattering and for Pomeron exchange. Near to the optical point, all of these amplitudes are predominantly imaginary, with a small real part and with an even smaller C-odd contribution, that makes the $\rho_0(s)$ different for elastic $pp$ and $p\bar p$
collisions, due to the contribution of the C-odd Odderon exchange.

\begin{figure}[hbt]
	\centering
	\includegraphics[width=0.8\linewidth]{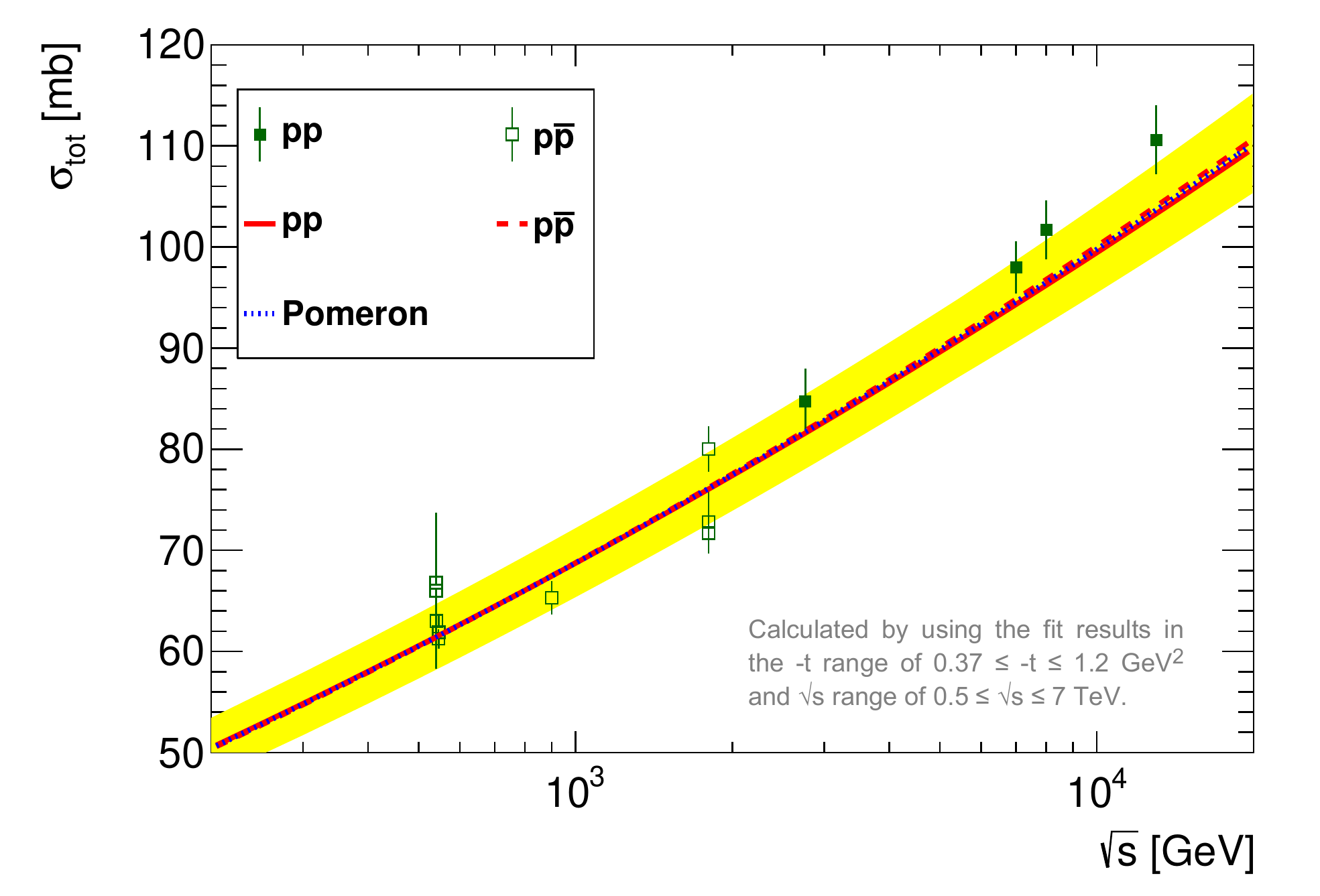}
	\caption{
	    Excitation function of the total cross-section for elastic $pp$, $p\bar p$ collisions and for the amplitude of Pomeron exchange, as
		evaluated from  the log-linear excitation functions of the opacity parameters $\alpha^{pp}(s)$ and  $\alpha^{p\bar p}(s)$ as well as that of the scale parameters, $R_q(s)$, $R_d(s)$, $R_{qd}(s)$, corresponding to Table ~\ref{tab:excitation_pars}. The yellow band indicates
		our conservative estimates on the systematic errors of the total cross-section of the Pomeron exchange. 
		}
	\label{fig:sigmatot-pp-pbarp-Pomeron}
\end{figure}

Fig.~\ref{fig:sigmatot-pp-pbarp-Pomeron} indicates the total cross-sections, as evaluated with the help of eq.~(\ref{eq:total_cross_section}), for
the elastic $pp$ and $p\bar p$ scattering as well as for the Pomeron exchange. The difference between the excitation functions for the total cross-sections of $pp$ and $p\bar p$ scattering seems to be less than the currently very small, of the order of 2 \% relative experimental error on the total cross-section measurements at LHC energies.

\begin{figure}[hbt]
	\centering
	\includegraphics[width=0.8\linewidth]{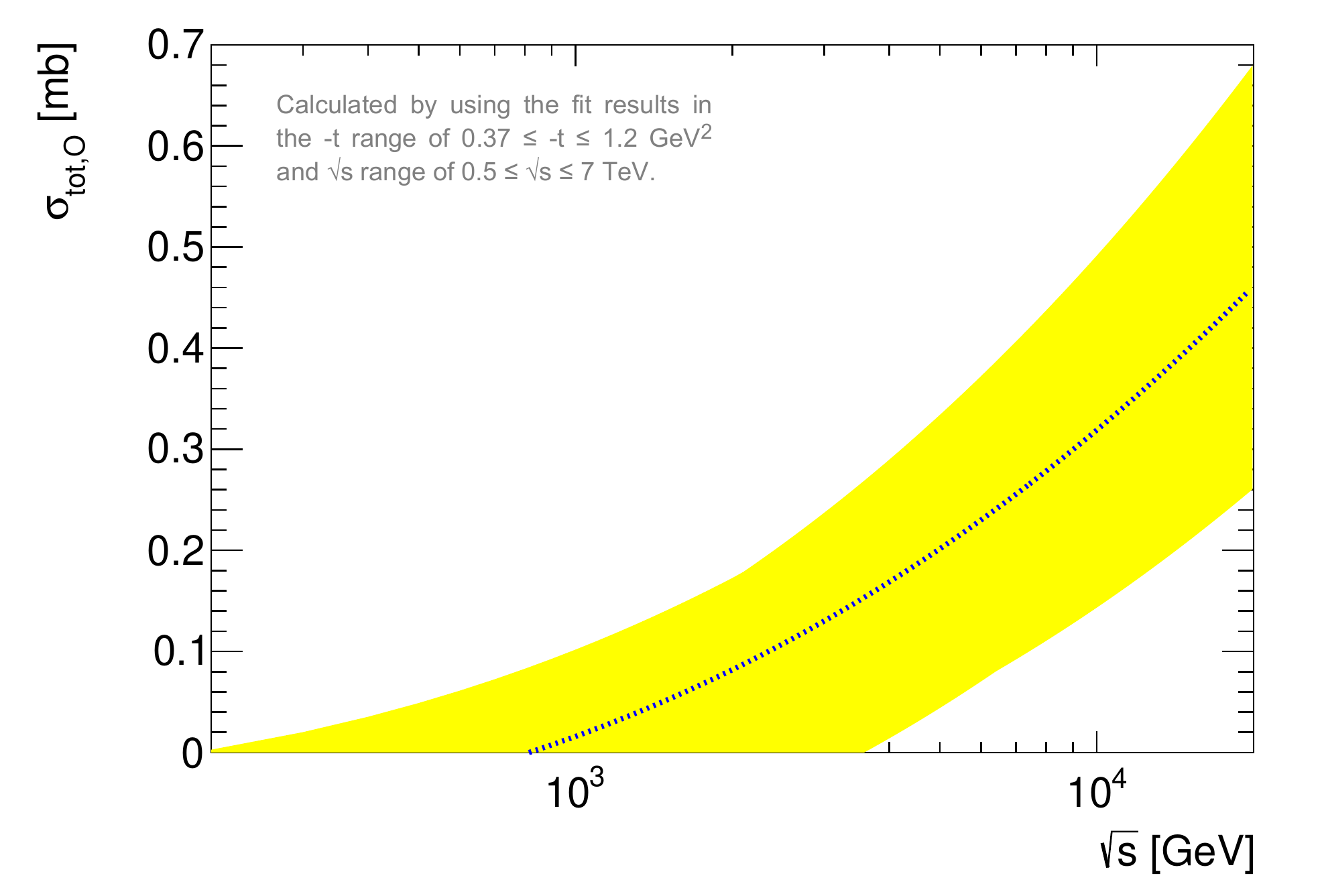}
	\caption{
        Excitation function of the total cross-section obtained from the optical theorem using the ReBB model amplitude of Odderon exchange, as
		evaluated from  the log-linear excitation functions of the opacity parameters $\alpha^{pp}(s)$ and  $\alpha^{p\bar p}(s)$ as well as that of the scale parameters, $R_q(s)$, $R_d(s)$, $R_{qd}(s)$, corresponding to Table~\ref{tab:excitation_pars}. The yellow band indicates
		our conservative estimate on the systematic errors of the total cross-section of this Odderon exchange. The result indicates that total cross-section of the Odderon exchange is sharply increasing in the few TeV energy range, but it is two orders of magnitude smaller than the contribution of the Pomeron exchange that is dominant at the same energy scale.
	}
	\label{fig:sigmatot-Odderon}
\end{figure}

Finally, Fig.~\ref{fig:sigmatot-Odderon} indicates the total cross-section corresponding to the Odderon component of the 
scattering amplitude, as evaluated with the help of eq.~(\ref{eq:total_cross_section}). This plot indicates that the Odderon cross-section
starts to increase in the $\sqrt{s} \ge 1$ TeV energy domain, but the total cross-section  of Odderon exchange
is at least two orders of magnitude smaller than the total cross-section for elastic $pp$ scattering in the TeV energy scale.
Actually we find $\sigma_{tot}^{\mathbb O} \leq 0.7 $ mb for $\sqrt{s} \leq 20 $ TeV.

Thus effectively, and within the framework of the ReBB model, we conclude that the Odderon occupies at least an order of magnitude smaller radius, as compared to the effective size of the Pomeron exchange. Thus we support the observations of ref.~\cite{Bartels:2019qho} and 
refs.~\cite{Gotsman:2018buo,Gotsman:2019dxj}, suggesting that the contribution of the Odderon exchange to the total $pp$ cross-section is rather small, of the order of 1 mb or less, even at the currently available largest LHC energies. Nevertheless, we also find that this currently rather small effect is
statistically significant, with a significance that is larger than the discovery treshold of $5 \sigma$, as detailed in the body of this manuscript.

\section{ISR energies and quadratic correctios to the excitation functions  \label{sec:appendix-d}}
\setcounter{equation}{0}

\begin{figure}[H]
	\centering
\includegraphics[width=0.8\linewidth]{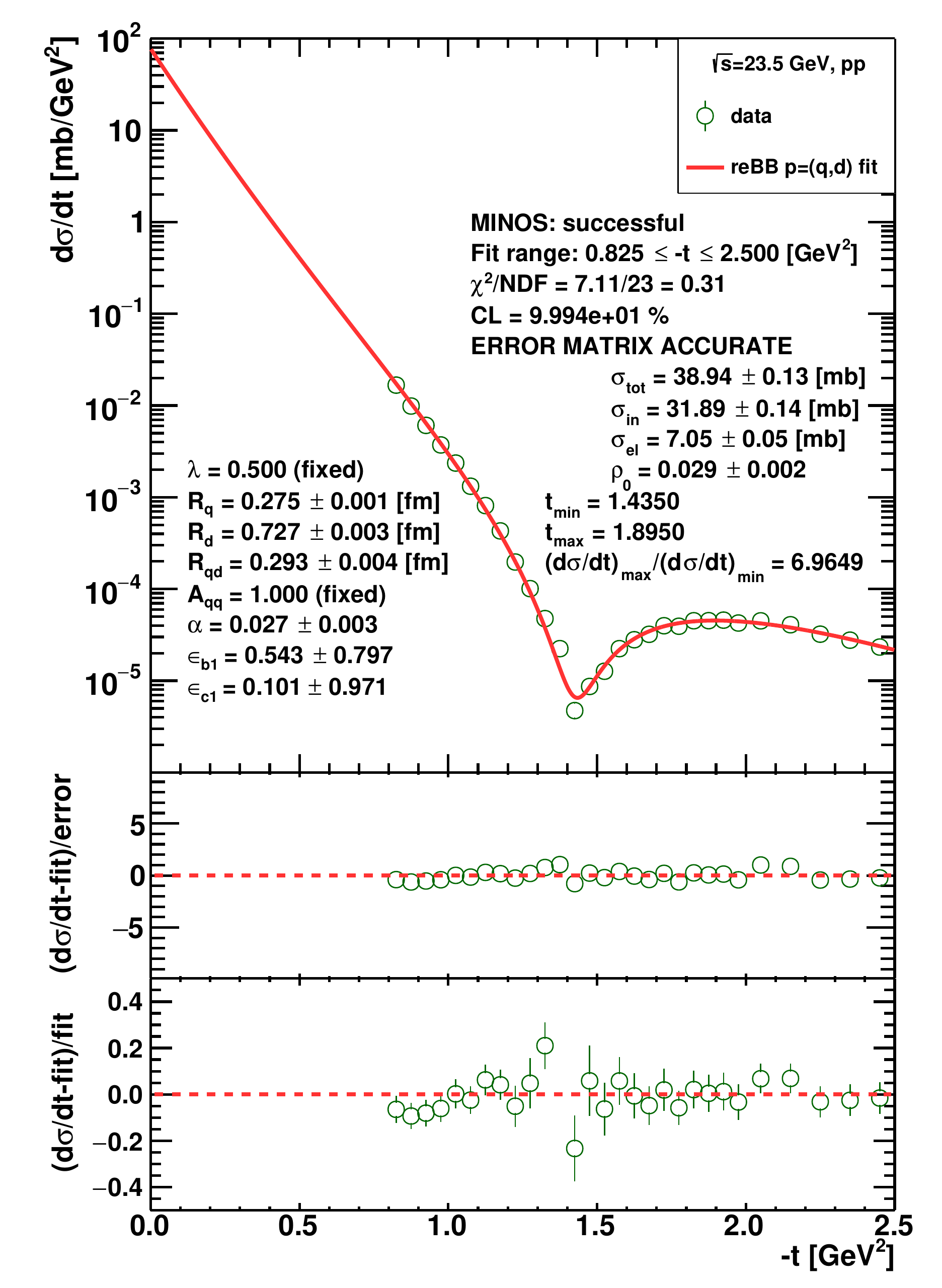}
	\caption{The fit of the ReBB model to the $pp$ ISR $\sqrt{s}=23.5$~GeV data in the range of $0.8\leq-t\leq2.5 $ GeV$^2$ \cite{Amaldi:1979kd}. The fit includes the $t$-dependent statistical (type A) and systematic (type B) uncertainties, the normalization (type C) uncertainty and the experimental values of the total cross section and parameter $\rho_0$ with their full error according to Eq.~(\ref{eq:chi2-final}). The fitted parameters are shown in the left bottom corner and their values are rounded up to three decimal digits.}
	\label{fig:reBB_model_fit_23_5_GeV}
\end{figure}

In this Appendix
we investigate the stability of the obtained linear logarithmic energy dependencies of the ReBB model parameters, discussed in Sec.~\ref{sec:excitation_functions}, for the case, when the energy range is extended towards lower values of $\sqrt{s}$. In order to do this, we refitted the ISR data \cite{Nagy:1978iw} at all the five available collision energy ($\sqrt{s}$ $=$ $23.5$, $30.7$, $44.7$, $52.8$ and $62.5$ GeV) in the squared momentum transfer range $0.8\leq-t\leq2.5 $ GeV$^2$ by using the $\chi^2$ definition determined by Eq.~\ref{eq:chi2-final}. The fits included the $t$-dependent (both vertical and horizontal) statistical (type A) and systematic (type B) errors, the normalization (type C) error and the experimental values of the total cross section and the parameter $\rho_0$ with their total uncertainties \cite{Amaldi:1979kd}. We have also tested the stability of the fit results for small variations of the fit range or the fitting method.
The only data set, where our results remained stable for the variation of the fit range around the selected range
and for small variations of the fitting procedure, and where the obtained results were both statistically and physically acceptable fit results describing not only the differential cross-section but the measured value of the total cross-section 
$\sigma_{tot}$ and the value of the real to imaginary ratio $\rho_0$ was the ISR dataset, measured at $\sqrt{s} = 23.5$ GeV. 
The result of this satisfactory fit is shown in Fig.~\ref{fig:reBB_model_fit_23_5_GeV}. Our other results were similar to
the results presented in ref.~\cite{Nemes:2015iia} and particularly resulted in a rather fluctuating description of the exctitation function of the $\alpha(s)$ at those ISR energies higher than 23.5 GeV. In the present study such fluctuating fits could not be used to establish the trends and the excitation functions.

\begin{figure}[H]
\centering
	\subfloat[Parameter $R_q$
	\label{fig:par_Rq_squa}]{%
		\includegraphics[width=0.5\linewidth]{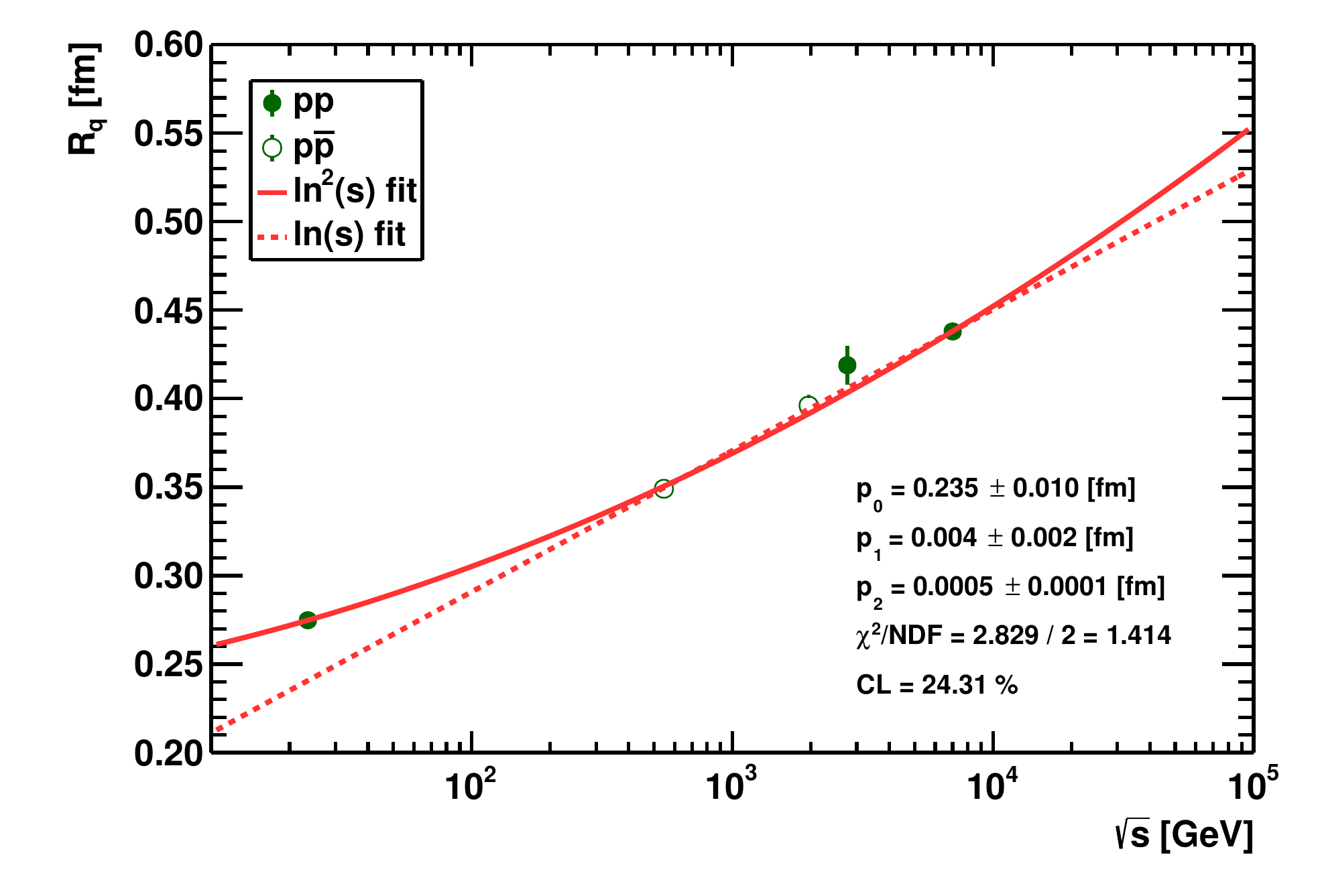}%
	}\hfill
	\subfloat[Parameter $R_d$\label{fig:par_Rd_squa}]{%
		\includegraphics[width=0.5\linewidth]{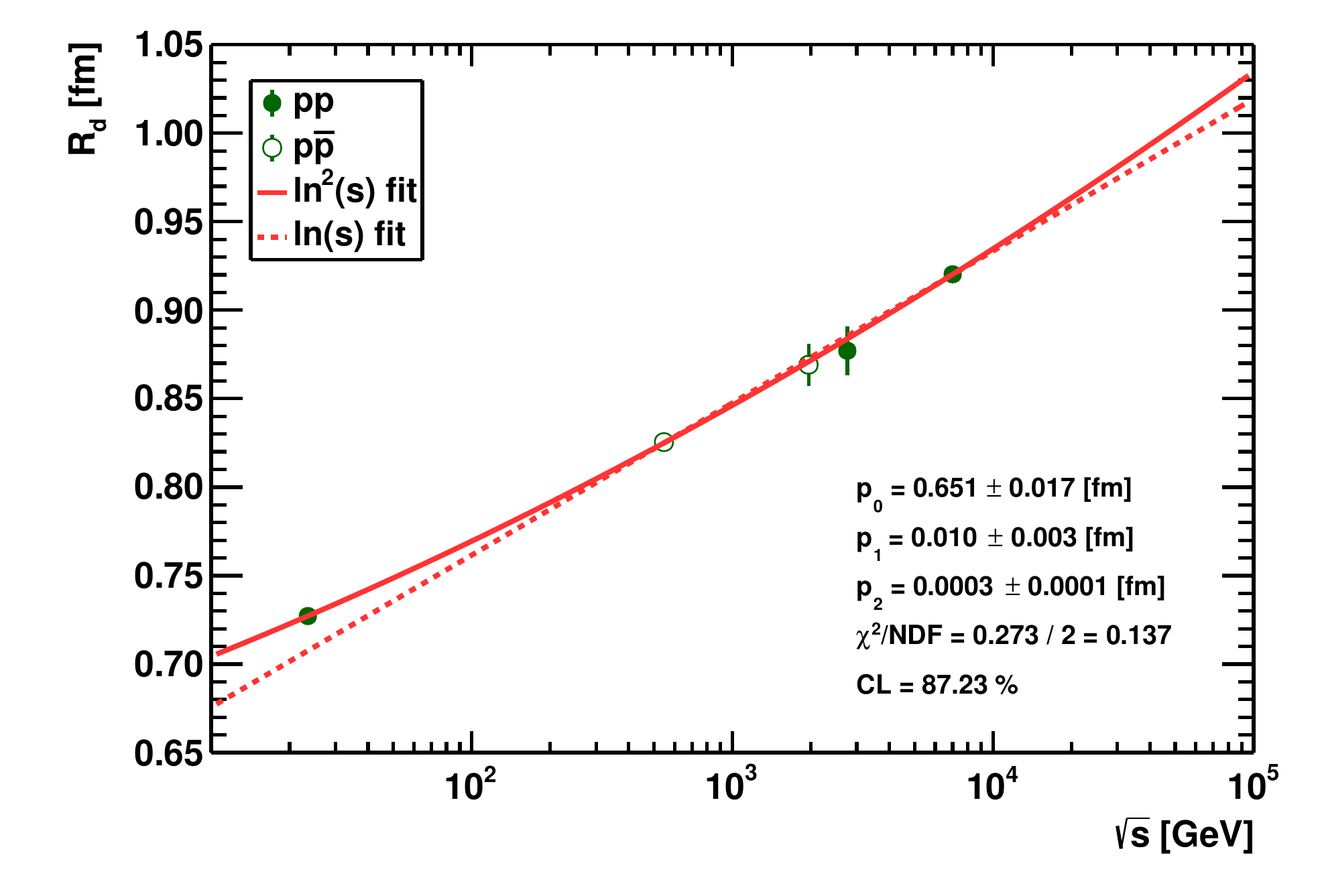}%
	}\hfill
	\subfloat[Parameter $R_{qd}$\label{fig:par_Rqd_squa}]{%
		\includegraphics[width=0.5\linewidth]{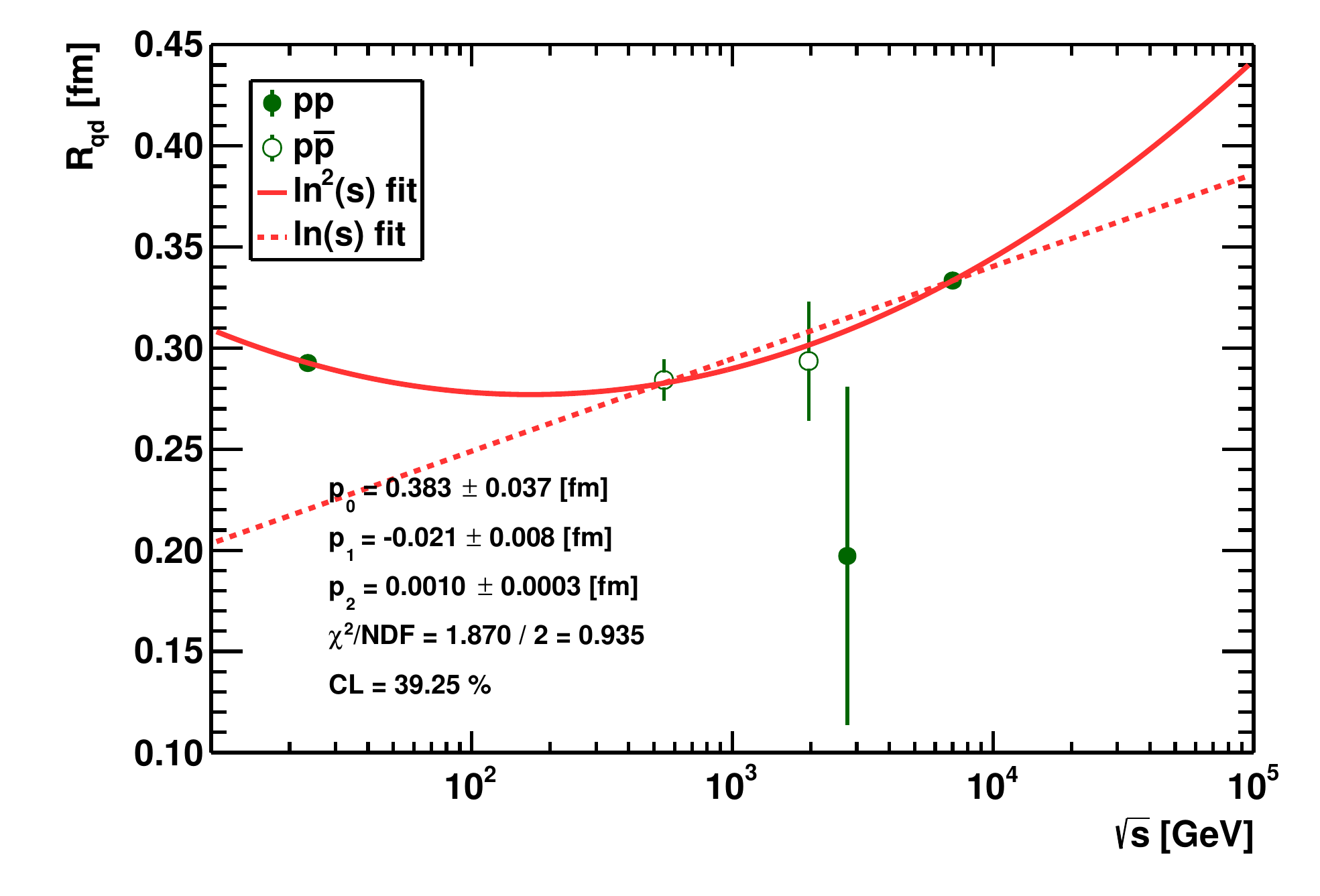}%
	}\hfill
	\subfloat[Parameter $\alpha$\label{fig:par_alpha_squa}]{%
		\includegraphics[width=0.5\linewidth]{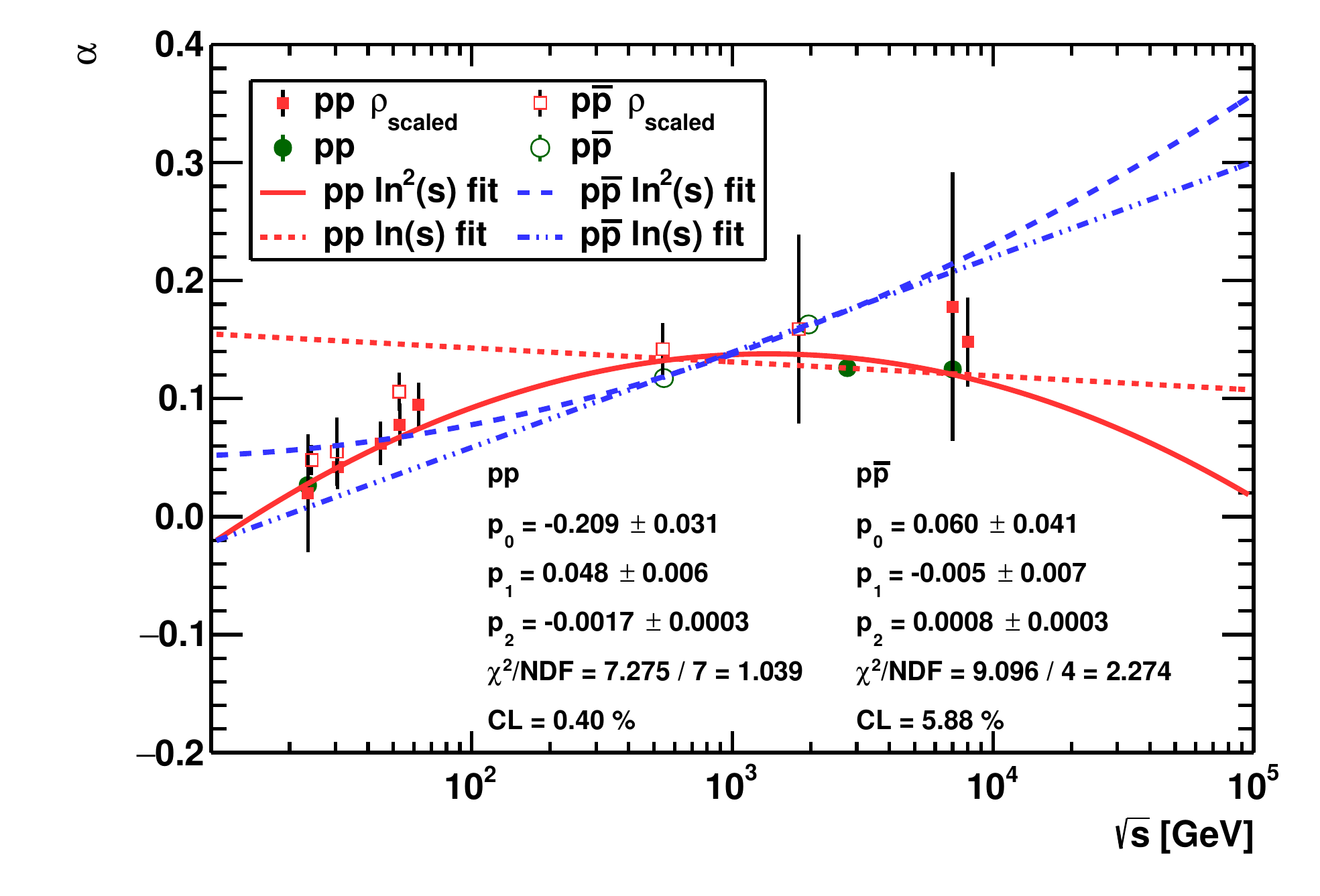}%
	}
	\caption{The energy dependence of the parameters of the ReBB model, $R_{q}$, $R_{d}$, $R_{qd}$ and $\alpha$, taken from Fig.~\ref{fig:reBB_model_fit_23_5_GeV} and Table~\ref{tab:fit_parameters}, determined by fitting a second order logarithmic polynomial, Eq.~(\ref{eq:parametrization_of_extrapolation_square}), to each of them one by one in the energy range of $23.5\leq\sqrt{s}\leq8000$ GeV. As a comparison these figures also show the result of the fit in the energy range of $546\leq\sqrt{s}\leq8000$ GeV with the linear logarithmic model determined by the parameters collected in Table~\ref{tab:excitation_pars}. It is clear that allowing for quadratic corrections
	does not change significantly the linear trends in the kinematic range of $0.5 \le \sqrt{s} \le 8$ TeV.
	\label{fig:reBB_model_log_square_extrapolation_fits}}
\end{figure}

\begin{figure}[H]
	\centering
	\includegraphics[width=0.8\linewidth]{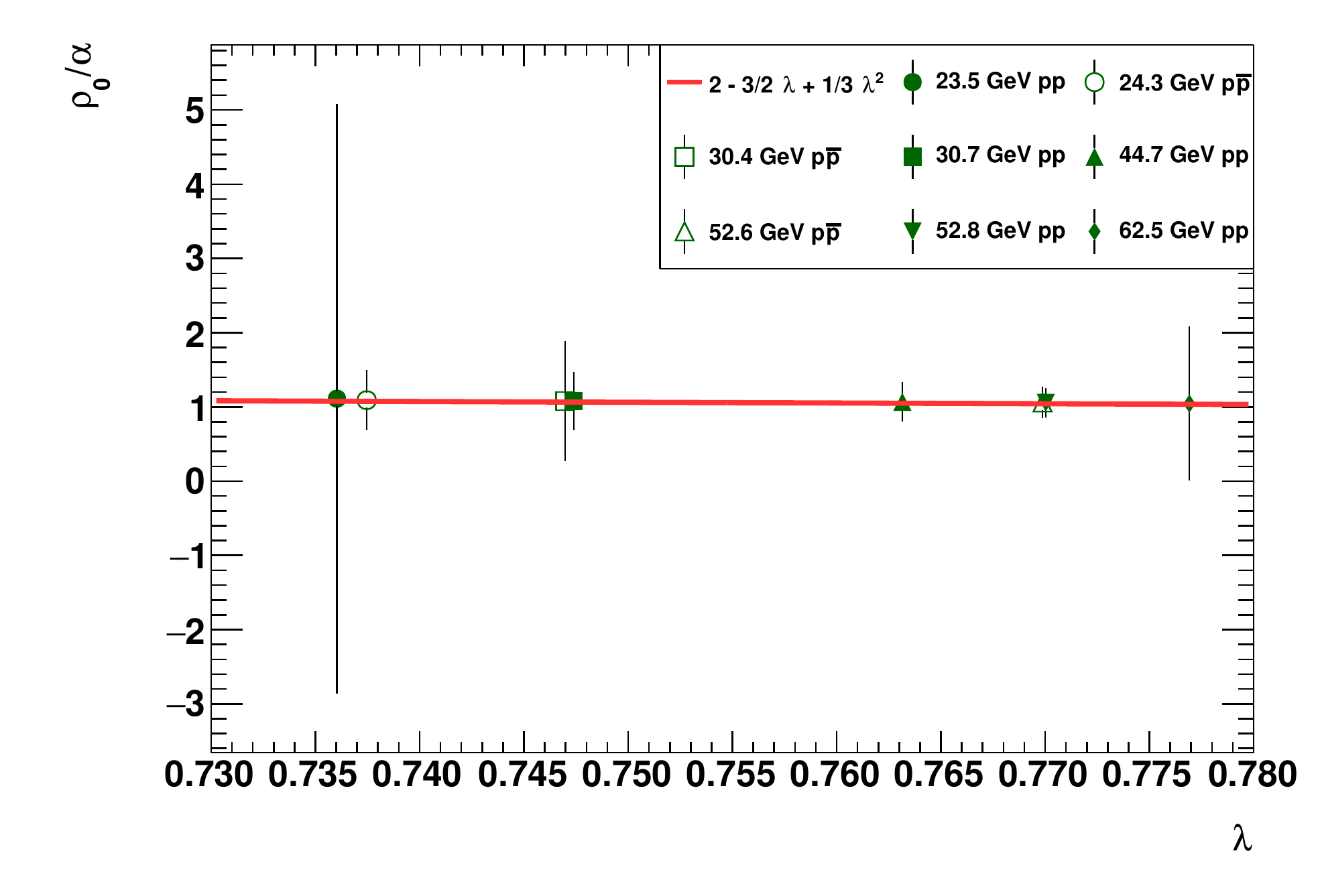}
	\caption{The dependence of $\rho_0/\alpha$ on $\lambda$ in the few tens of GeV energy range.
	Filled and empty symbols correspond to the $pp$ and $p\bar p$ cases, respectively.
	These values and the error-bars for  $\rho/\alpha$ are obtained from the ReBB model fits by using the excitation functions of the scale parameters $R_q(s)$, $R_d(s)$, $R_{qd}(s)$, shown in Figs.~\ref{fig:par_Rq_squa}-\ref{fig:par_Rqd_squa}  
	and summarized in Table~\ref{tab:excitation_pars}, as well as the experimentally measured ratio $\rho_0$ values. The red curve represents the analytic result, corresponding to 
	eq.~(\ref{eq:rho-vs-alpha-appendix}) in~\ref{sec:appendix-b},  showing a good agreement between the analytic considerations and the
	numerical results.}
		\label{fig:alpha-rho0-ISR}
\end{figure}

\begin{figure}[H]
	\centering
	\includegraphics[width=0.8\linewidth]{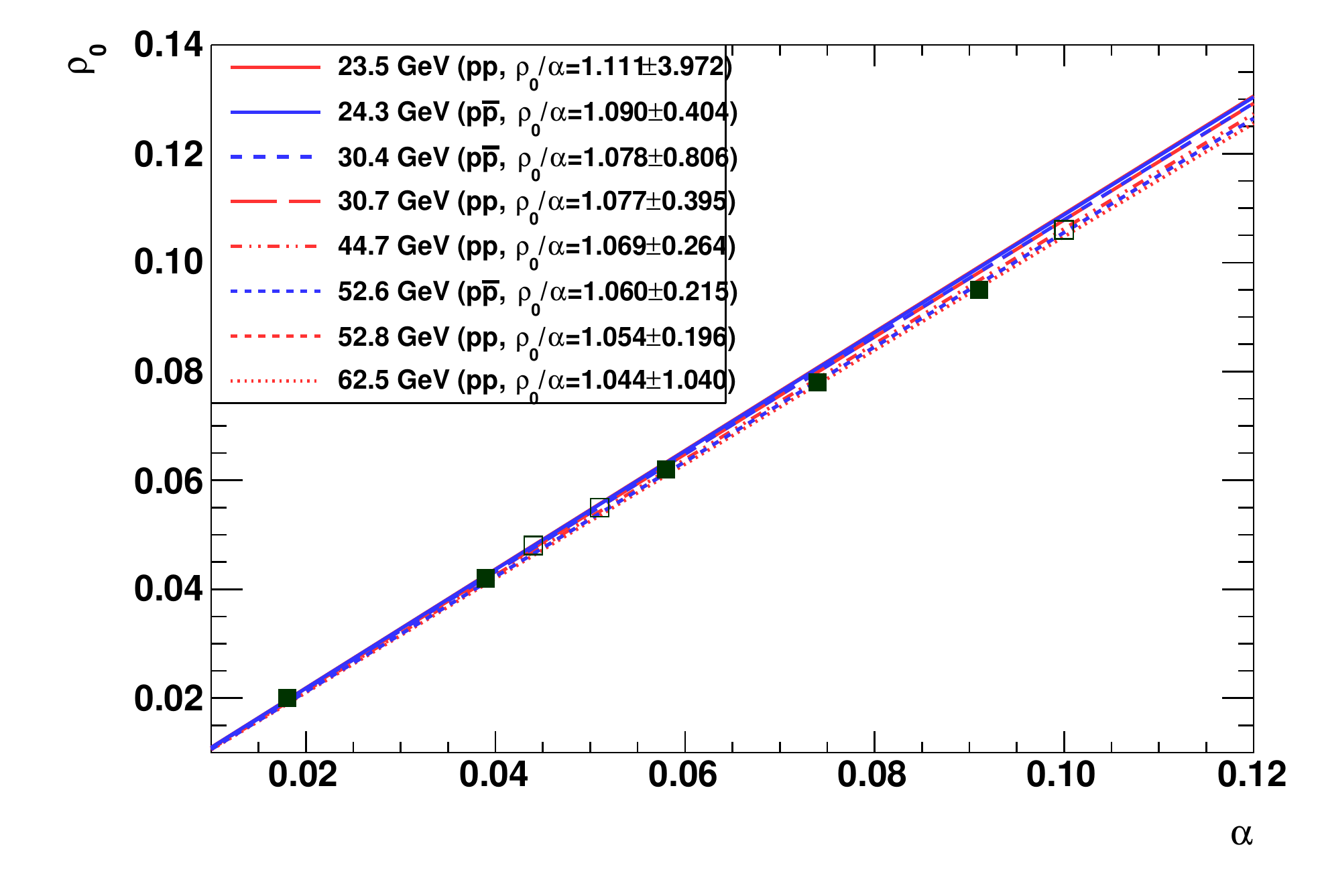}
	\caption{Linearity between the ratio $\rho_0$ and the $\alpha$ parameter of the ReBB model in the few tens of GeV energy region calculated from the trends of the scale parameters, $R_q(s)$, $R_d(s)$, $R_{qd}(s)$, corresponding to Figs.~\ref{fig:par_Rq_squa}-\ref{fig:par_Rqd_squa}. The square shaped markers in the figure are positioned to the experimentally measured $\rho_0$ values. In the ISR energy range, the ratio $\rho_0(s)/\alpha(s)$ is in an excellent agreement
	with the analytic approximations given by eq.~(\ref{eq:rho-vs-alpha-appendix}) of~\ref{sec:appendix-b}, as also illustrated on Fig.~\ref{fig:alpha-rho0-ISR}.
	}
	\label{fig:alpha-per-rho-ISR}
\end{figure}

Taking the restricted opportunities, we utilized the only reasonable ISR energy fit result, \textit{i.e.}, the result at 23.5 GeV to cross-check the compatibility of the linear logarithmic trends obtained in Sec.~\ref{sec:excitation_functions} with the lower energy region. When the $\sqrt{s} = 23.5 $ GeV energy data are included to those summarized in Table~\ref{tab:fit_parameters}, the energy dependence of the model parameters can be determined satisfactorily if model parameters are fitted one by one by applying a quadratic polynomial as a function of $\ln(s/s_0)$,
\begin{equation} 
P(s)=p_{0} + p_{1}\, \ln{(s/s_{0})} + p_{2}\, \ln^2{(s/s_{0})}, \ \ P\in{\{R_{q},R_{d},R_{qd},\alpha\}},
\label{eq:parametrization_of_extrapolation_square}
\end{equation}
where $p_0$, $p_1$, $p_2$ are free parameters and $s_{0}$ is fixed at 1 GeV$^{2}$. The obtained results are summarized in Fig.~\ref{fig:reBB_model_log_square_extrapolation_fits}. The parameters of the excitation functions are indicated on the subplots of Fig.~\ref{fig:reBB_model_log_square_extrapolation_fits}  and also summarized in Table~\ref{tab:excitation_pars_quadratic}. To fit the $\alpha$ parameter we used the same procedure described in Sec.~\ref{sec:excitation_functions}, \textit{i.e.}, utilizing also the measured and rescaled $\rho_0$ values. As seen in Figs.~\ref{fig:alpha-per-rho-ISR} and \ref{fig:alpha-rho0-ISR} the linear dependence of the ratio $\rho_0$ on the parameter $\alpha$ is satisfied at ISR energies as well.

In Fig.~\ref{fig:reBB_model_log_square_extrapolation_fits} the dotted curves show the result of the fits in the energy range of $546\leq\sqrt{s}\leq8000$ GeV with the linear logarithmic model determined by the parameters collected in Tab.~\ref{tab:excitation_pars} and discussed in Sec.~\ref{sec:excitation_functions}. Investigating Fig.~\ref{fig:reBB_model_log_square_extrapolation_fits}, one can conclude that although the energy dependence is not linear logarithmic if the data at the ISR energy region are included, the linear approximation in the energy region of $0.546\leq\sqrt{s}\leq8.0$ TeV is completely valid.

\clearpage

\begin{table*}[tbh]
\caption { Summary of the parameter values which determine the energy dependence according to the quadratic dependence in $\ln(s)$ by  Eq.~(\ref{eq:parametrization_of_extrapolation_square}). The values of the parameters are rounded up to three valuable decimal digits except for $p_2$ that are rounded up to four valuable decimal digits. These parameters 
 are also shown on the panels of Fig.~\ref{fig:reBB_model_log_square_extrapolation_fits} .
 For $R_q$, $R_d$ and $R_{qd}$, the values  of the parameters $p_0$, $p_1$ and $p_2$ are given in units of femtometers (fm). For the parameters 
 $\alpha(pp)$ and $\alpha(p\bar p)$, the parameters $p_0$, $p_1$ and $p_2$ are dimensionless.
}
\label{tab:excitation_pars_quadratic}
\begin{center}
	{\begin{tabular}{|c|c|c|c|c|c|} \hline
	Parameter      & $R_{q}$ [$fm$]  & $R_{d}$ [$fm$]  & $R_{qd}$ [$fm$]  & $\alpha$ ($pp$)      &$\alpha$ ($p\bar p$)  \\ \hline
	$\chi^{2}/NDF$ & $2.829/2$       & $0.273/2$       & $1.870/2$ 	      & $0.760/2$             &$1.212/2$        \\	
	CL [\%]		   & 24.31	         & 87.23           & 39.25 	          & 0.68                 &54.54            \\	
	$p_{0}$ & $0.235\pm0.010$ & $0.651\pm0.017$ & $0.383\pm0.037$  &                       $-0.209\pm0.031$      & $0.060\pm 0.041$  \\ 
	$p_{1}$ & $0.004\pm0.002$ & $0.010\pm0.003$ & $-0.021\pm0.008$  &                     $0.048\pm0.006$     &$-0.005\pm0.007$ 	\\   
	$p_{2}$ & $0.0005\pm0.0001$ & $0.0003\pm0.0001$ & $0.0010\pm0.0003$  &                 $-0.0017\pm0.0003$     &$0.0008\pm0.0003$ 	\\   \hline
	\end{tabular}}
	\end{center}
\end{table*}

\clearpage

\bibliographystyle{utcaps}
\bibliography{mybibfile}
\end{document}